\documentclass[structabstract]{aa}

\usepackage{lmodern}

\makeatletter
  \newcommand{\miniscule}{\@setfontsize\miniscule{5}{6}}
\makeatother

\usepackage{subcaption}
\usepackage{graphicx}
\usepackage{txfonts}
\usepackage{txfonts}
\usepackage{lscape}
\newcommand{\lsim}{\ \raise -2.truept\hbox{\rlap{\hbox{$\sim$}}\raise
    5.truept\hbox{$<$}\ }} \newcommand{\gsim}{\ \raise
  -2.truept\hbox{\rlap{\hbox{$\sim$}}\raise 5.truept\hbox{$>$}\ }}
\newcommand{\nodata}{...}  

\newcommand{\ui}{$(u{-}i)$\ }
\newcommand{\ik}{$(i{-}K_s)$\ }
\newcommand{\feh}{$[Fe/H]$\ }

\begin{document} 

   \title{A VST and VISTA study of globular clusters in NGC\,253}

   \subtitle{}

   \author{Michele Cantiello\inst{1}\fnmsep\thanks{This work is based
       on observations taken at the ESO La Silla Paranal Observatory
       within the VST Science Verification Programme ID 60.A-9286(A)
       and VISTA Science Verification Programme ID 60.A-9285(A)}
Aniello Grado\inst{2}\and
Marina Rejkuba\inst{3}
Magda Arnaboldi\inst{3}
Massimo Capaccioli\inst{4}\and
Laura Greggio\inst{5}\and
Enrica Iodice\inst{2}\and
Luca Limatola\inst{2}
   }

\institute{INAF Osservatorio Astronomico di Teramo, via Maggini, I-64100,
 Teramo, Italy \email{cantiello@oa-teramo.inaf.it}
 \and INAF Osservatorio Astronomico di Capodimonte Napoli, Salita Moiariello, I-80131, Napoli, Italy
  \and European Southern Observatory, Karl-Schwarzschild-Str. 2, D-85748, Garching bei M\"unchen, Germany
  \and Dip. di Fisica, Universit\'a di Napoli Federico II, C.U. di Monte  Sant'Angelo, Via Cintia, I-80126 Napoli, Italy
 \and  INAF, Osservatorio Astronomico di Padova, Vicolo dell’Osservatorio 5, I-35122, Padova, Italy}

  \date{}
               
   \date{Received ; accepted }

\abstract
{Globular clusters (GCs) are key to our understanding of the Universe,
  as laboratories of stellar evolution, fossil tracers of the past
  formation epoch of the host galaxy, and effective distance indicators
  from local to cosmological scales.}
{We analyze the properties of the sources in the NGC\,253 with the aim
  of defining an up to date catalog of GC candidates in the
  galaxy. Given the distance of the galaxy, GCs in NGC\,253 are
  ideal targets for resolved color-magnitude diagram studies of
  extragalactic GCs with next-generation diffraction limited
  ground-based telescopes.}
   {Our analysis is based on the science verification data of two ESO
     survey telescopes, VST and VISTA. Using $ugri$ photometry from VST and
     $JK_s$ from VISTA, GC candidates were selected using as reference
     the morpho-photometric and color properties of spectroscopically
     confirmed GCs available in the literature. The strength of the
     results was verified against available archival HST/ACS data from
     the GHOSTS survey: all but two of the selected GC candidates
     appear as star clusters in HST footprints.}
{The adopted GC selection leads to the definition of a sample of
  $\sim350$ GC candidates. At visual inspection, we find that 82 
    objects match all the requirements for selecting GC candidates and
  155 are flagged as uncertain GC candidate; however, 110 are unlikely
  GCs, which are most likely background galaxies. Furthermore, our
    analysis shows that four of the previously spectroscopically
    confirmed GCs, i.e., $\sim20\%$ of the total spectroscopic sample,
  are more likely either background galaxies or high-velocity Milky
  Way stars. The radial density profile of the selected best
  candidates shows the typically observed $r^{1/4}$-law radial
  profile. The analysis of the color distributions reveals only
    marginal evidence of the presence of color bimodality, which is normally
    observed in galaxies of similar luminosity. The GC luminosity
  function does not show the typical symmetry, mainly because of the
  lack of bright GCs. Part of the bright GCs missing might be at very large
  galactocentric distances or along the line of sight of the galaxy
  dusty disk. As an alternative possibility, we speculate that a
  fraction of low luminosity GC candidates might instead be metal-rich,
  intermediate age clusters, but fall in a similar
  color interval of old, metal-poor GCs.}
{Defining a contaminant-free sample of GCs in extragalactic systems is
  not a straight forward exercise. Using optical and near-IR
  photometry we purged the list of GCs with spectroscopic membership
  and photometric GC candidates in NGC 253. Our results show
  that the use of either spectroscopic or photometric data only does not
  generally ensure a contaminant-free sample and a combination of both
  spectroscopy and photometry is preferred.}


   \maketitle
%

\section{Introduction}

Globular clusters (GCs) are a key tool for understanding the formation
and evolution of galaxies
\citep{harris01,brodie06,peng08,georgiev10,harris13,durrell14}.

Extragalactic, unresolved GCs are possibly the simplest class of
astrophysical objects beyond stars as, to a first approximation, they
host a simple, single age and single metallicity, stellar
population. Nevertheless, in the last decade such a classical paradigm
has been demonstrated as invalid for a fraction of Milky Way (MW) GCs
and for some of the clusters in the Magellanic Clouds
\citep{gratton04,piotto08,carretta09}. In spite of that, doubtless GCs
are the simplest stellar aggregates to be found in galaxies.

Two other properties make GCs very useful for extragalactic studies:
old age and high luminosity.  In the few galaxies, beyond the MW,
where spectroscopic or multiband photometric studies of the GCs have
been carried out, the results almost uniformly revealed a population
with mean ages comparable to the GC system of the MW, older than
$\sim$10 Gyr
\citep[e.g.,][]{cohen98,cohen03,strader05,chiessantos11}. This makes
GCs fossil tracers of the formation of the host galaxy. In addition,
extragalactic GCs appear as bright clumps of light on the otherwise
smooth light profile of the galaxy and, under typical observing
conditions from the ground, they appear as point-like sources.

The compactness and high contrast with respect to the background light
from galaxy and sky, make GCs observable out to large
distances. Photometric studies with the Hubble Space Telescope (HST) have
been carried out for GC systems at $z\sim0.2$ \citep[$d\sim800$ Mpc,
  ACS data,][]{alamo13} and at $z\sim0.3$ \citep[$d\sim1250$ Mpc,
  ACS and WFC3 data,][]{janssens17}. Furthermore, very recently
\citet{vanzella17} have reported the discovery of several compact extremely
young objects at redshift $z>3$, observed with Very Large Telescope/Multi Unit
Spectroscopic Explorer (VLT/MUSE), some of which have inferred physical
properties consistent with those expected in proto-GCs.

Beyond their role as laboratories for the validation and calibration
of stellar evolution from the data of the closest resolved GCs
\citep{salaris05}, the systematic study of extragalactic GC systems
has revealed a wealth of properties used to trace the physical
characteristics of the GC system and the host galaxy; for example, the luminosity
function of GCs (GCLF), spatial distribution, projected
surface density, radial color profiles, specific frequency,
kinematical properties, and color-magnitude relations  are
effective tracers of the past formation and evolution history of the
galaxy, its physical distance, possible merging events, mass
distribution, etc. \citep{harris01,brodie06}.

Studies of extragalactic GCs based on photometry are hampered by a
varying level of contamination from foreground MW stars and background
galaxies. On the other hand, spectroscopic observations can provide
cleaner GC samples, but they are feasible only for the nearest,
brightest GC systems, even with the largest 8-10 m class telescopes
\citep{brodie14}.

In this paper we use the science verification imaging data in the
$ugriJK_s$ bands, taken with the VST and VISTA ESO survey telescopes,
to obtain an updated catalog of GCs in NGC\,253, thereby taking
advantage of the large wavelength range covered.

The galaxy, the brightest in the Sculptor group
\citep{karachentsev03}, at a distance of $\sim3.5$ Mpc \citep{rs11},
is an edge-on disk galaxy and, together with M\,82, is one of the two
brightest, closest starburst galaxies \citep{mccarty87}. Two different
spectroscopic studies, by \citet{beasley00} and by \citet{olsen04},
have targeted the GC system of the galaxy, leading to a total of 21
spectroscopically confirmed GCs, which we use as reference sample.
Other photometric studies exist in the literature
\citep[][]{liller83,blecha86}. The advantage of the catalog we present
in this work is the wider wavelength coverage and larger angular
coverage of the data analyzed. \citet{galleti04} announced 380
globular cluster candidates in NGC\,253 selected with photometric
data. The follow-up study based on VIMOS/VLT spectroscopic data is
still unpublished. In general, both spectroscopic and photometric
studies of GCs in the Sculptor group are consistent with low
metallicity, \feh$\lsim-1$, and with a cumulative luminosity function
(LF) consistent with the Milky Way. However, for NGC\,253, the
conclusions are tempered by the small sample size of confirmed GCs and
by the large contamination in photometric samples. We aim to remedy
that by using VST and VISTA high-resolution multiwavelength imaging
data.

\vskip +0.2cm
\noindent The paper is organized as follows: The next section
describes the observations, procedures for data reduction and for
deriving the photometry of the sources in the optical and near-IR
frames. The selection of GCs candidates is described in Section
\ref{sec_sel}.  Section \ref{sec_final} provides the final catalog and
a brief discussion of our results. A summary of the main results
closes the paper.

\section{Observations, data reduction, and analysis}
\label{sec_data}

The data presented in this paper were collected as part of science
verification of VST \citep{iodice12} and VISTA
\citep{arnaboldi10,arnaboldi12}. In particular we analyzed the $ugri$
data from VST and the $JK_s$ from VISTA.

The VST data were processed with VST-Tube \citep{grado12}, a pipeline
specialized for the data reduction of VST-OmegaCAM executing
prereduction (bias subtraction and flat normalization), illumination and
(for the $i$ band) fringe corrections, and photometric and astrometric
calibration. 

The data reduction for VISTA was carried out with the VISTA data flow
system at the Cambridge Astronomy Survey Unit \citep[see][for more
  details]{irwin04,greggio14,iodice14}.

\begin{figure}
  \centering
  \includegraphics[width=0.8\hsize]{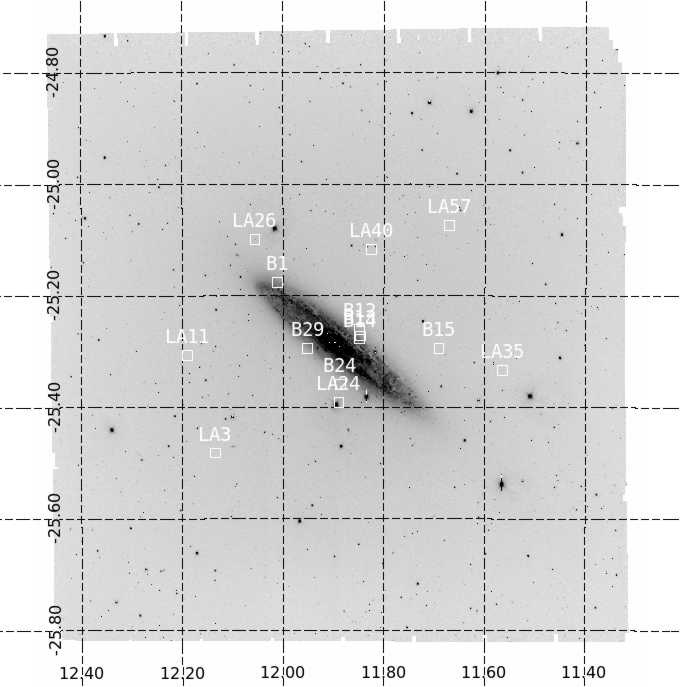}
   \caption{Full field of view observed with VST ($g$ band is shown,
     other bands are nearly identical). White squares indicate the 14
     spectroscopic confirmed GCs from \citet[][]{beasley00}, labeled
     as in Table 2 (Col. 6) in that paper.}
   \label{ngc253_g}
  \end{figure}

\begin{figure}
  \centering
  \includegraphics[width=0.8\hsize]{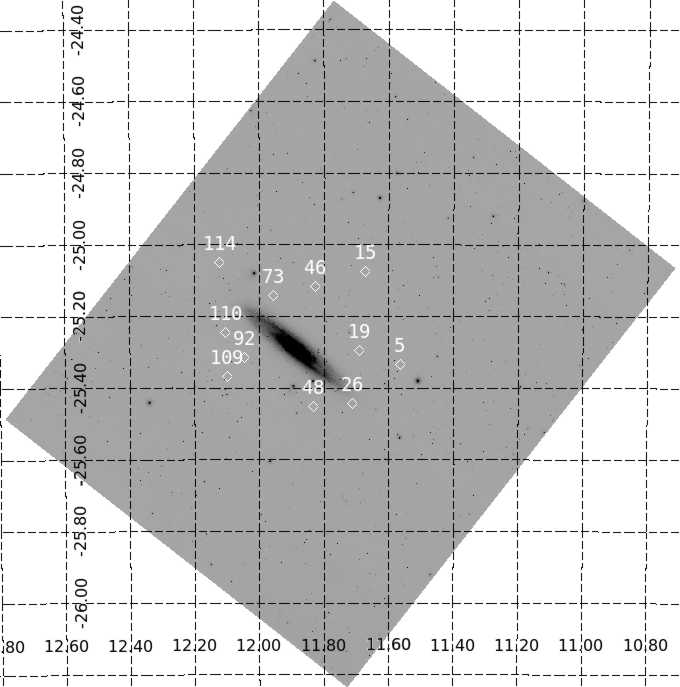}
   \caption{Full field of view observed with VISTA ($J$ band is shown,
     $K_s$ band is nearly identical). White diamonds indicate the 11
     spectroscopic confirmed GCs from \citet[][]{olsen04}, labeled
     as in Table 3 (Col. 1) in that paper.}
   \label{ngc253_j}
  \end{figure}

A summary of the observations is provided in Table \ref{tab_props},
together with a list of properties of the target galaxy.  The final
image size was $\sim1.04^{\circ}\times1.09^{\circ}$ for VST at
$0.21\arcsec/pixel$ resolution with the $ugri$ pointings centered on
the galaxy. For VISTA, the image size was
$\sim1.2^{\circ}\times1.5^{\circ}$, at $0.339\arcsec/pixel$ resolution
centered at R.A.=00:46:59 and Dec.=-25:17:26 with the photometric
center of the galaxy slightly offset eastward of the pointing center.
Figure \ref{ngc253_g} and Figure \ref{ngc253_j} show the full field of
view observed by VST and VISTA in $g$ and $J$ band, respectively. The
final combined field of view is of $\sim 1.05 \deg^2$, over which we
have photometry in all six bands.

We need to minimize contamination due to the light from the galaxy to
study GCs. To model and subtract the surface brightness profile of
NGC\,253, we used the ISOPHOTE/ELLIPSE task in IRAF/STSDAS
\citep{jedrzejewski87}\footnote{IRAF is distributed by the National
  Optical Astronomy Observatory, which is operated by the Association
  of Universities for Research in Astronomy (AURA) under cooperative
  agreement with the National Science Foundation.}.  The geometric
parameters of the galaxy light fits are on average position angle
$\sim51\deg$ and ellipticity $\sim0.78$ in all inspected bands, which
is consistent with previous results \citep{iodice14}.

After modeling and subtracting the profile of the galaxy, to produce a
complete catalog of all sources in the VST and VISTA field of view, we
independently ran SExtractor \citep{bertin96} on the
galaxy-model-subtracted frame for each filter. We obtained aperture
magnitudes within a diameter aperture of eight pixels ($\sim1\farcs7$
at OmegaCAM/VST resolution and $\sim2\farcs7$ for the VIRCAM/VISTA), and
applied aperture correction to infinite radius. The aperture
correction is derived from the analysis of the curve of growth of
bright isolated point-like sources. The aperture correction terms
derived are $ap.corr.$=0.43, 0.38, 0.40, 0.25, 0.19, 0.14 mag in $u$,
$g$, $r$, $i$, $J$, $K_s$, respectively, with typical uncertainty of
$\sim0.01$ mag. We assumed constant Galactic extinction on the
frames with the \citet{sf11} recalibration of the \citet{sfd98}
infrared-based dust maps.

The VST images are calibrated in the SDSS photometric system using
several \citet{landolt92} standard fields with calibrated SDSS
photometry.  VISTA is instead calibrated against 2MASS photometry. We
independently verified the calibrations by comparing the $ugri$
magnitudes with $\sim500$ objects with photometry available from
APASS. For the $u$\footnote{We transformed the APASS $B$-band
  photometry to $u$, using the equations given in
  \url{https://www.sdss3.org/dr8/algorithms/sdssUBVRITransform.php} },
$g,$ and $r$ bands the median difference between our VST and APASS
magnitudes is $\leq$0.04 mag, which is smaller than the $rms$ scatter
in every case. For the $i$ band, we found a small color independent
offset of $\sim0.1$ mag that is still consistent with zero within the
estimated $rms$ ($\langle m_i^{VST}-m_i^{APASS}\rangle=0.10\pm0.11$
mag). The same behavior of VST and APASS photometry in $i$ band was
also found on completely different targets, from the Fornax Deep
Survey \citep{iodice16,dabrusco16,cantiello17} and with a data
reduction tool independent from VST-Tube (i.e., with AstroWise; Aku
Venhola, priv. communication). Furthermore, the comparison of $i$-band
photometry for VST data with data in the literature for other targets
in the VEGAS survey \citep[e.g., NGC\,3115;][]{cantiello15} did not
show any peculiar offset in this band. Hence, our conclusion is that a
small difference exists in the system throughput and image quality at
$i$-band wavelengths between the two telescope-instrument
combinations, leading to the observed increased offset and scatter.

For the near-IR bands, we checked the photometry with an independent
comparison to 2MASS point sources photometry. The agreement is
satisfactory for both bands with an offset $\leq0.04$ mag and $rms$
nearly twice as large.

The photometric catalogs in the six bands were matched adopting
$1\farcs2/1\farcs4$ matching radius for VST/VISTA. The full catalog of
$\sim 1.200.000$ sources is available on the VEGAS project
web-page\footnote{Project page: \url{http:
 //www.na.astro.it/vegas/VEGAS/VEGAS_Targets.html.}} and on
the CDS archive. Sources with matched photometry in all six bands, or
with missing detections in one or more filters, are included in the full
catalog.

\begin{figure}
  \centering
  \includegraphics[width=1.0\hsize]{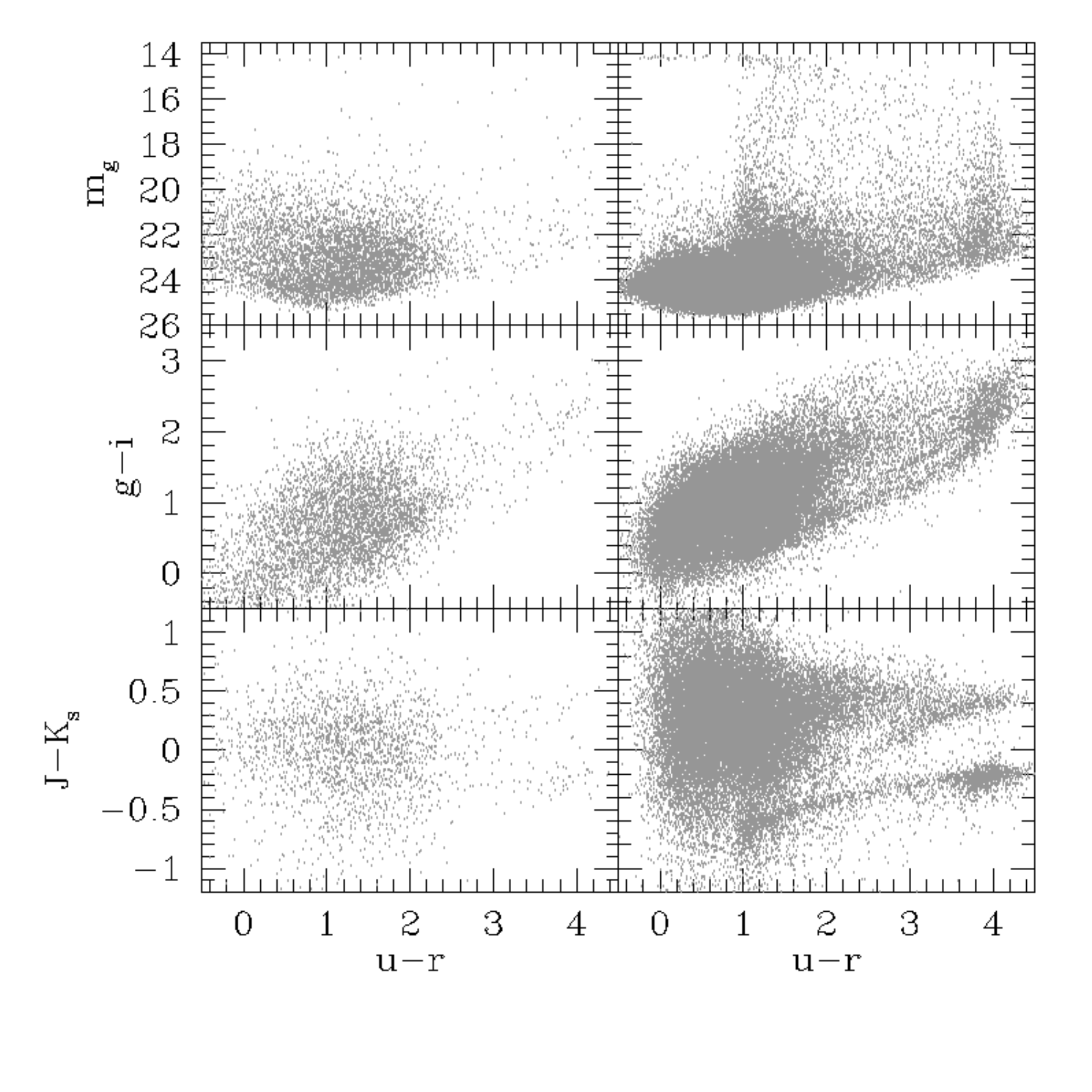}
   \caption{Color magnitude and color-color diagrams for the full
     matched catalog of sources detected. Left panels: Sources
     detected at galactocentric radii with moderate-to-high galaxy
     surface brightness background, $\mu_g\leq 23.6~mag/arcsec^2$, are shown. Right
     panels: Sources detected in regions with low galaxy background are shown.}
   \label{cmdcol}
  \end{figure}

A color magnitude diagram and some color-color diagrams of the full
matched catalog are shown in Figure \ref{cmdcol}. In the figure, we
show separately the $\sim70.000$ sources detected in areas with high
and highly variable galaxy backgrounds (i.e., in regions where
$\mu_g\leq 23.6~mag/arcsec^2$, left panels) and those detected
where the galaxy background is negligible (right panels).

%
\begin{table}
\caption{\label{tab_props} Properties of the target}
\centering
\begin{tabular}{lcc}
\hline
\hline
R.A. (J2000) &  00h 47m 33.1s  & (1)\\
Dec. (J2000) & -25d 17m 18s  & (1) \\
$cz$ (km/s)&  243$\pm$2   & (2) \\
$M_V^{tot}$(mag) & $\sim$-21$\pm$0.5 & (2) \\
$E(B{-}V)$      & 0.017 & (1) \\
Type         &   SABc       &  (2) \\
$T_{type}$    &  $5.1\pm0.4$ & (2) \\
$\sigma$ (km/s)  & $97\pm18$  & (2) \\
$(m{-}M)$      & 27.70$\pm$0.07 & (3) \\
\hline
\hline
\multicolumn{3}{c}{Details on observations}\\
\hline
Passband     & exp. time (s) & $\langle FWHM\rangle$ \\ 
\hline
\multicolumn{3}{c}{VST}\\
 $u$ (s)&  29008   & $0\farcs94$ \\
 $g$ (s)&  2100    & $0\farcs83$\\
 $r$ (s)&  3718    & $0\farcs89$ \\
 $i$ (s)&  1500   &  $0\farcs75$ \\
\hline
\multicolumn{3}{c}{VISTA}\\
$J$   &  $\sim 80000$    & $0\farcs97$ \\
 $K_s$ & $\sim 2100$ & $1\farcs29$ \\
\hline 
\end{tabular}
\tablefoot{Listed properties are taken from: (1) the NASA
  Extragalactic Database, \url{http://ned.ipac.caltech.edu}, (2) the
  HyperLeda archive, \url{http://leda.univ-lyon1.fr/}, and (3)
  \citet{rs11}.}
\end{table}

\section{Selection of GC candidates}
\label{sec_sel}

To select GC candidates we applied the photometric, morphometric, and
color selection criteria already used in \citet{cantiello17}, with
some differences explained below.

\begin{figure*}
  \centering
  \includegraphics[width=0.43\hsize]{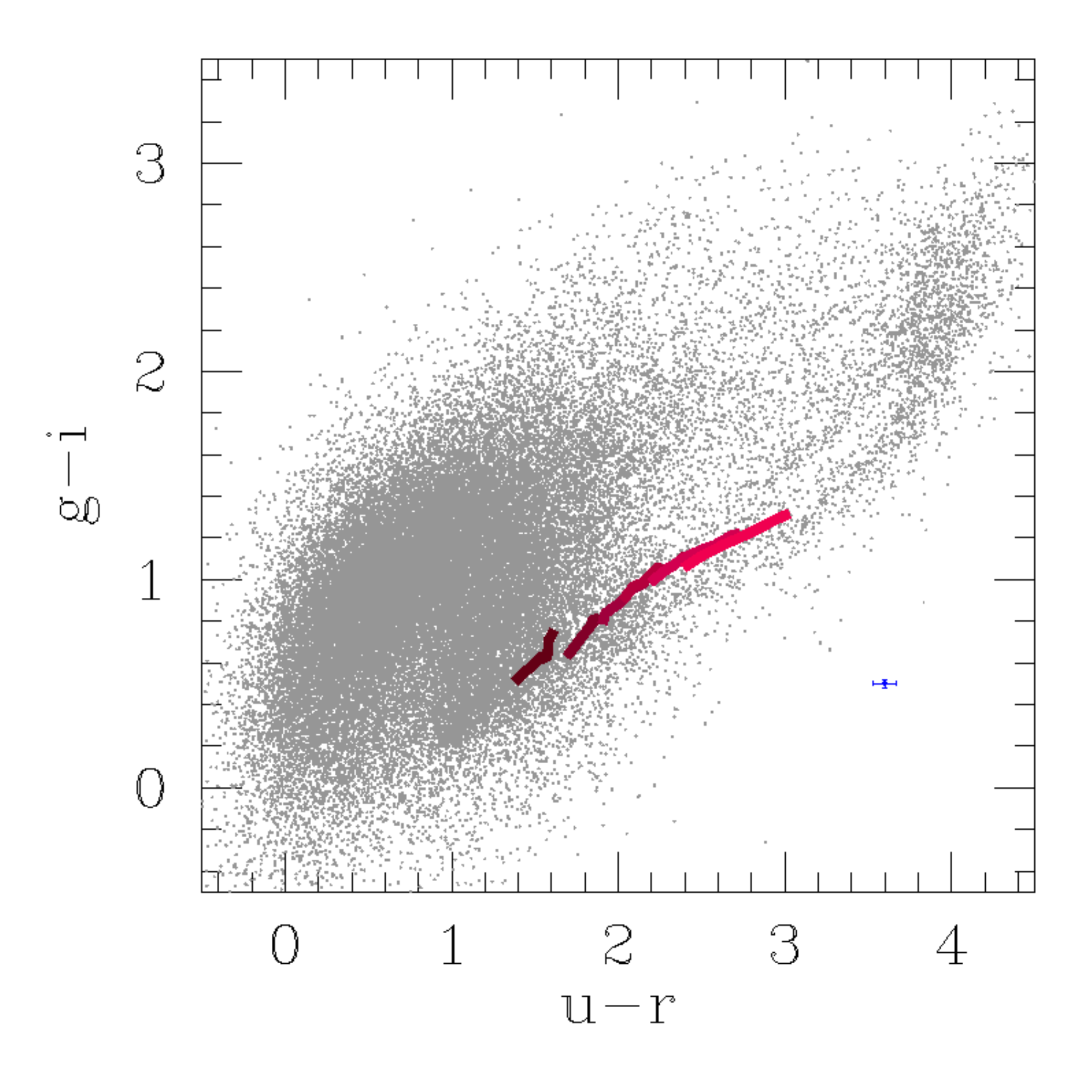}
  \includegraphics[width=0.43\hsize]{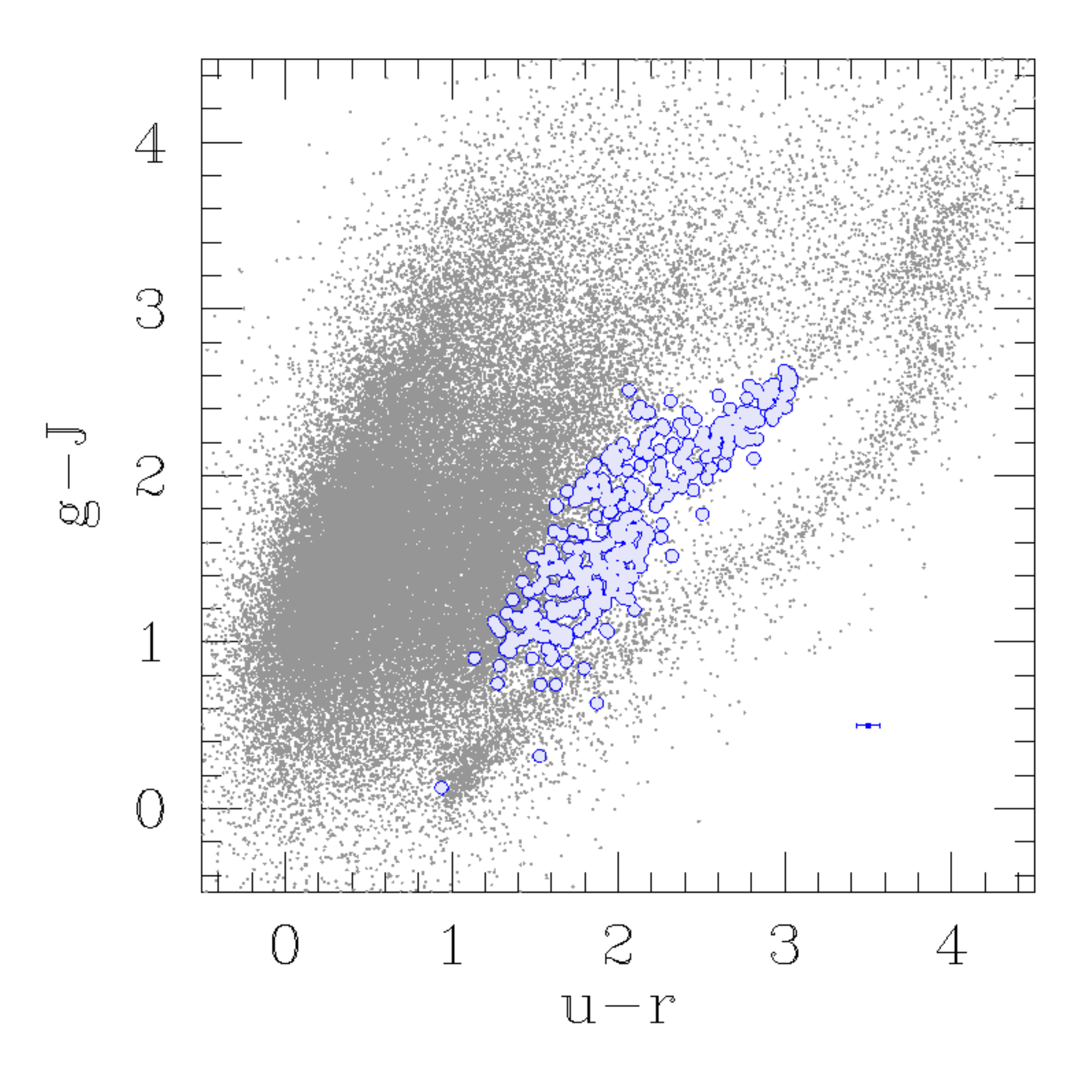}
  \includegraphics[width=0.43\hsize]{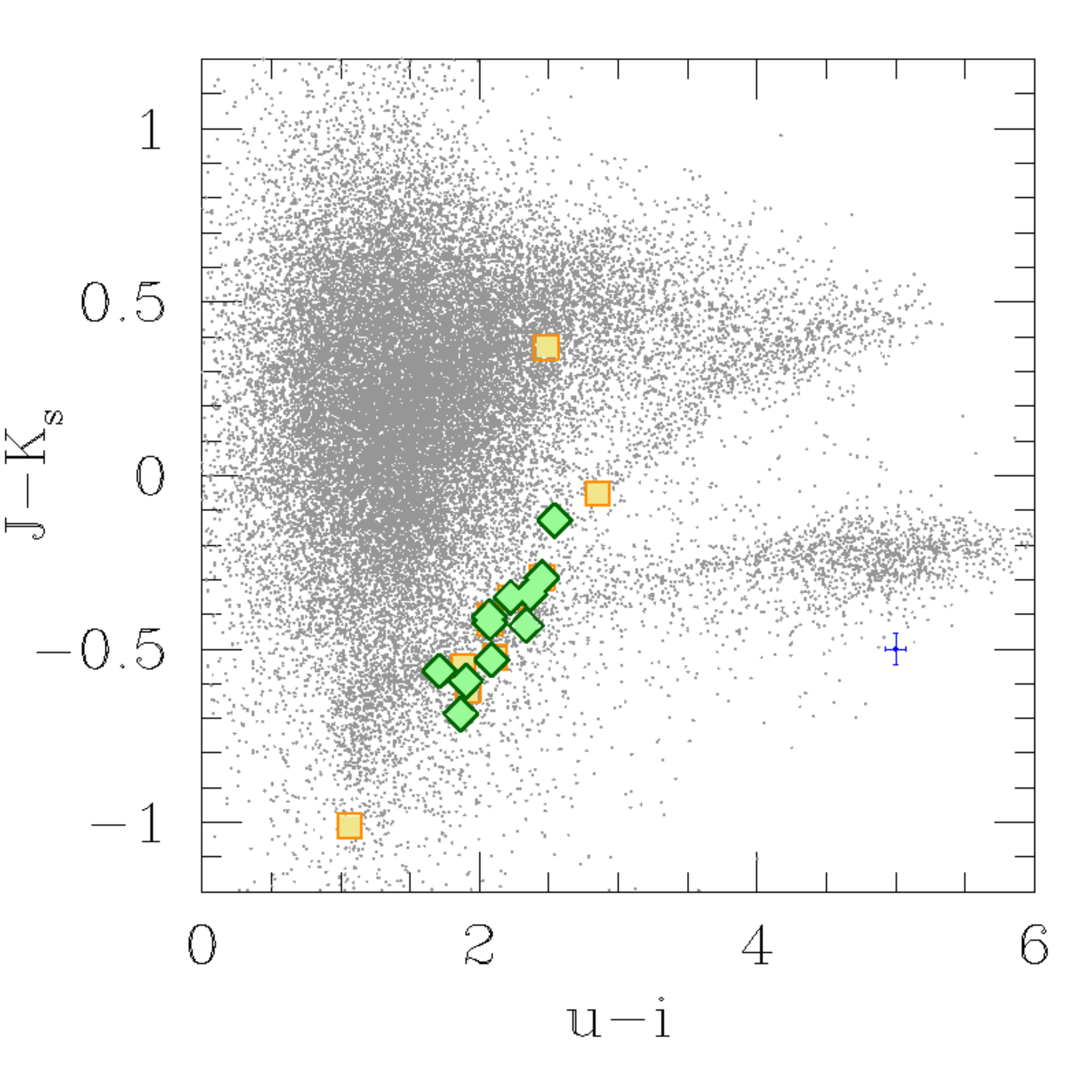}
  \includegraphics[width=0.43\hsize]{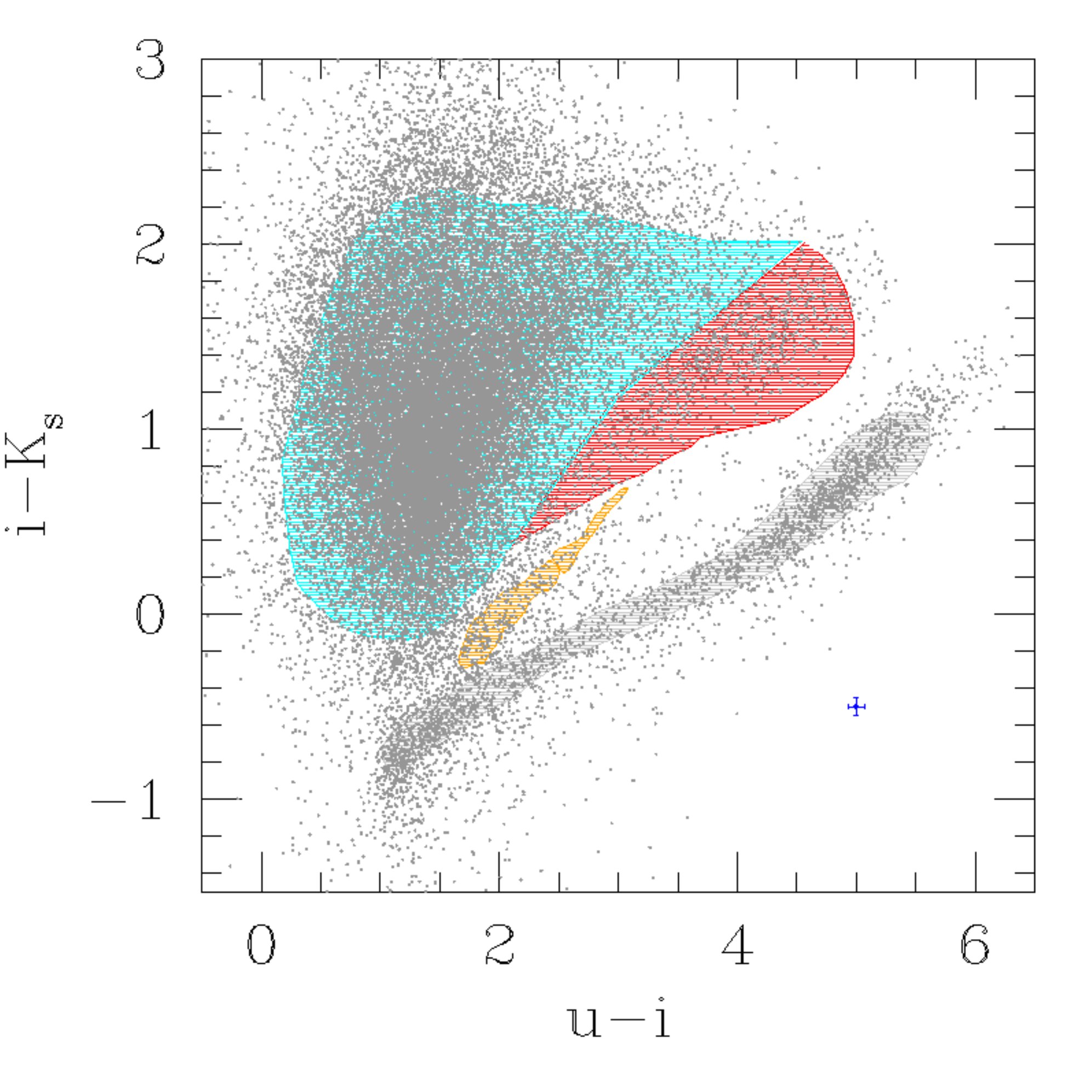}
  \caption{Selection of color-color diagrams used for selecting
    GC candidates. Gray dots show the full sample of matched sources.
    In each panel, in addition to the full sample, single elements of
    the GC selection process are highlighted.  Upper left panel: SSP
    models for ages between 6 and 14 Gyrs and \feh between $-$2.3 and
    $+$0.3 are plotted with solid lines in various shades of red
    (darker for more metal-poor; at fixed \feh, older ages have redder
    colors). Upper right: Blue dots show the sample of $\sim350$ GC
    candidates selected using color-color, photometric, and
    morphometric selection criteria described in text. The median
    error bars are indicated (lower right blue symbols), calculated as
    the sample median from GC candidates and shown in other
    panels as well. Lower left panel: Yellow squares and green diamonds indicate
    the location of \citet{beasley00} and \citet{olsen04}
    spectroscopically confirmed GCs. Lower right panel: We highlight
    the approximate MW stars sequence (gray), the area occupied by
    background galaxies (passive in red and blue star-forming galaxies in
    cyan; see also Appendix \ref{appendix}), and the area identified
    by \citet{munoz14} as the locus of GCs (orange).}
  \label{colors}
  \end{figure*}

We started by applying color-color selections using all available
pairs of colors.  Figure \ref{colors} shows some examples of the
color-color diagrams.  In each panel, besides showing the full
  sample of matched sources in light gray, we highlight some stages of
  the GCs selection procedure adopted.  For the purely optical
  color-color diagram (upper left panel), we plot the locus of simple
  stellar population models from the Teramo-SPoT group
  \citep{raimondo05,raimondo09}\footnote{Models available at the
    URL:\url{http://www.oa-teramo.inaf.it/spot}. Old GCs are expected
  to match the models sequence. The final sample of selected GC
  candidates is shown in the upper right panels with filled blue
  circles. The sample of 21 spectroscopically confirmed GCs from
  \citet{beasley00} and \citet{olsen04} are plotted in the lower left
  panel.  The two spectroscopic databases contain 14 \citep{beasley00}
  and 11 GCs \citep{olsen04}, respectively, and four sources are
  common to both.  Depending on the plotted colors, the
  various sequences of MW stars, passive and star-forming galaxies
  appear relatively well defined (see also Appendix \ref{appendix}).
  In the \ui versus \ik plot ($uiK_s$ hereafter) we highlighted the
  approximate position of such areas, adding the locus of GC
  candidates as found by \citet{munoz14} properly shifted to take into
  account the different $u$ bands between the two works.}

For each pair of colors, we selected only the sources falling within
the color-color area defined by confirmed GCs and SSP models as good
GC candidates, extended by $\sim$20\% on each side; i.e., no less than
0.2 mag for the colors spanning an interval lower than 1 mag, such as
$r{-}i$. The choice of $\sim$20\%, adopted after several tests, was
motivated by the need for an area large enough to include the largest
percentage of confirmed GCs, but not too wide to avoid exceedingly
large contamination from obvious non-GCs sources. We adopted as lower
age limit for SSP models $t=6$ Gyr, based on the comparison of the
range of colors from confirmed GCs. Such limit might be rather low for
the ages, generally $\gsim10$ Gyr \citep[e.g.,][]{puzia05}, typical
for old GCs. However, given the age-metallicity degeneracy for the
optical colors \citep{worthey94b}, younger SSP models with higher \feh
overlap with the sequence of older SSP models with lower \feh. In any
case, the main constraint to the color-color selections adopted here
comes from empirical data. We used the SSP models as a countercheck,
as they are confined to a very narrow region of the optical
color-color planes, while empirical data are more scattered and also
include the selection with near-IR bands.

To give an idea of the efficiency of the color-color selection
adopted, we highlight that, starting from a sample of $\sim 1.200.000$
matched sources, the sample of color-color preselected GCs includes
$\sim 1.500$ objects.

To further narrow down the sample of reliable GC candidates, we used
other photometric and morphological properties of the sample, as
described in \citet{cantiello17}. We measured the {\it magnitude
  concentration index} \citep{peng11}, obtained as the difference
between the magnitude measured at 6 pixel aperture diameter and at 12
pixel, $\Delta X_{6-12}\equiv mag_{X,6pix}-mag_{X,12pix}$ , where X is
one of the optical $ugri$ bands and aperture corrected magnitudes are
used. For point-like sources, after applying the aperture correction
to the magnitudes at both radii, $\Delta X_{6-12}$ should be
statistically consistent with zero. Hence, $\Delta X_{6-12}$ is an
ideal tool to identify point-like sources, such as stars and
extragalactic GCs in very distant galaxies as they appear
unresolved. In the case of NGC\,253, given the spatial resolution and
FWHM of our dataset, GCs appear as slightly resolved extended sources
and, consequently, their magnitude concentration index is larger than
zero. By analyzing the sample of confirmed GCs, we find $\Delta
X_{6-12}>0.1$ mag in all optical bands (with three exceptions
discussed below), and a median of $\Delta X_{6-12}\sim0.2$ mag. For
the morphologic selection we did not use VISTA data as the pixel
resolution is lower than for VST.  Moreover, the $J$-band image is
much more crowded than optical images; given the depth of the frame,
stars in the field of NGC\,253 are also detected \citep{greggio14}.

\begin{figure*}
  \centering
  \includegraphics[width=0.4\hsize]{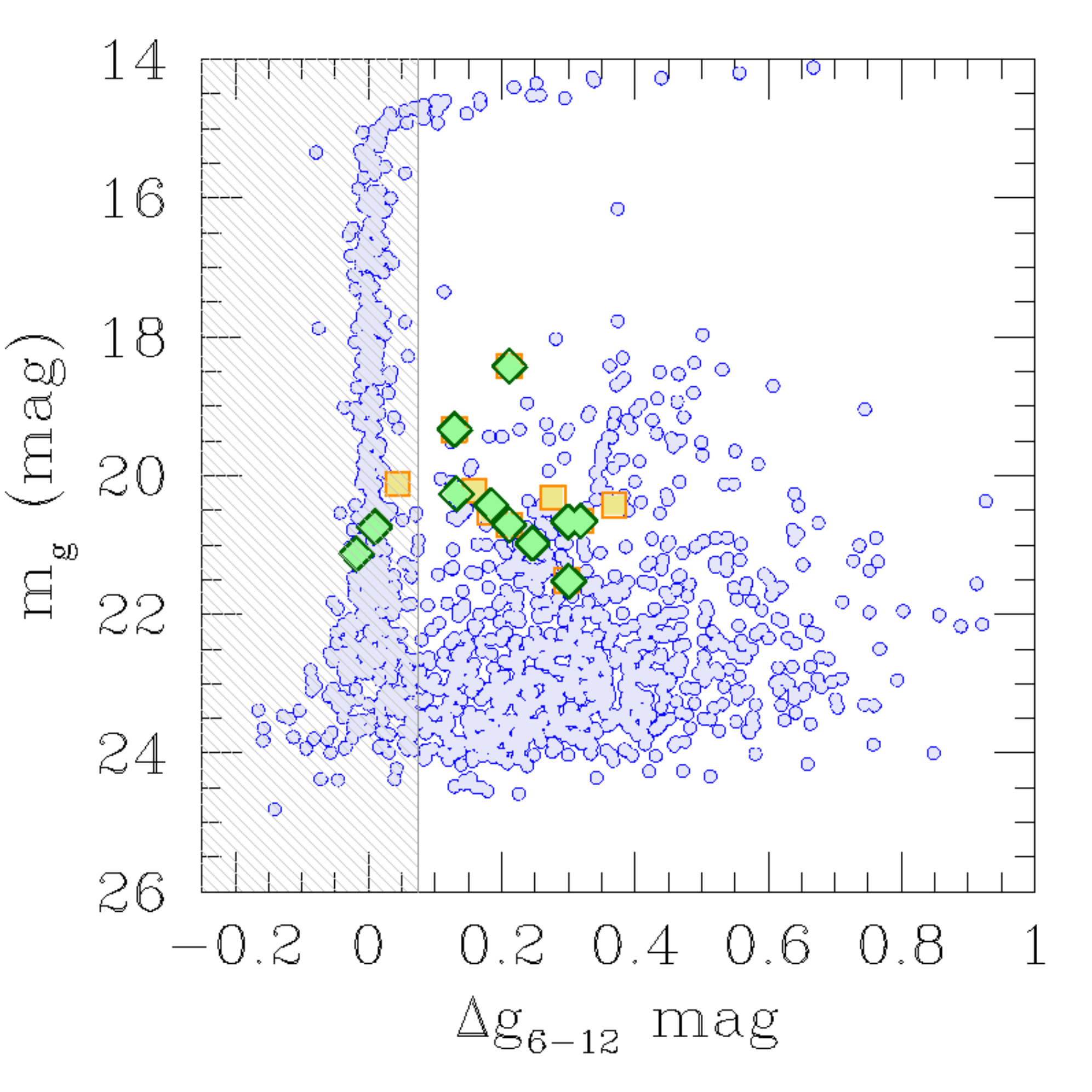}
  \includegraphics[width=0.4\hsize]{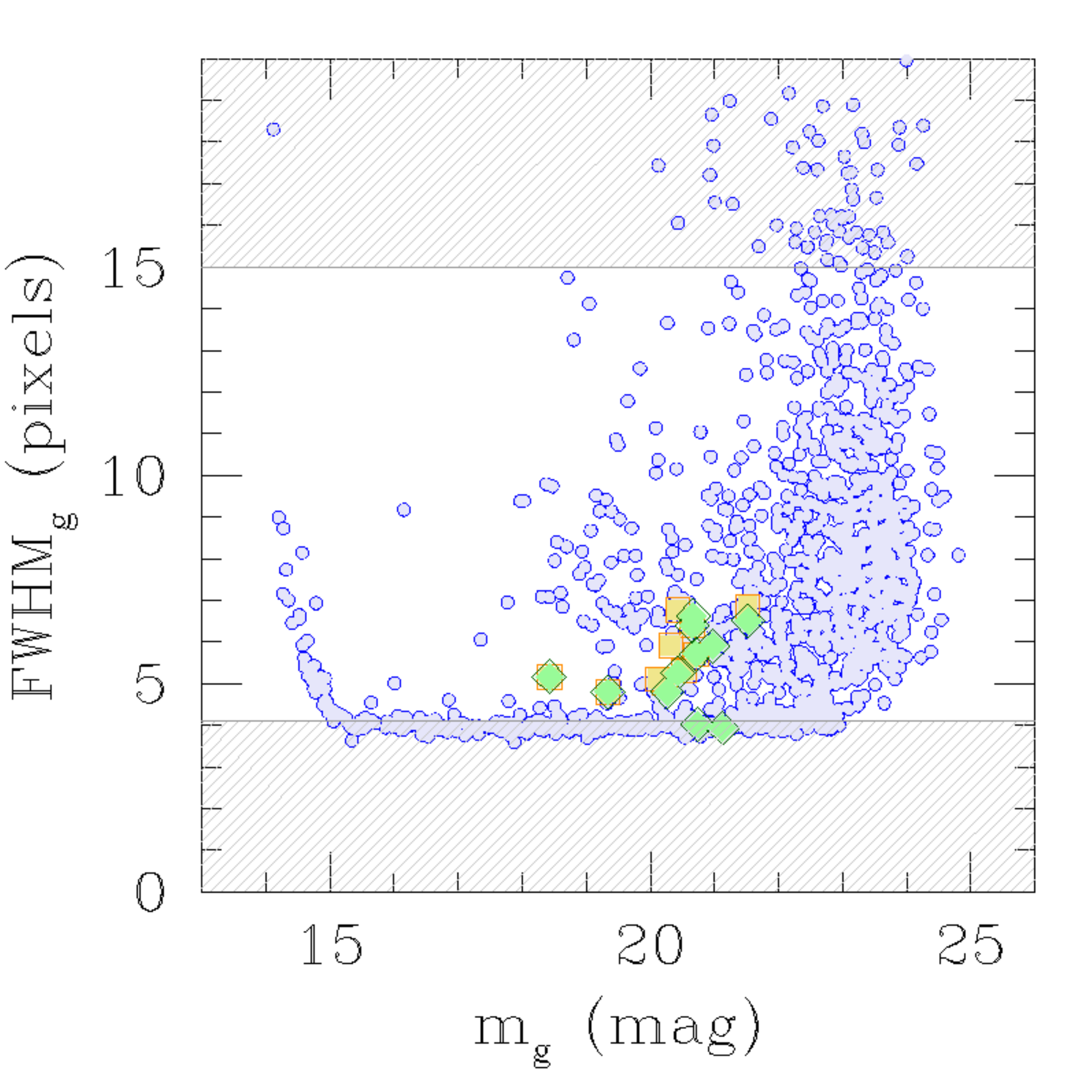}
  \includegraphics[width=0.4\hsize]{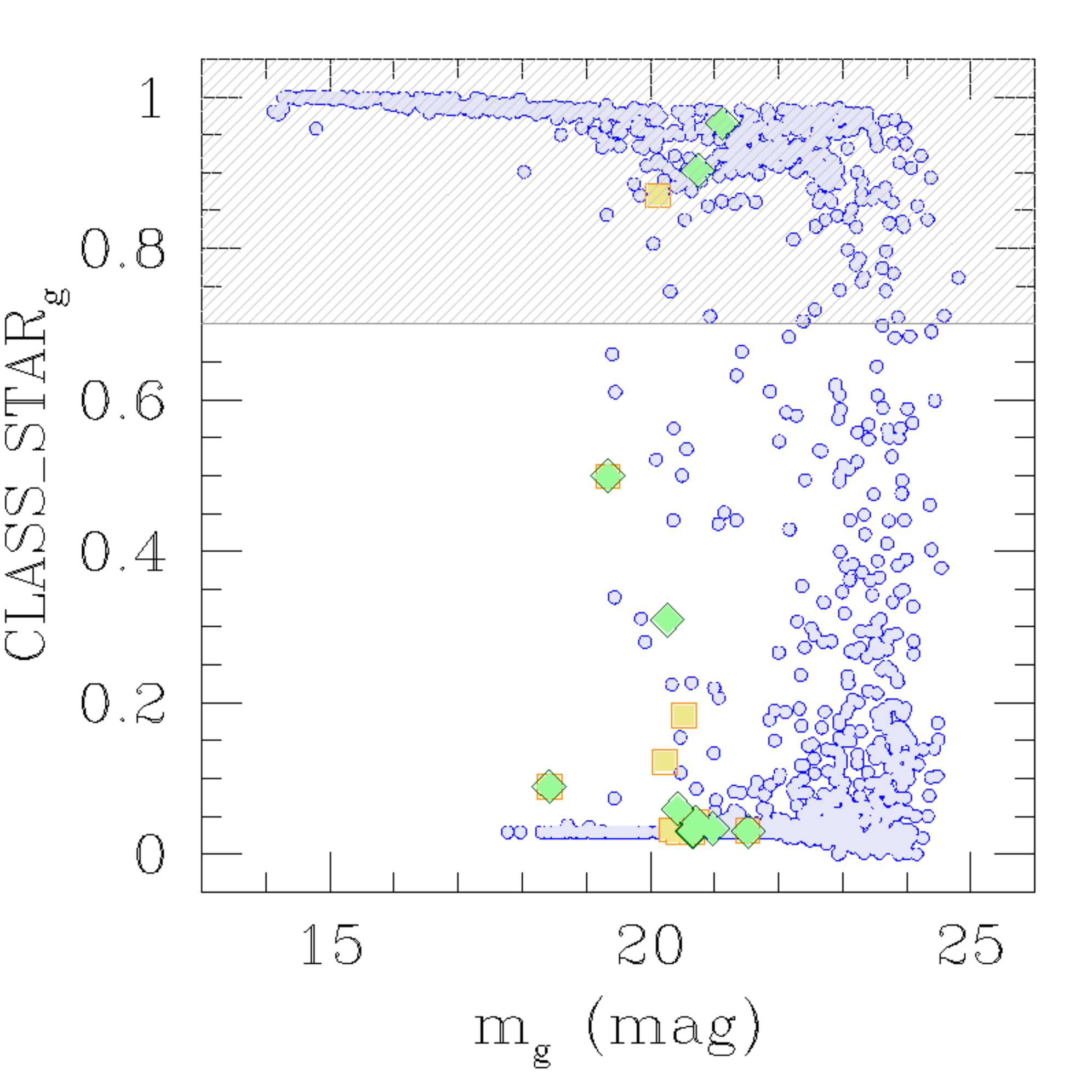}
  \includegraphics[width=0.4\hsize]{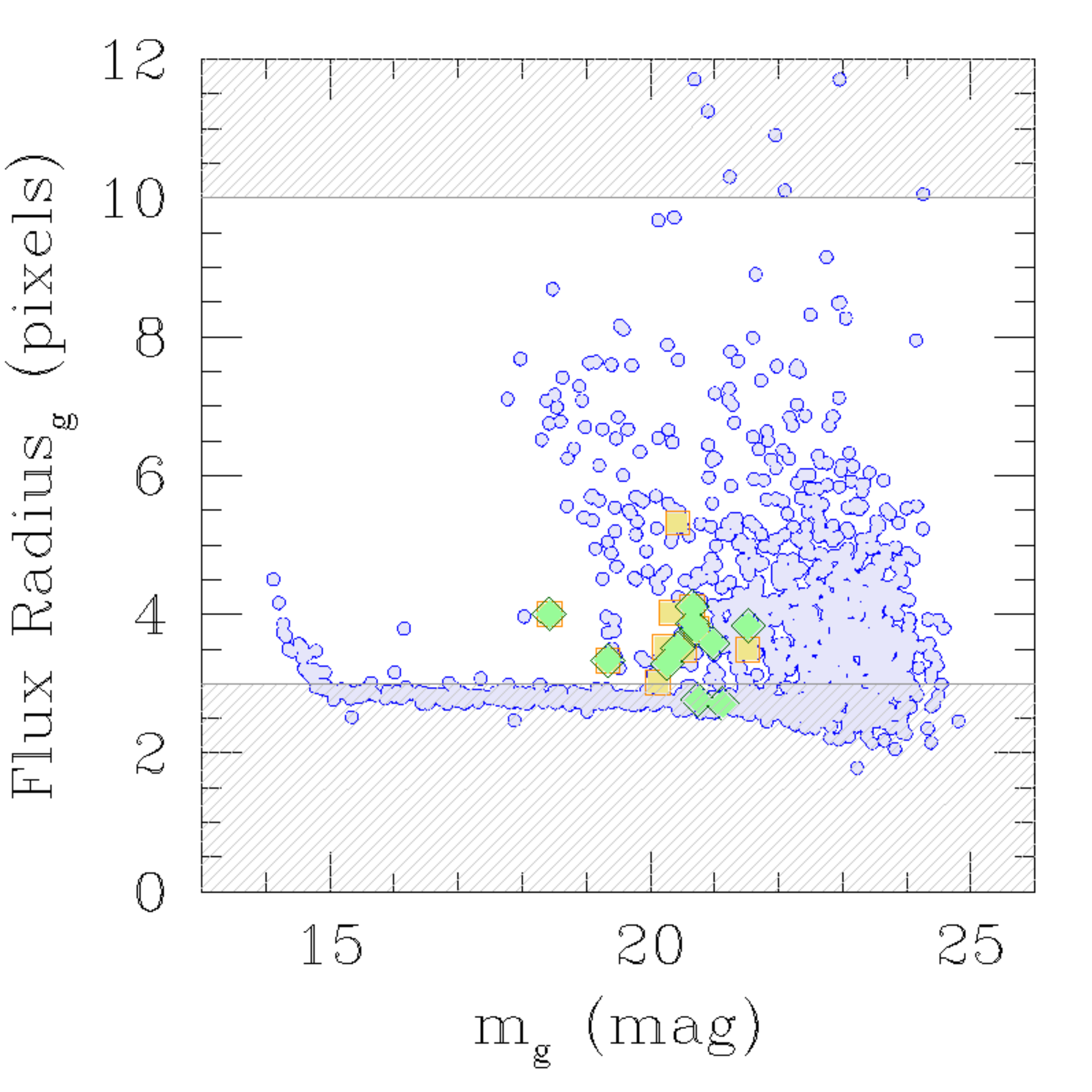}
  \caption{Upper left: $g$-band magnitude concentration index for the
    sample of color-selected GC candidates, shown in blue. As in
    Figure \ref{colors}, yellow squares and green diamonds indicate the
    location of \citet{beasley00} and \citet{olsen04} spectroscopic
    confirmed GCs, respectively. The gray shaded area defines the
    region in which sources are rejected as reliable GC candidates. Upper
    right: $g$-band FWHM from SExtractor vs. magnitude. Symbols are
    as upper left panel. Lower left: As upper right, but SExtractor
    CLASS\_STAR parameter is plotted instead of FWHM. Lower right: As
    upper right, but SExtractor Flux Radius (i.e., half-light radius)
    parameter is plotted instead of FWHM. The saturation that is
    manifested as the rightward tail at bright magnitudes in the upper
    left panel, similar to the tail that points upward in the upper
    right and lower right panels, appears at a magnitude brighter
    than the brightest GCs (e.g., $m_g<15.5$ mag; see also Table
    \ref{tab_selparam}), and therefore does not affect our GC candidate
    selection process.}
  \label{sel_g}
  \end{figure*}

The upper left panel of Figure \ref{sel_g} shows the $g$-band
magnitude concentration index, $\Delta g_{6-12}$, for the sample of
color-color selected GC candidates. In the panel, where the confirmed
GCs are also reported, the stellar sequence at $\Delta
g_{6-12}\sim0.0$ mag is easily recognized, as well as the positive,
$\geq0.1$ mag, values for all but three confirmed GCs. The three
sources at $\Delta g_{6-12}\leq0.075$ mag, which we adopted as
threshold for reliable candidates, are the candidates with ID \#109
and \#114 from \citet{olsen04}, and LA11 from \citet{beasley00}. For
the first two sources, we observe that the magnitude concentration
index is consistent with zero in all optical bands and, similarly, the
SExtractor and Ishape (see section \ref{sec_ishape}) output parameters
described below, are all consistent with the stellar nature of the two
sources. Analyzing the two candidates in more detail, we also find
that both have FWHM that is locally indistinguishable with all
confirmed stellar sources (selected by colors and the other
morpho-photometric criteria). The case of LA11 is described in more
detail later in this section.

As additional selection criteria, for the sole optical bands, we used
a selection of SExtractor output parameters (i.e., FWHM, CLASS\_STAR,
minor-to-major axis ratio $b/a$, and flux radius; see definitions in
Table \ref{tab_selparam}), the limits of the GCLF, and a maximum
photometric error. The SExtractor selection criteria were derived by
comparison with the same parameter for confirmed GCs; for the minimum
axis ratio, $b/a$, we conservatively assumed $ b/a\geq$0.67, which is
comparable to the observed minimum for MW and Magellanic Clouds GC
systems \citep{vdb84,harris96,cantiello09}.  The FWHM, CLASS\_STAR,
and flux radius selections are also shown in Figure \ref{sel_g}.

There is some level of degeneracy for some of the adopted
morphometric quantities. Our intention, in adopting such large set of
parameters, is to exclude anomalous or peculiar sources that might be
more efficiently detected with one parameter rather than others, hence
minimizing any contamination from non-GC sources.

The bright and faint magnitude cuts were derived from the GCLF as
follows. We adopted the RGB-tip distance modulus given in Table
\ref{tab_props} and then we used the results from \citet[][]{villegas10}
for the GCLF turnover magnitude (TOM or $M_{TOM}$, hereafter) and for
the GCLF dispersion, $\sigma_{GCLF}$. In particular, we adopted
$M^g_{TOM}=-7.4$ mag and estimated $\sigma_{GCLF}=1.1$ mag (assuming a
total magnitude of NGC\,253 of $M_Z\sim-22$ mag). Finally, we adopted
as magnitude cuts $\pm3 \sigma_{GCLF}$ brighter and fainter than the
TOM. For sake of simplicity, the TOM in $u$, $r,$ and $i $ bands, were
derived from the $M^g_{TOM}$ band reported above, and from the median
$u{-}g$, $g{-}r,$ and $g{-}i$ of known GCs in the sample, i.e., $\sim
1.4,~0.6,$ and 0.8 mag, respectively.  Furthermore, to be most
inclusive as possible, the magnitude cuts were rounded off to the closest
more conservative semi-entire magnitude (e.g., we
adopted $m^{bright}=16$ for the $r$ band rather than 16.2 mag, and $m^{faint}=23$
rather than 22.8 mag).

%
\newpage
\begin{table}
\caption{\label{tab_selparam} Photometric and morphometric selection criteria adopted}
\centering
\tiny
\begin{tabular}{lcc}
  \hline\hline
Quantity        & Passband & Range adopted \\
                &          & for selection \\
\hline
$\Delta X_{6-12}$ (mag) & All    & $\geq 0.075$                    \\   
CLASS\_STAR     & $u$      & $\leq0.8$                            \\            
CLASS\_STAR     & $g$      & $\leq0.7$                            \\            
CLASS\_STAR     & $r$      & $\leq0.95$                            \\            
CLASS\_STAR     & $i$      & $\leq0.95$                            \\            
PSF FWHM (pixels) & $u$      & $\geq 4.5$                   \\        
PSF FWHM (pixels)       & $g$      & $\geq 4.1$ \& $<15$     \\        
PSF FWHM (pixels)       & $r$      & $\geq 4.25$ \& $<15$    \\        
PSF FWHM (pixels)       & $i$      & $\geq 3.9$ \& $<15$     \\        
Flux Radius (pixels)    & $u$      & $\geq 3.2$ \& $<10$      \\
Flux Radius (pixels)    & $g$      & $\geq 3$ \& $<10$        \\
Flux Radius (pixels)    & $r$      & $\geq 3$ \& $<10$       \\
Flux Radius (pixels)    & $i$      & $\geq 2.5$ \& $<15$      \\
Axis Ratio b/a  & All      & $\geq 0.67$                           \\                
$\Delta$ mag    & All      &  $\leq0.25$                      \\                  
$m^{bright}$-$m^{faint}$ (mag)& $u$& 18-25                    \\               
$m^{bright}$-$m^{faint}$ (mag)& $r$& 16.5-23.5                \\               
$m^{bright}$-$m^{faint}$ (mag)& $g$& 16-23                    \\               
$m^{bright}$-$m^{faint}$ (mag)& $i$& 16-23                    \\               
\hline 
\end{tabular}
\tablefoot{Explanation of listed parameters. $\Delta X_{6-12}$: Threshold for the
  magnitude concentration index. SExtractor parameters: CLASS\_STAR:
  Neural-Network-based star/galaxy classifier; PSF FWHM: point spread
  function full width at half maximum; Flux Radius: half light radius;
  axis-ratio: semi-minor over semi-major axis ratio \citep[see][and
  references therein for more details]{bertin96}. Other selection
  parameters. $\Delta$ mag: maximum error on magnitude;
  $m^{bright}$-$m^{faint}$ bright and faint magnitude cuts (see
  text).}
\end{table}

All the morpho- and photometric selection criteria adopted are
summarized in Table \ref{tab_selparam}. The final sample of selected GC
candidates, passing through all adopted selections, contained
$\sim350$ sources.

\begin{figure*}
  \centering
  \includegraphics[width=0.43\hsize]{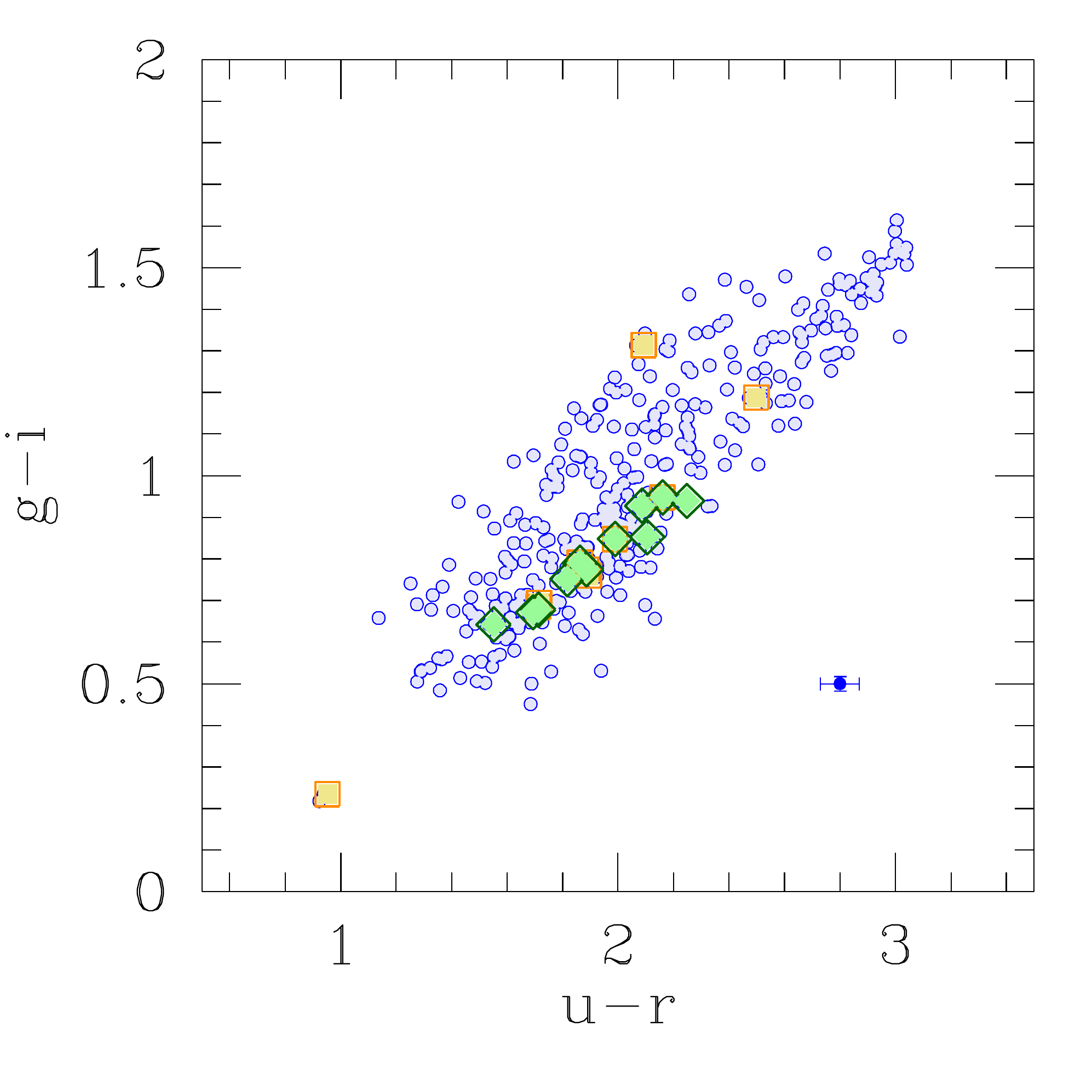}
  \includegraphics[width=0.43\hsize]{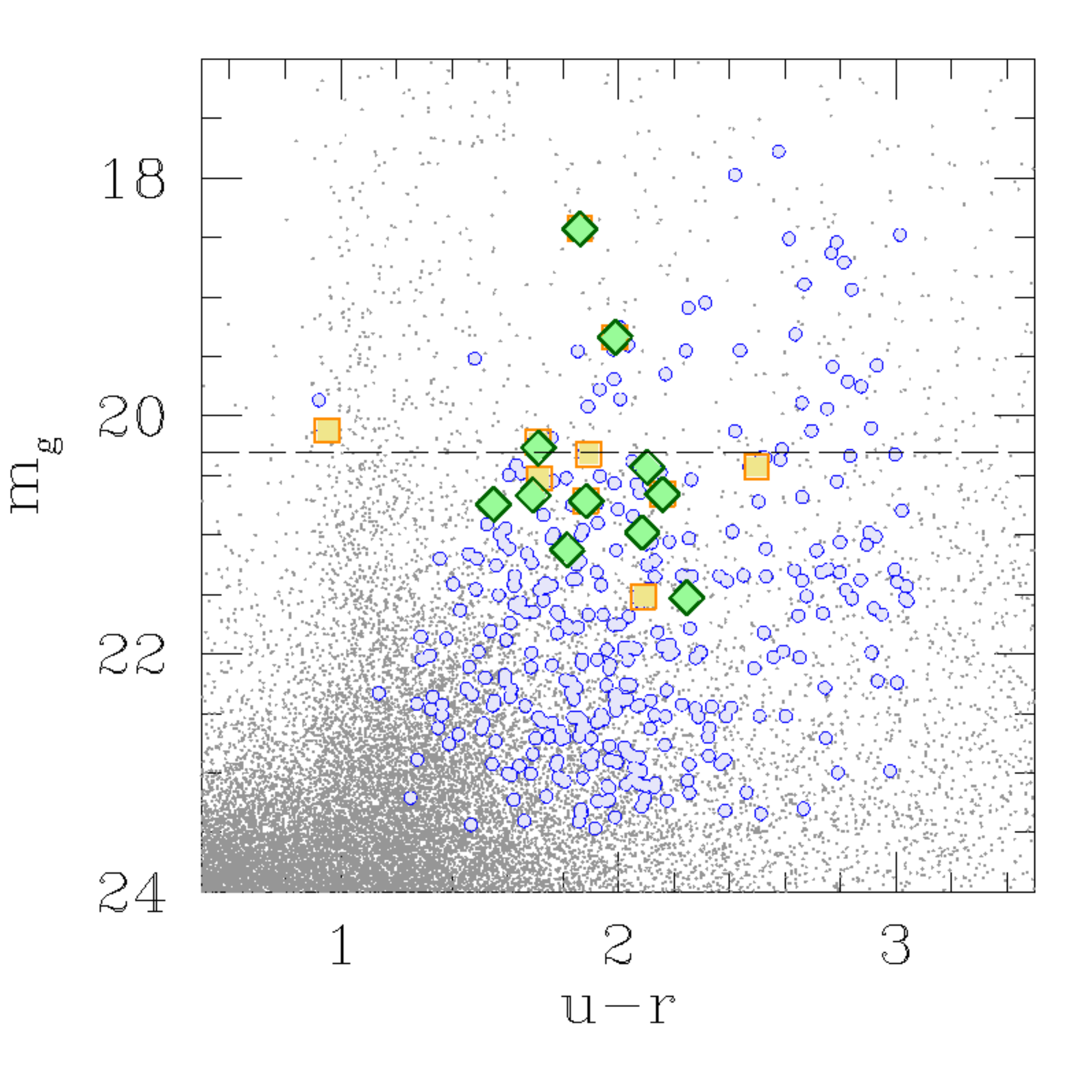}
  \includegraphics[width=0.43\hsize]{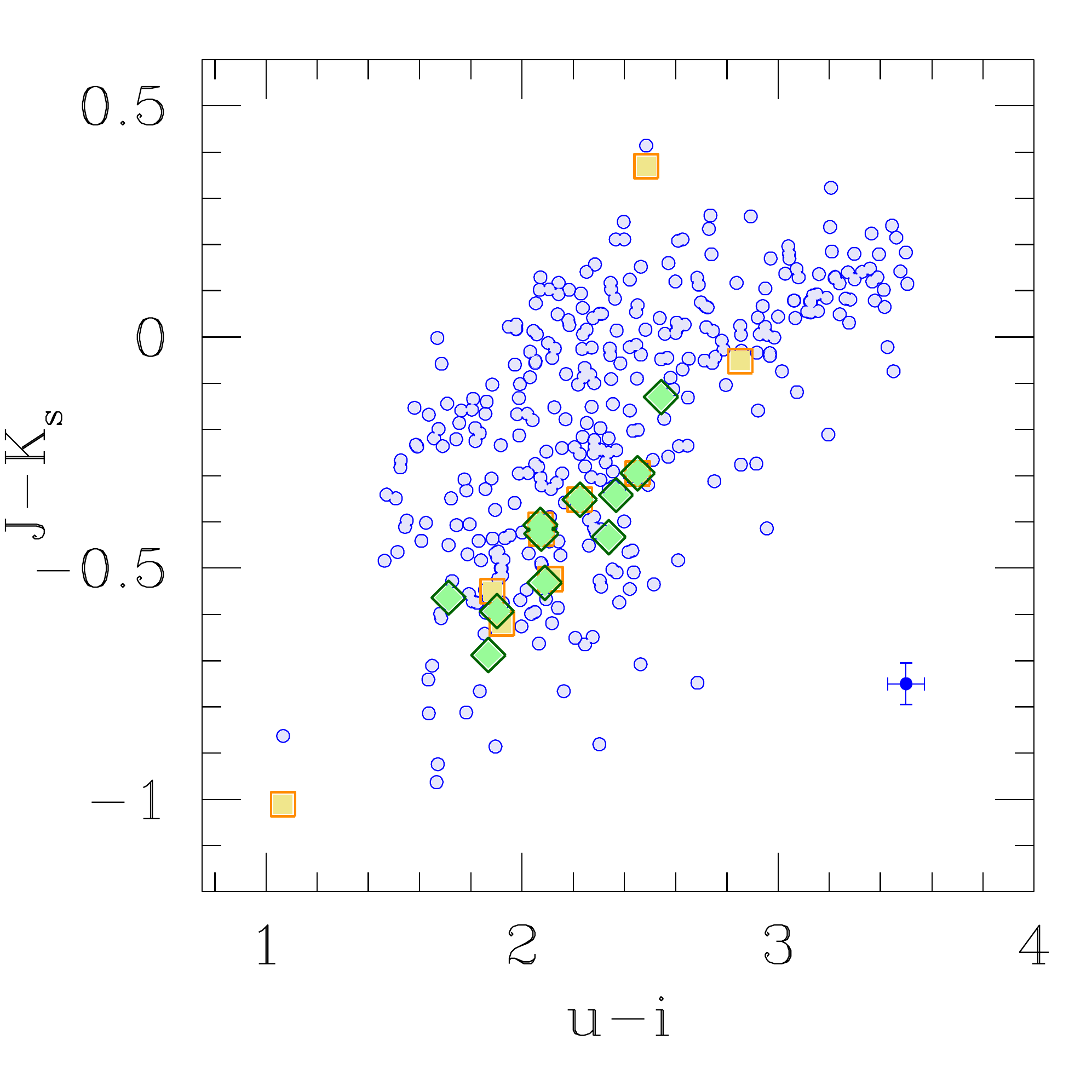}
  \includegraphics[width=0.43\hsize]{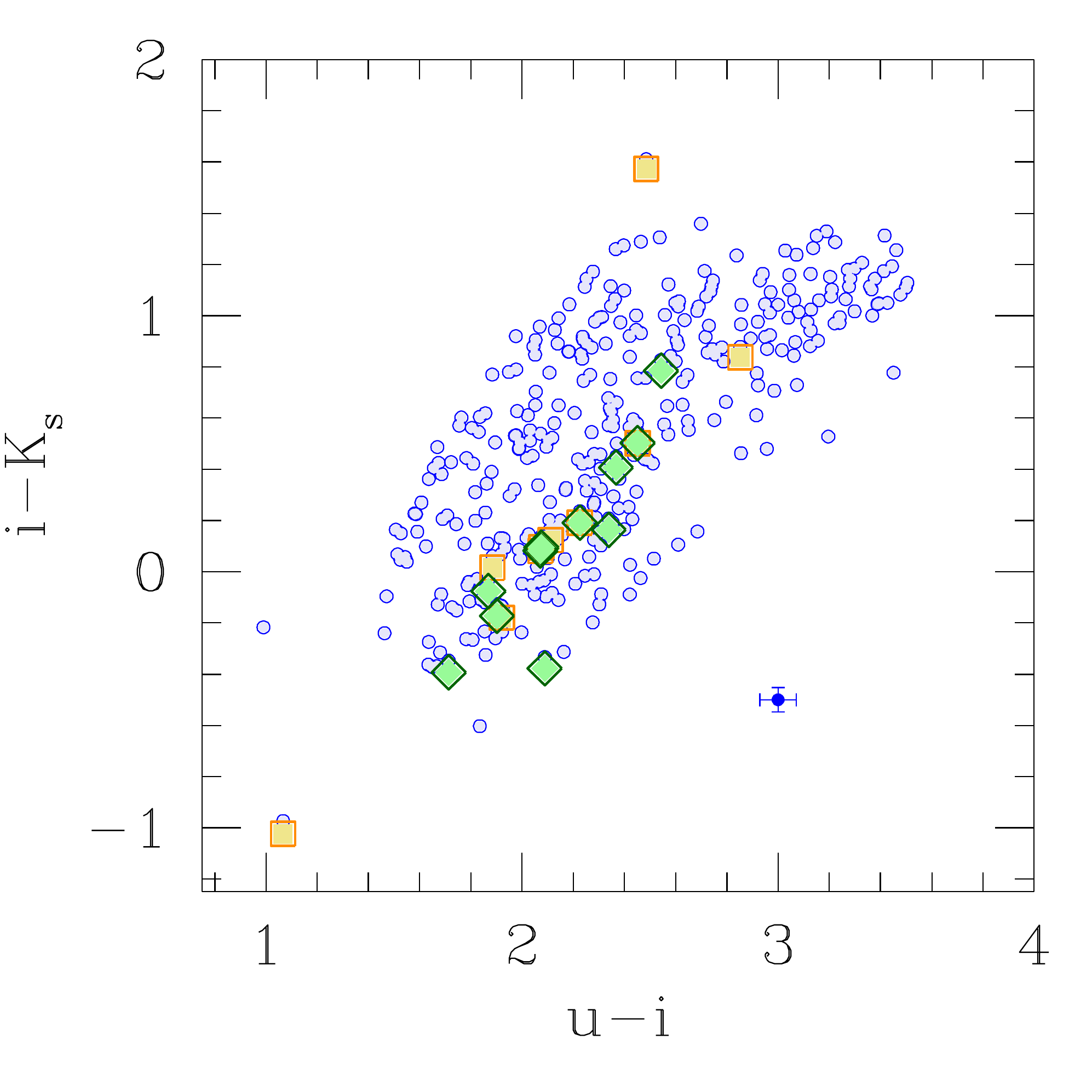}
  \caption{Color-color (upper left and lower panels) and
    color-magnitude (upper right) diagrams of the $\sim350$ selected
    GC candidates (blue circles) and the spectroscopically confirmed
    GCs. Yellow squares and green diamonds indicate the
    \citet{beasley00} and \citet{olsen04} samples, respectively. For
    the color-magnitude diagram, the full sample of matched sources
    (gray dots) and the position of the turnover magnitude
    (long-dashed horizontal black line) are also shown.}
  \label{zoom}
  \end{figure*}

In Figure \ref{zoom} we show some of the color-color diagrams already
shown in Figure \ref{colors}, but this time plotting only the
$\sim350$ GCs candidates selected using the photometric, morphometric,
and color selection criteria described above. Overplotted in green and
yellow symbols are spectroscopically confirmed GCs. In addition, a
color magnitude diagram (upper right panel) is reported.

  \begin{figure*}
  \centering
  \includegraphics[width=0.45\hsize]{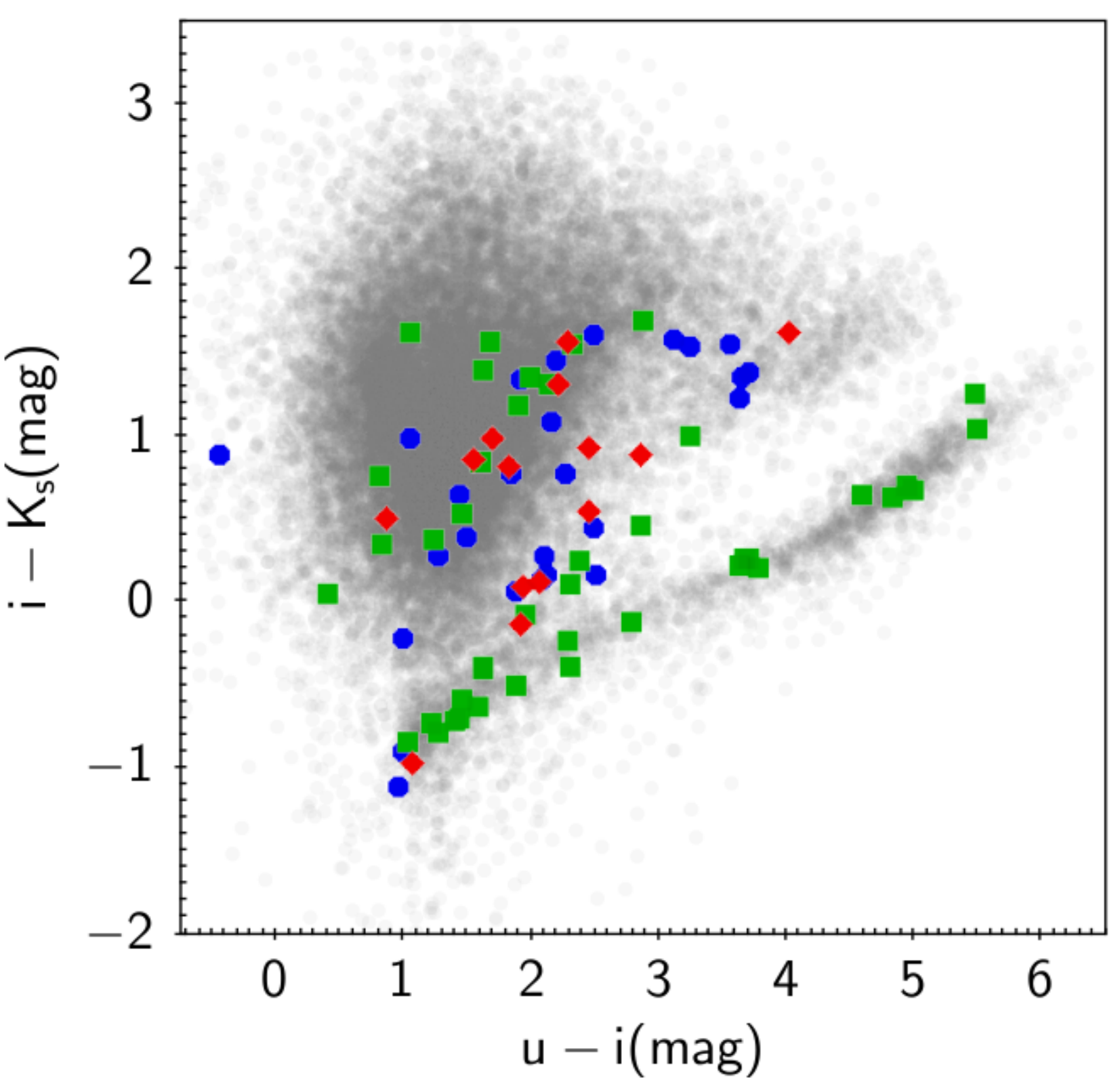}
  \includegraphics[width=0.45\hsize]{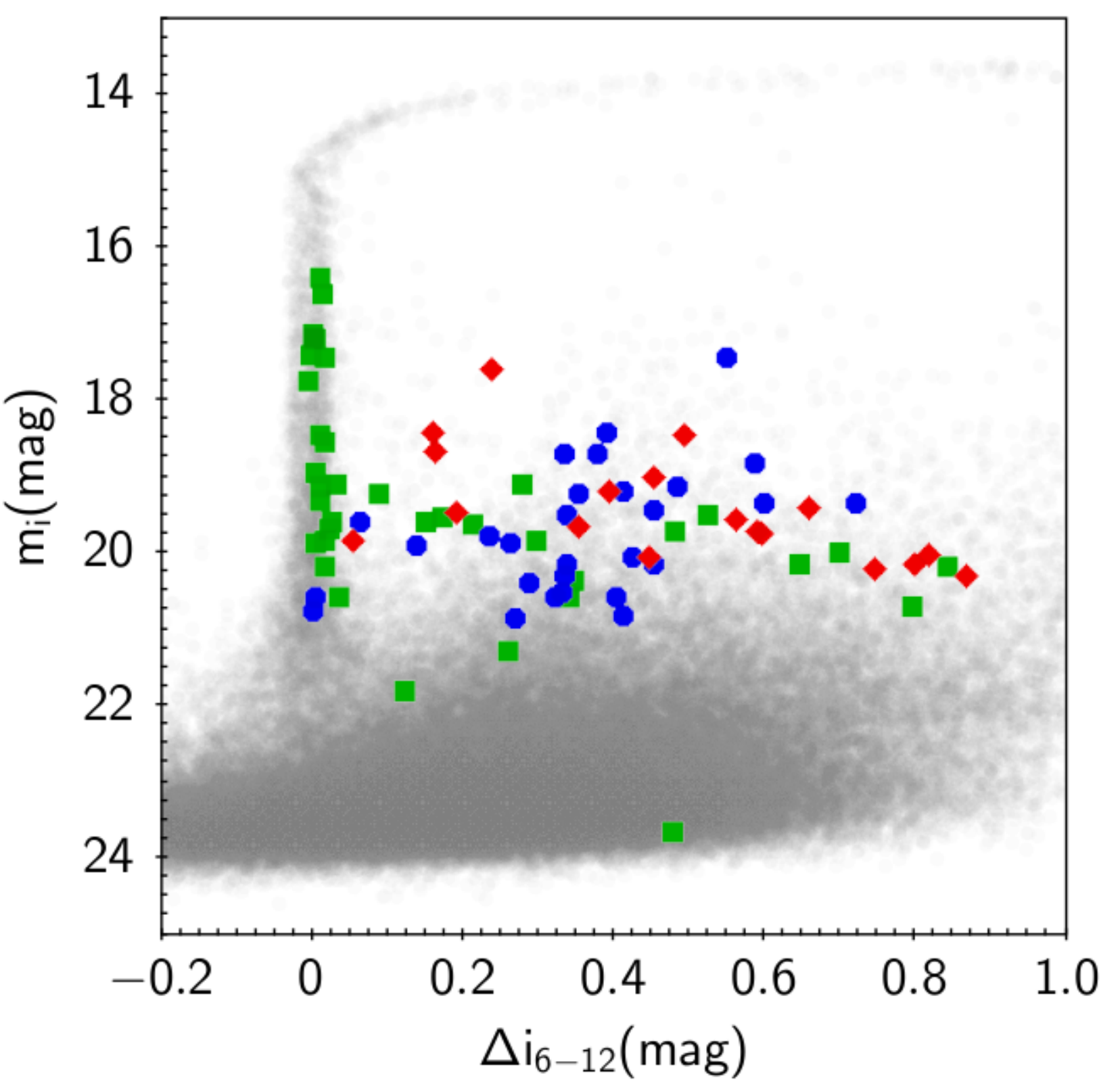}
  \caption{Color-color ($uiK_s$, left) and $i$- band concentration
    index ($\Delta i_{6-12}$, right) diagrams for the full matched
    catalog (gray symbols) and the photometric GC candidates from the
    literature. Red diamonds, blue circles, and green squares indicate
    \citet{liller83}, \citet{blecha86}, and \citet[Table 6]{beasley00} GC
    candidates, respectively. }
  \label{previous}
  \end{figure*}

\subsection{Comparison with spectroscopic and photometric GC samples}

Our sample of $\sim350$ color-color, morpho- and photometric selected
GC candidates does not contain some of the spectroscopically confirmed
sources by either \citet{beasley00} or \citet{olsen04}. We already
anticipated the cases of 2 out of 11 GCs from \citet[][IDs \#109
  and \#114 from their Table 3]{olsen04}, which are consistent with
being foreground stars in all bands, including $J$ and $K_s$, as they
are coherently consistent with stellar morpho-photometric
parameters. The two objects have line-of-sight velocity of $177\pm
5~km/s$ and $192\pm17~km/s$, which are relatively high and explain
their classification as GCs in NGC\,253, which has $cz=243\pm2$
km/s. Nevertheless, the photometric properties of the couple indicate
they are likely high-velocity MW stars \citep[e.g.,][]{xue08}. The
remaining 9 sources from \citeauthor{olsen04} are correctly selected
as GCs in our final sample.

As for the sample of confirmed GCs by \citet{beasley00}, the clusters
with IDs LA11, LA24, B1, B13, B14, and B29 from their Table 2 (we
adopted the alternative IDs given by the authors), are not
selected. Because LA24 is very close to a bright star ($m_V<9$ mag),
this GC is undetected in some passbands, while B1 is undetected in the
$K_s$ band because it is faint.

The other missing four candidates are excluded from our sample because
of their colors (LA11 also for its concentration index, see Figure
\ref{sel_g}).  The rejected candidates are shown in the RGB thumbnail
of Figure \ref{bs00}. The candidate LA11, in addition to the non-GC
colors, shows the presence of obvious features in all the imaging data
from VST; for B29, its red colors are consistent with a background
early-type galaxy; this possibility is also supported by its
elongation, which exceeds our adopted $b/a$ limit of 0.67, with both
the SExtractor and Ishape analyses. Hence, for both the latter
objects, our analysis rather supports the non-GC nature of the two
sources.

Sources B13 and B14 appear deeply enshrouded in the dust of the
galaxy. Hence, because of host-galaxy extinction, the colors of the
sources were off the color-color areas we adopted.

All such missed sources are in any case included in the final table of
GC candidates, properly commented in our classification scheme.

As for the comparison with previous photometric catalogs of GCs, in
Figure \ref{previous} we plot some properties of our full catalog of
matched sources with the samples of photometric candidates from
\citet{liller83}, \citet{blecha86}, and \citet[][]{beasley00}. In
addition to the spectroscopic sample used here, the latter authors
presented a sample of $\sim 90$ photometrically selected GCs. The
figure highlights that a substantial number of selected candidates are
indeed stars or background galaxies, both because of their colors or
the concentration index, or both. The improved efficiency of the
analysis presented here is due to a combination of the larger
inspected area, which is a factor of $\sim$3 to $\sim10$ with respect
to previous studies, the better seeing conditions, from 10\% better to
300\%, and the much wider wavelength coverage; other studies are based
on only $B$ or $B$ and $V$ photometry.

\subsection{Globular cluster sizes}
\label{sec_ishape}

At the distance of NGC\,253, and with the seeing conditions of our
observational dataset, the half-light radii $R_h$ of GCs can be
derived.  Size measurements can be very challenging, especially with
ground-based imaging data. In spite of this, angular sizes and
intrinsic shapes have been obtained for a large sample of slightly
resolved star clusters in different environments and with various
ground- and space-based telescopes
\citep[e.g.,][]{larsen99,larsen03,jordan04b,cantiello07c,caso13,puzia14,cantiello15}.

  \begin{figure*}
  \centering
  \includegraphics[width=\hsize]{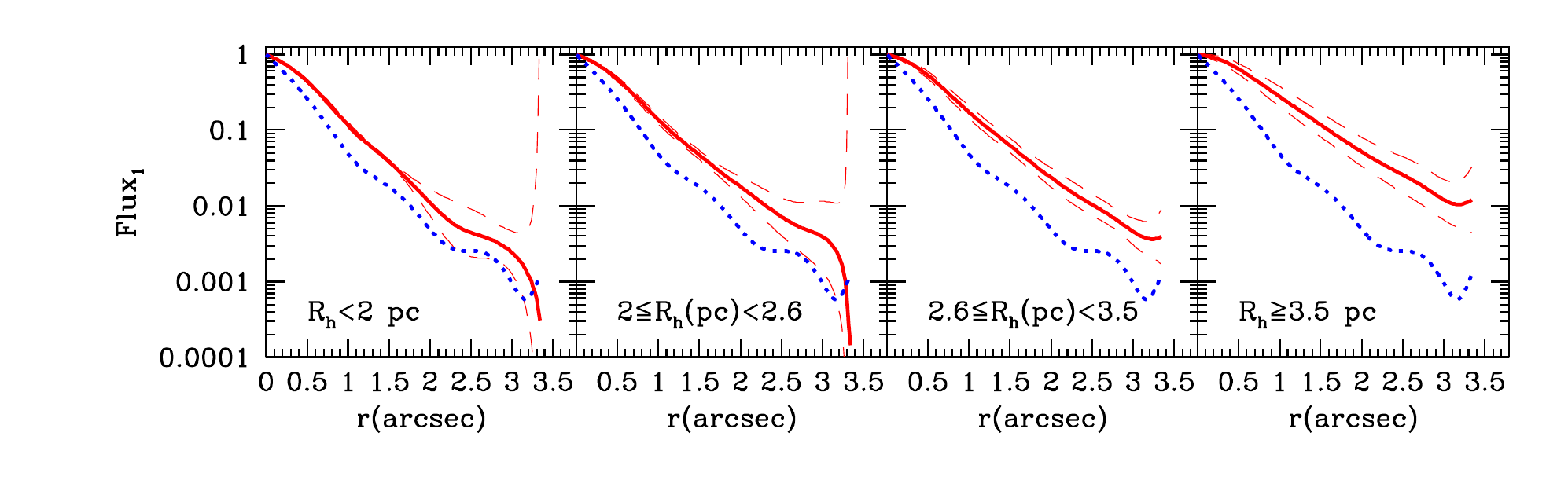}
  \caption{Radial flux profiles of GCs in $g$ band. The various
    panels show the average flux profile (solid red lines, normalized
    to one at center) for GC candidates with measured $R_h$ within the
    labeled interval. Thin dashed lines show the standard deviation
    of the mean for the average profile; blue dotted line indicates the
    PSF profile.}
  \label{radialgc}
  \end{figure*}

To estimate the intrinsic size of a source exceeding some
instrumental-dependent size limit, specific tools have been designed
and implemented to analyze the light profiles of sources with
intrinsic sizes comparable or slightly smaller than the instrumental
point spread function (PSF).  We adopted Ishape\footnote{The software can be downloaded at
  http://baolab.astroduo.org/. For the present work we used the
  release 0.94.1e.} to obtain structural parameters (in particular
$R_h$ and the minor-to-major axis ratio $b/a$) of candidate
GCs. Ishape is optimized for modeling the light distribution for
marginally resolved sources down to 1/10 of the FWHM of the PSF
\citep{larsen99,larsen00}. In such context, the VST dataset of
NGC\,253 is very attractive. At the adopted distance modulus
(corresponding to $\sim3.47$ Mpc), and given the FWHM of the images
(Table \ref{tab_props}), Ishape can be used to determine the physical
extent of objects with $R_h\geq1.3$ pc. For reference, excluding
highly extincted GCs, with $E(B-V)\geq0.5$, the MW hosts two GCs with
$R_h\lsim1.2$ pc and five with $R_h\leq1.5$ pc \citep{harris96}.  The
median is $R_h=3.22$ pc. Since the measurement of source sizes
  below the FWHM is particularly demanding in terms of signal-to-noise
  ratio and image quality, we limited the analysis of GC radii to
  $gri$ band data. Using Ishape, we fitted all sources with a King
  profile with concentration index c=15 \citep{larsen99,larsen00}. The
  final $R_h$ and $b/a$ values are derived from the weighted average
  of the three bands.

Figure \ref{radialgc} shows the $g$-band radial flux profile of our
{\it bona fide} GC candidates (see next section; flux is normalized to
peak one), compared with the radial profile of the PSF in the same
band. In this panel we plot the average profile of GCs with
estimated effective radii within the labeled $R_h$ intervals. The
width of the $R_h$ intervals is chosen to contain similar numbers of
GC candidates ($\sim20$) per $R_h$ bin. The figure shows the
significant differences between PSF (i.e., stellar) and GCs light
profiles even for the most compact candidates, reported in the left
panel. Hence, unlike typical studies of extragalactic GCs
\citep[e.g.,][]{durrell14}, MW stars represent a minor source of
contamination in our GC catalog because of the combined effect of
galaxy distance, GC physical size, and good image quality.

Finally, we specifically run Ishape on the two sources from
\citet{olsen04} that we identified as stars (IDs \#109 and \#114,
mentioned in previous section). The results confirm the single star
origin of the two sources, as their $R_h$ are consistent with zero in
all three inspected bands, and the $\chi^2$ for the fit to an extended
source does not improve with respect to the $\chi^2$ obtained modeling
a compact, stellar source.


  \begin{figure*}
  \centering
  \includegraphics[width=0.22\hsize]{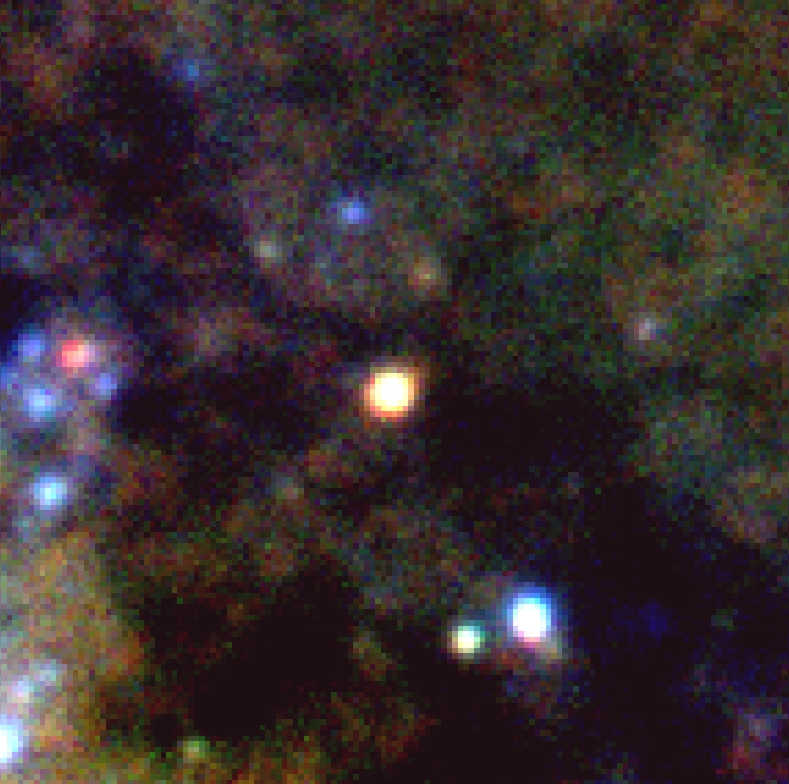}
  \includegraphics[width=0.22\hsize]{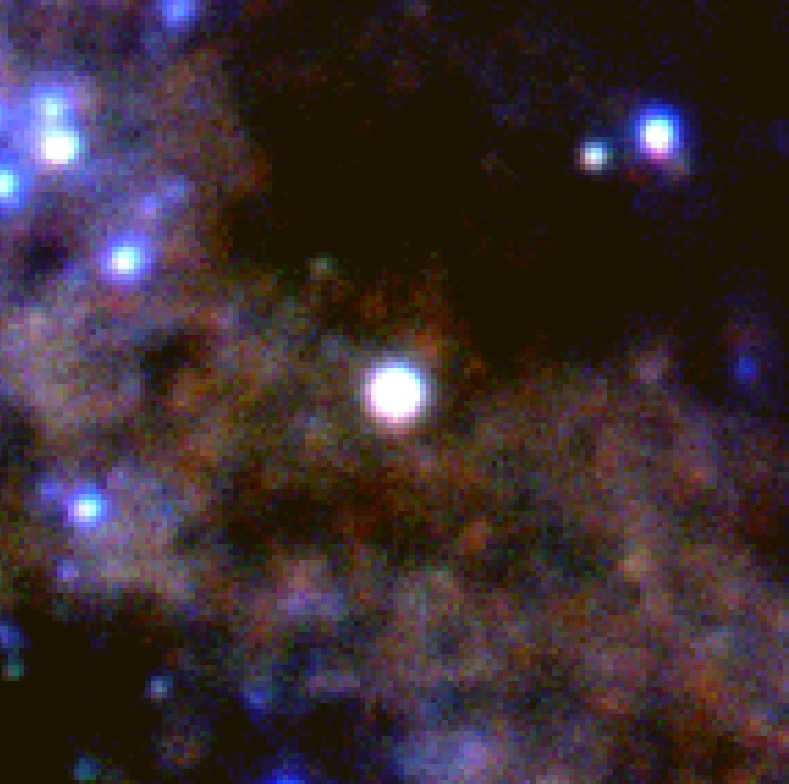}
  \includegraphics[width=0.22\hsize]{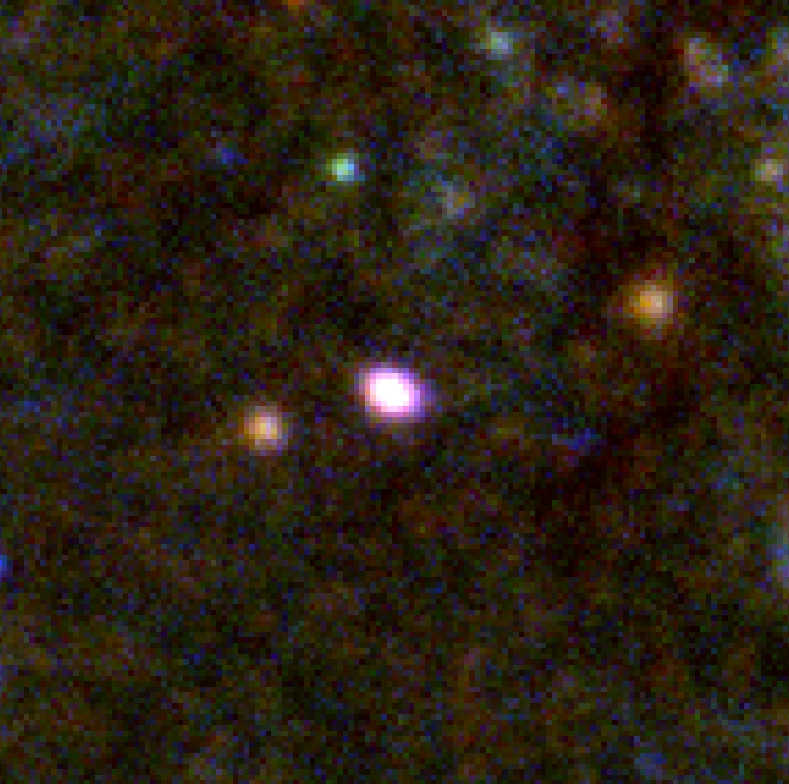}
  \includegraphics[width=0.22\hsize]{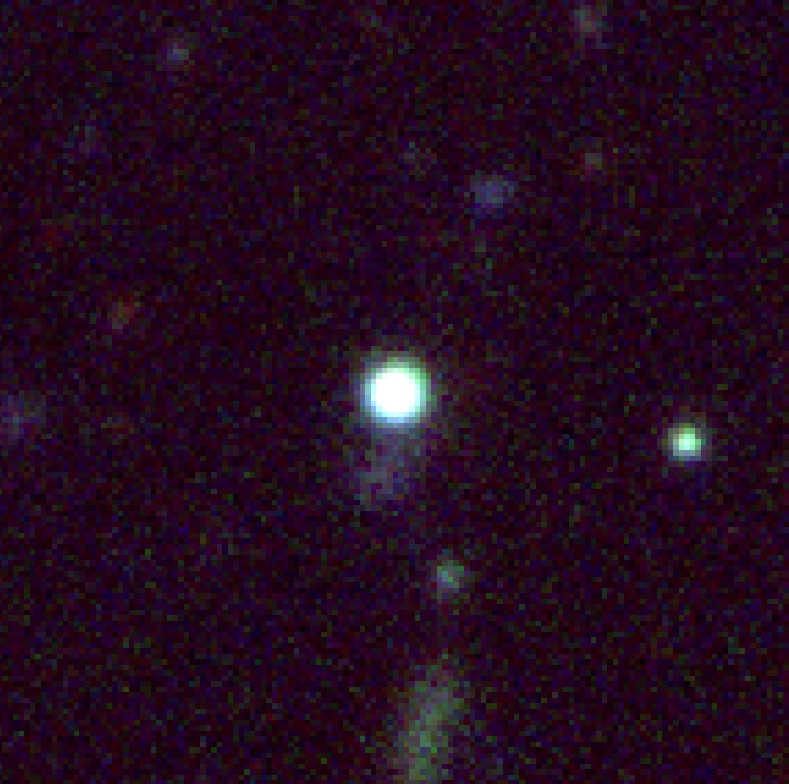}
   \caption{RGB thumbnails from VST data of star cluster candidates
     from \citet{beasley00} that did not pass our selection criteria
     for GC candidates. Starting from left: B13 and B14 (close to
     galaxy dusty regions), and B29 and LA11 (likely background
     galaxies). Thumbnails are 15$\arcsec$ on each side.}
   \label{bs00}
  \end{figure*}

  \begin{figure*}
    \centering

\begin{subfigure}{\textwidth}
    \centering
  \includegraphics[width=0.135\hsize]{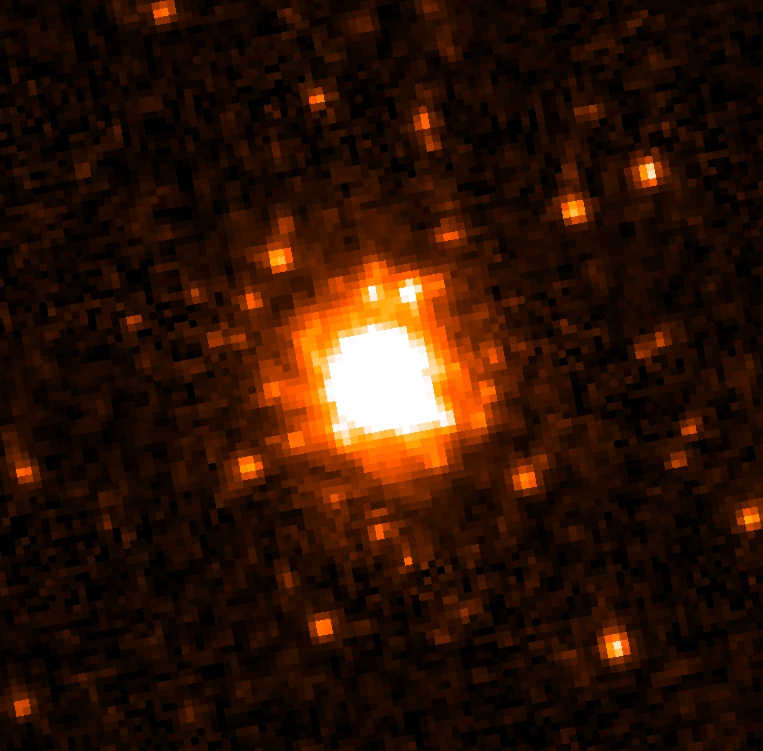}
  \includegraphics[width=0.135\hsize]{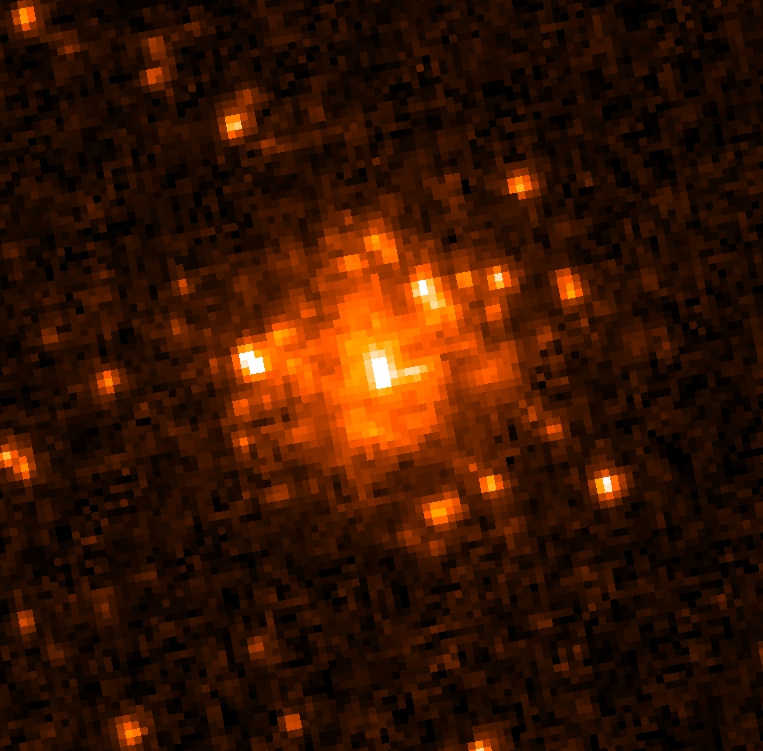}
  \includegraphics[width=0.135\hsize]{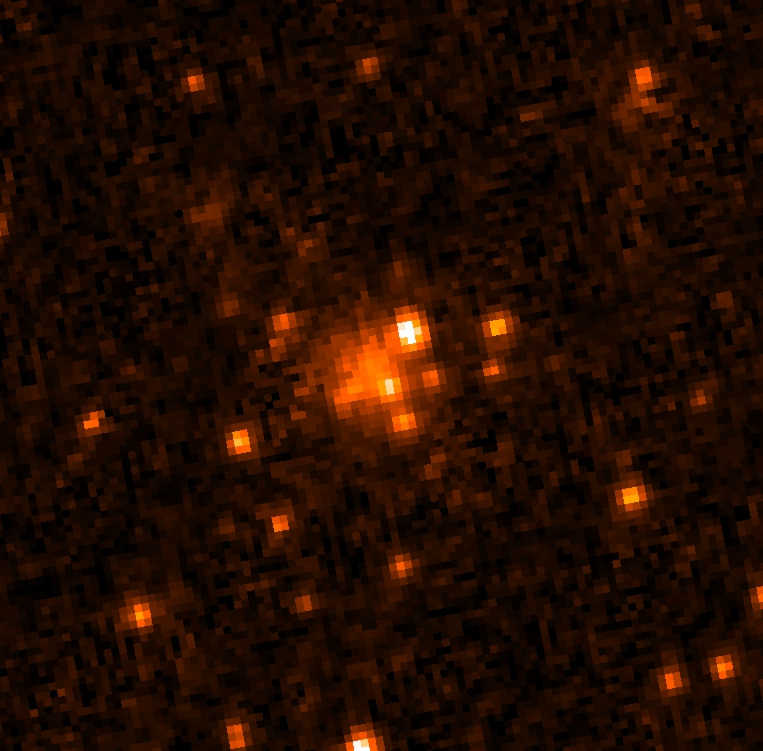}
  \includegraphics[width=0.135\hsize]{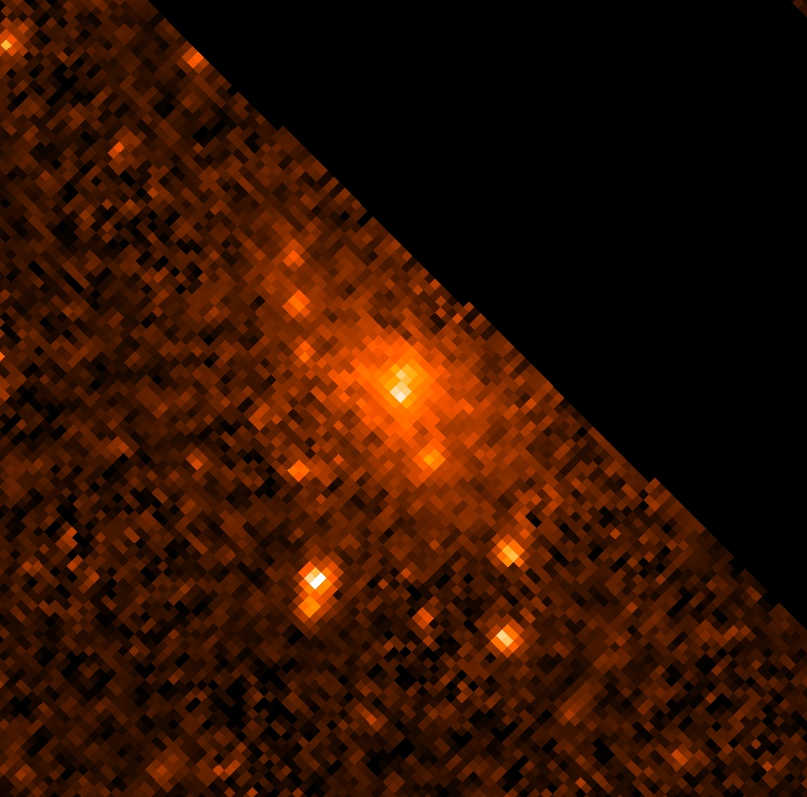}
  \includegraphics[width=0.135\hsize]{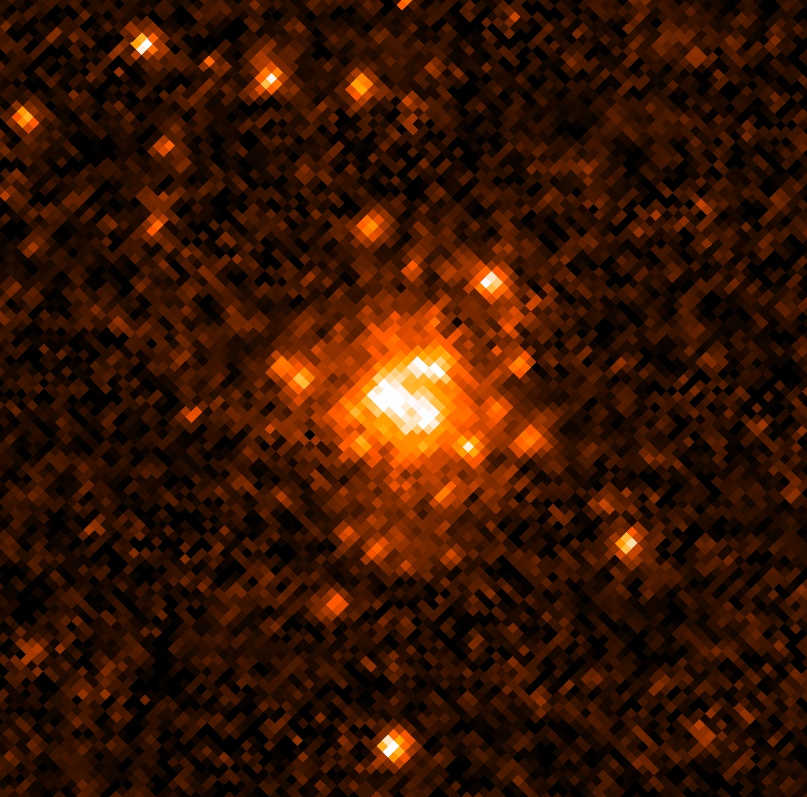}
  \includegraphics[width=0.135\hsize]{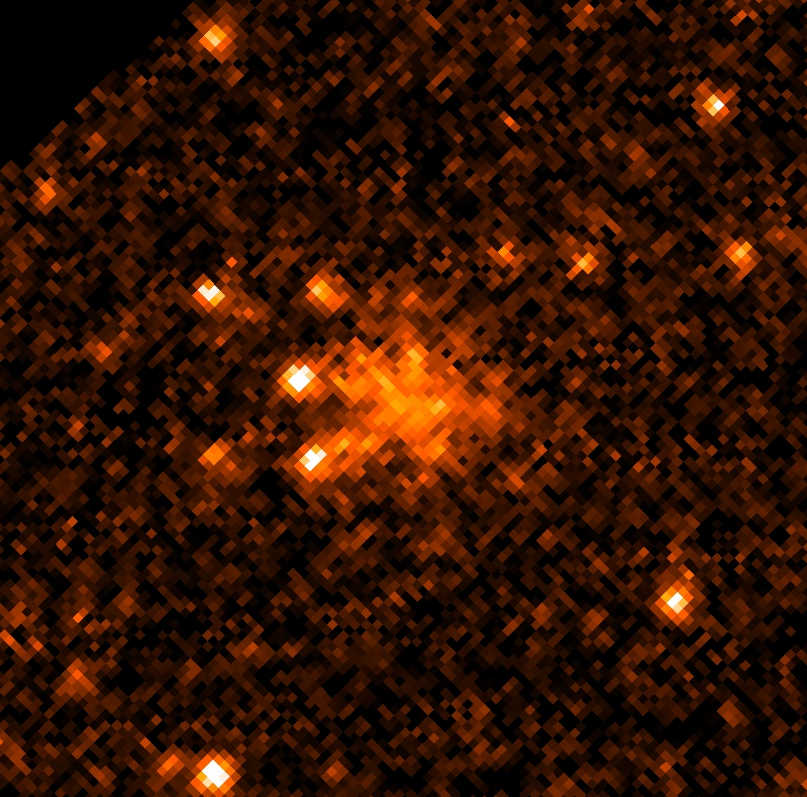}
\caption{From left to right: Cluster candidates number \#99, \#109,
  \#111, \#124, \#128 and \#141 in Table \ref{tab_final}.} 
\end{subfigure}
\begin{subfigure}{\textwidth}
    \centering
  \includegraphics[width=0.135\hsize]{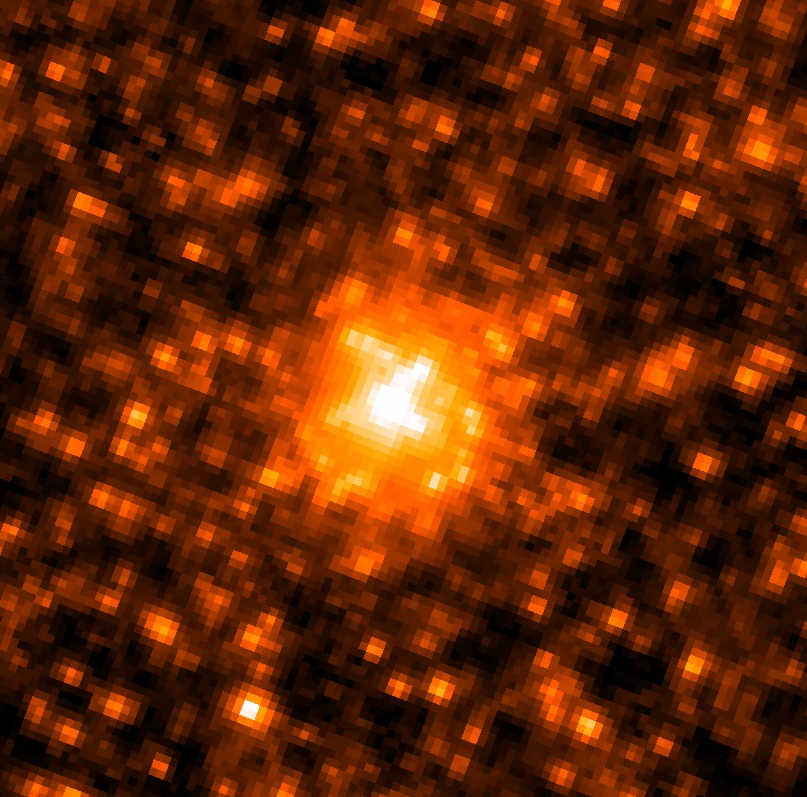}
  \includegraphics[width=0.135\hsize]{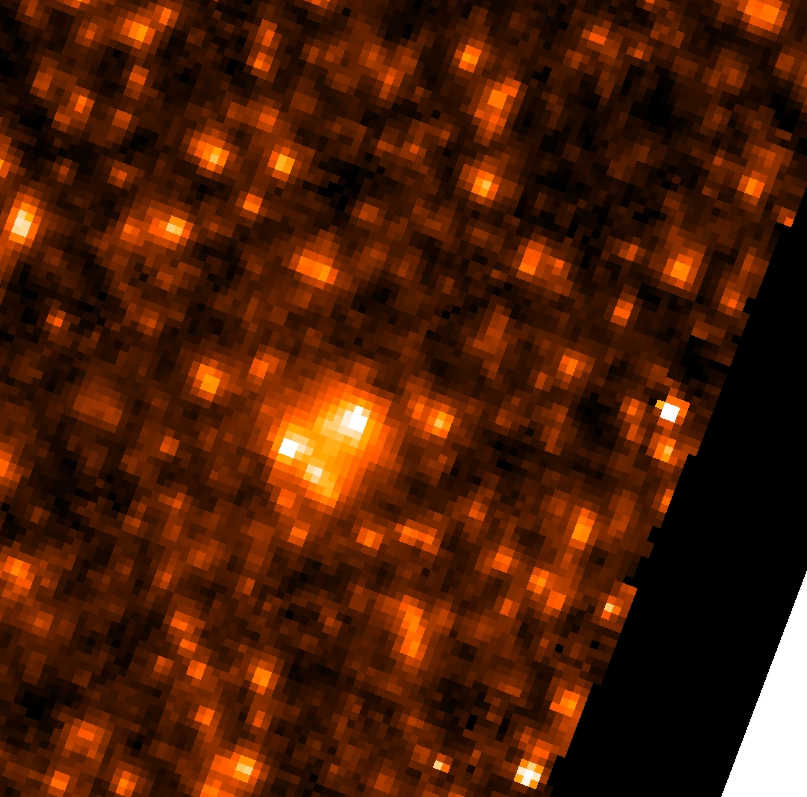}
  \includegraphics[width=0.135\hsize]{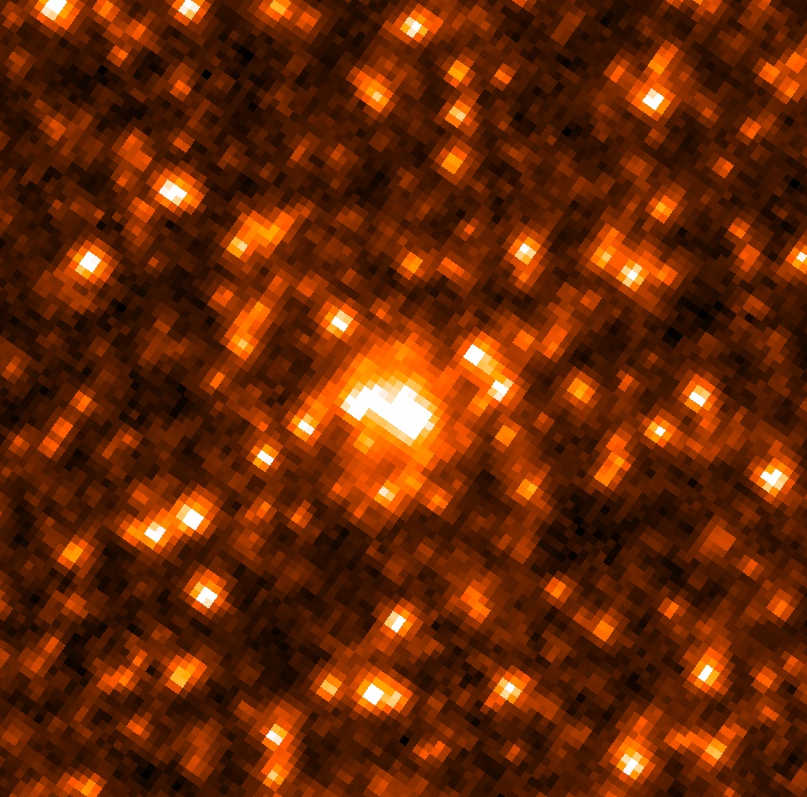}
  \includegraphics[width=0.135\hsize]{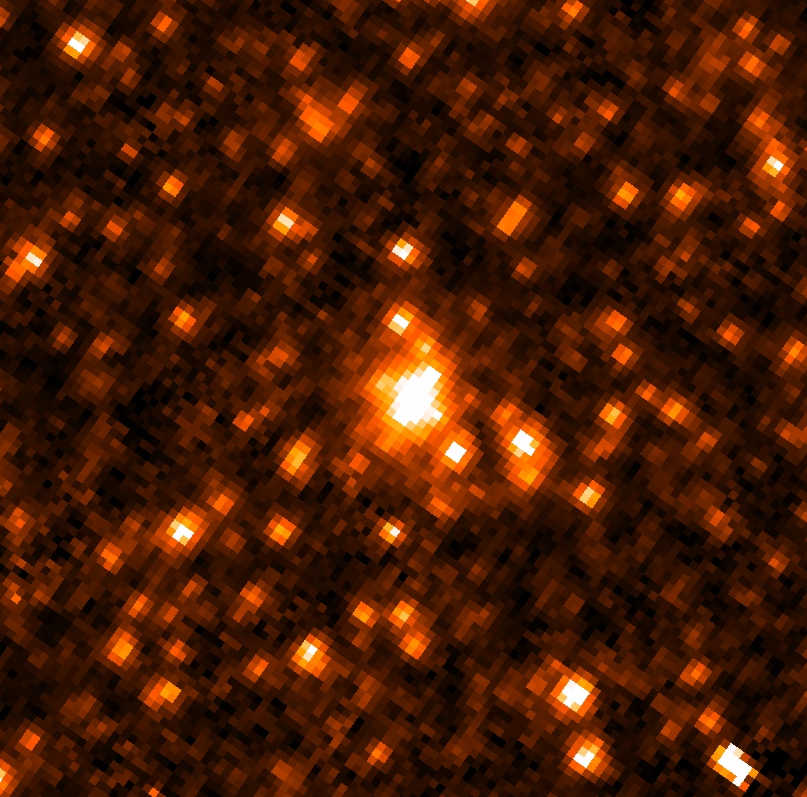}
  \includegraphics[width=0.135\hsize]{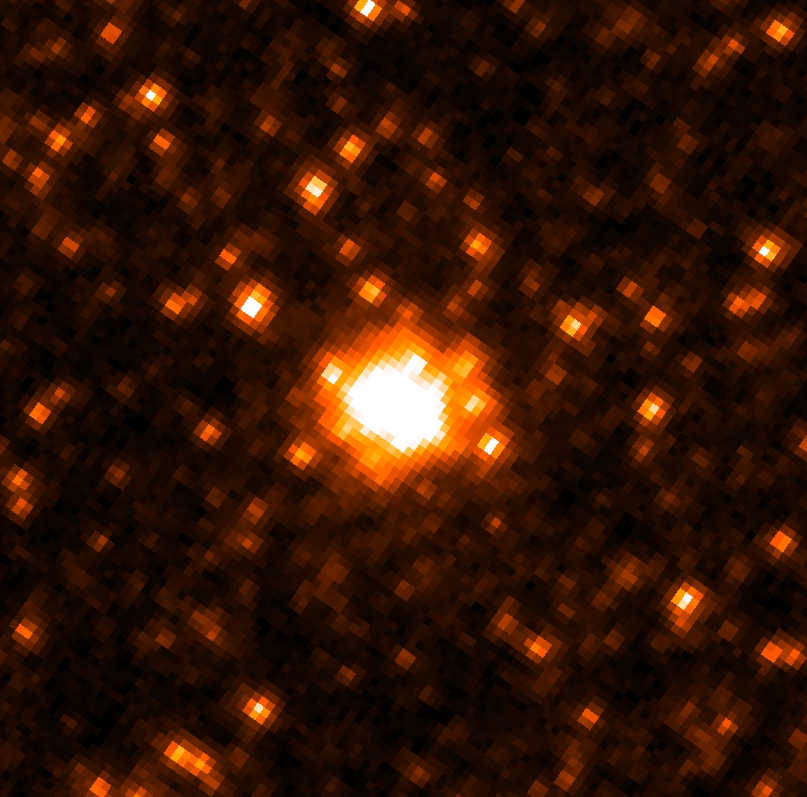}
  \includegraphics[width=0.135\hsize]{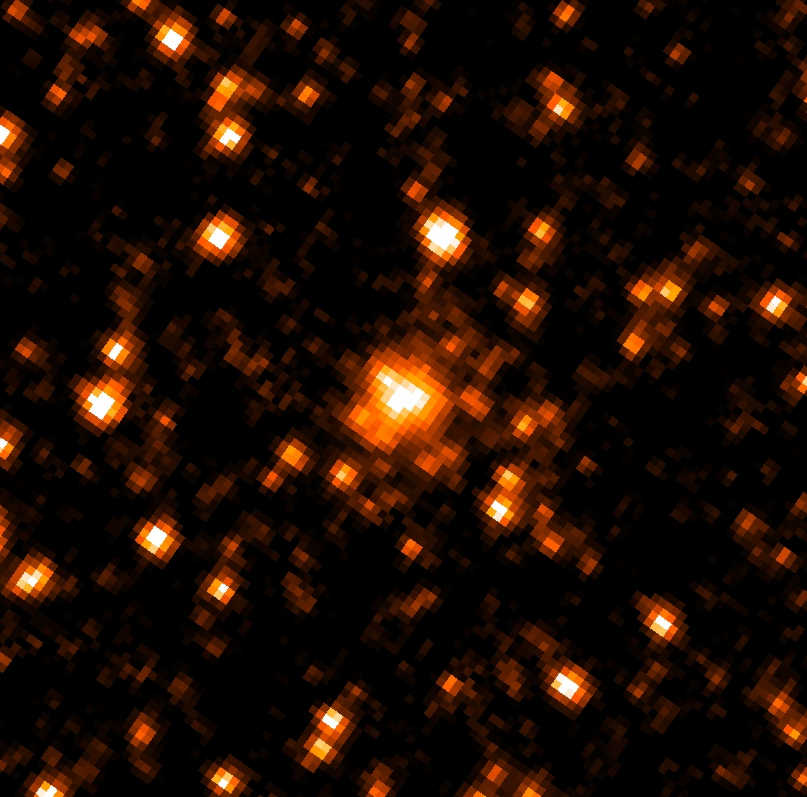}
\caption{From left to right: Cluster candidates number \#178, \#186,
  \#200, \#204, \#205 and \#207, in Table
  \ref{tab_final}.} 
\end{subfigure}
\begin{subfigure}{\textwidth}
    \centering
  \includegraphics[width=0.135\hsize]{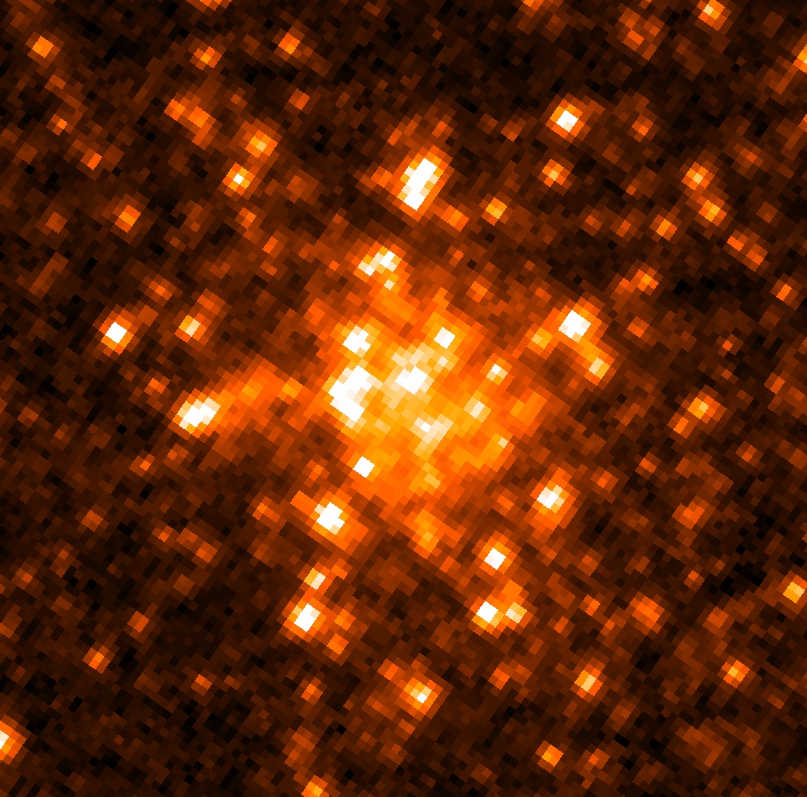}
  \includegraphics[width=0.135\hsize]{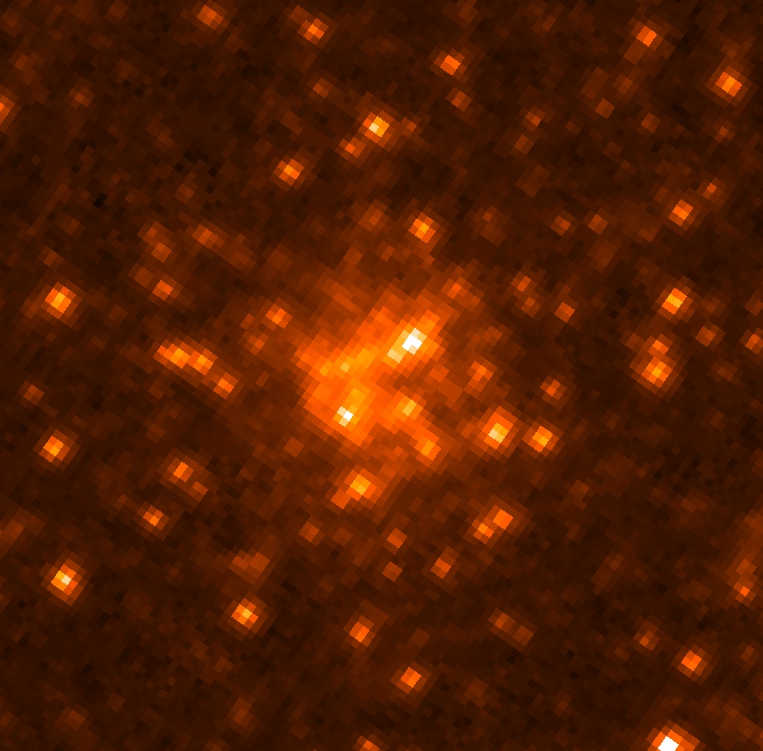}
  \includegraphics[width=0.135\hsize]{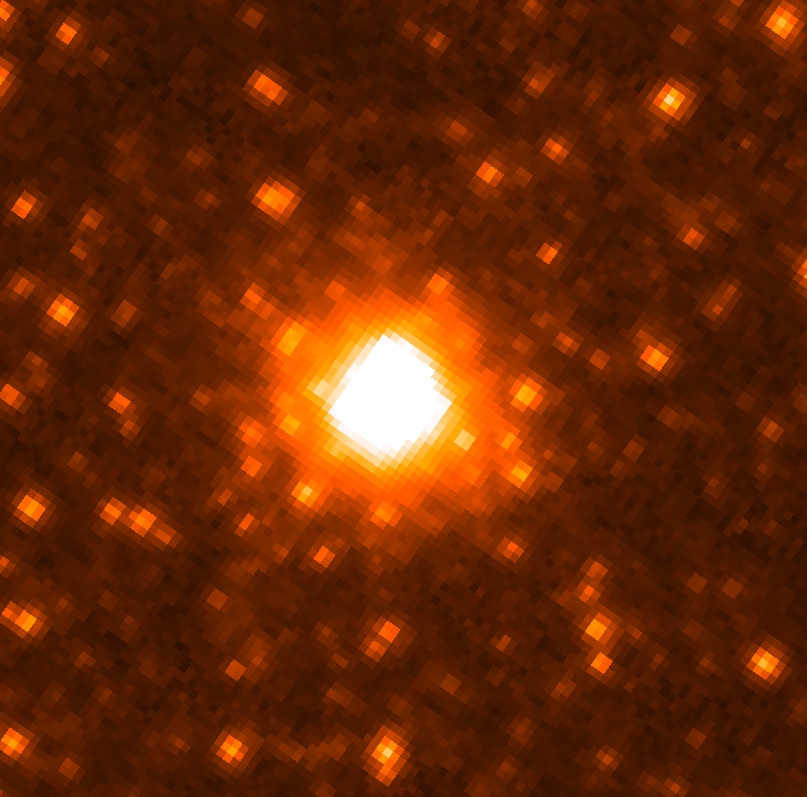}
  \includegraphics[width=0.135\hsize]{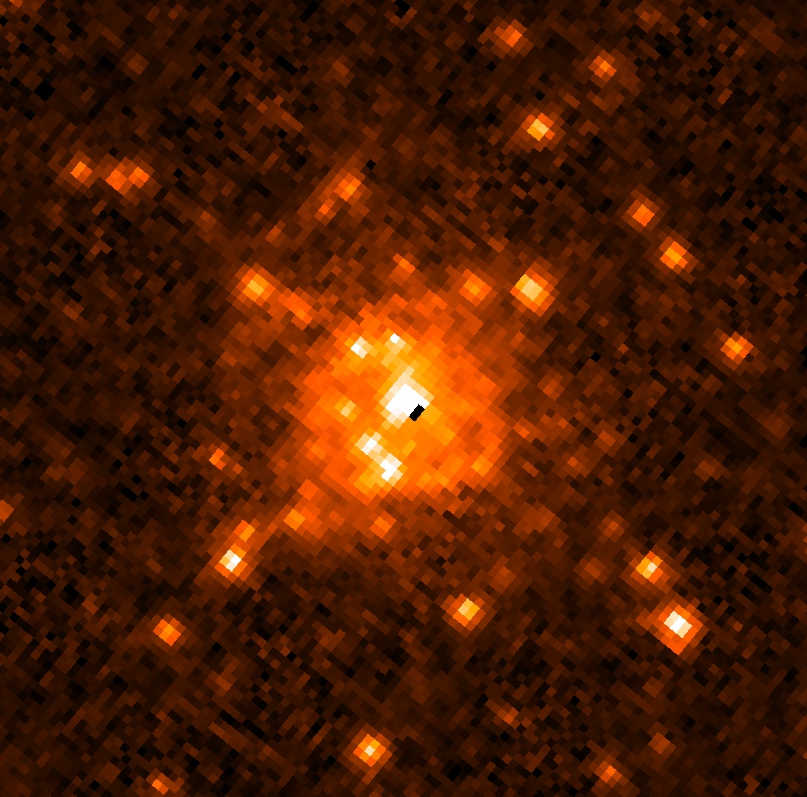}
  \includegraphics[width=0.135\hsize]{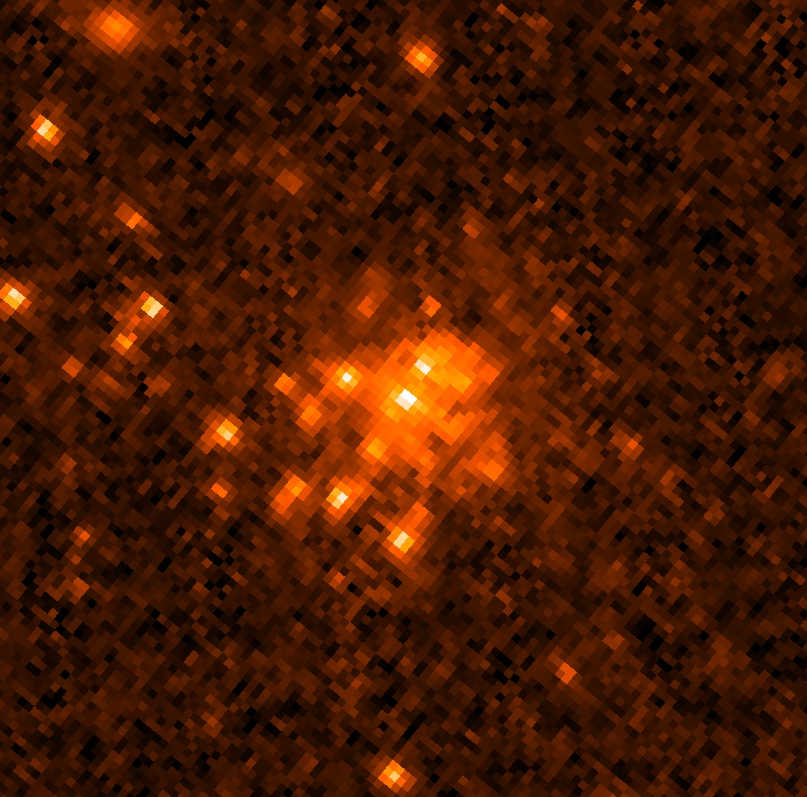}
  \includegraphics[width=0.135\hsize]{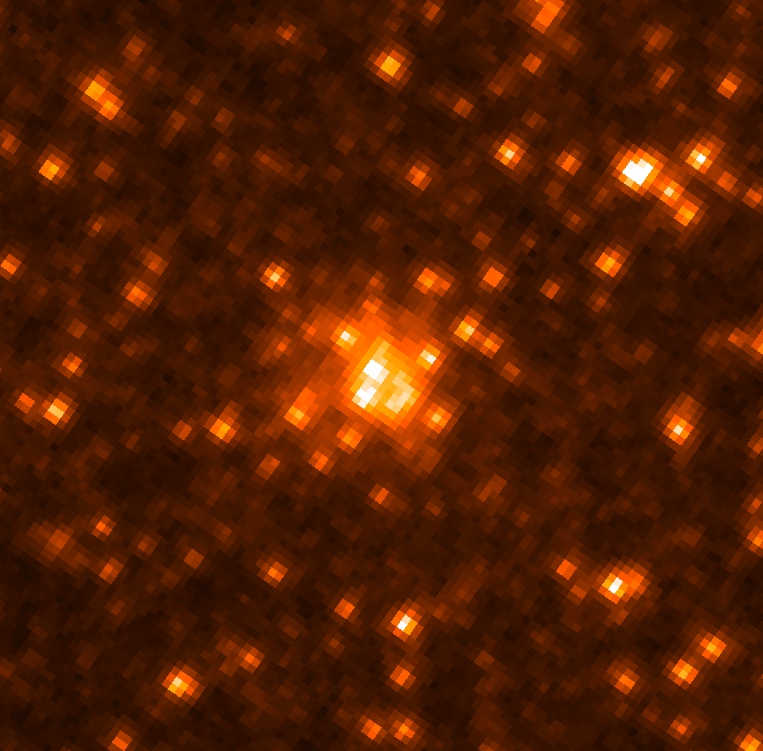}
\caption{From left to right: Cluster candidates number \#209, \#210,
  \#211, \#218, \#227 and \#325 in Table
  \ref{tab_final}.} \label{ghost1}
\end{subfigure}
\begin{subfigure}{\textwidth}
     \centering
  \includegraphics[width=0.135\hsize]{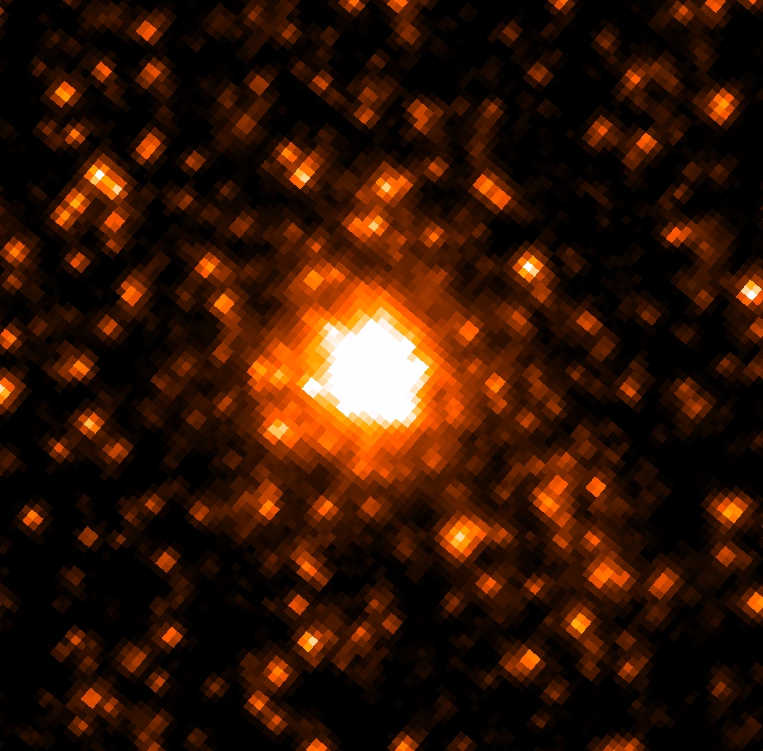}
  \includegraphics[width=0.135\hsize]{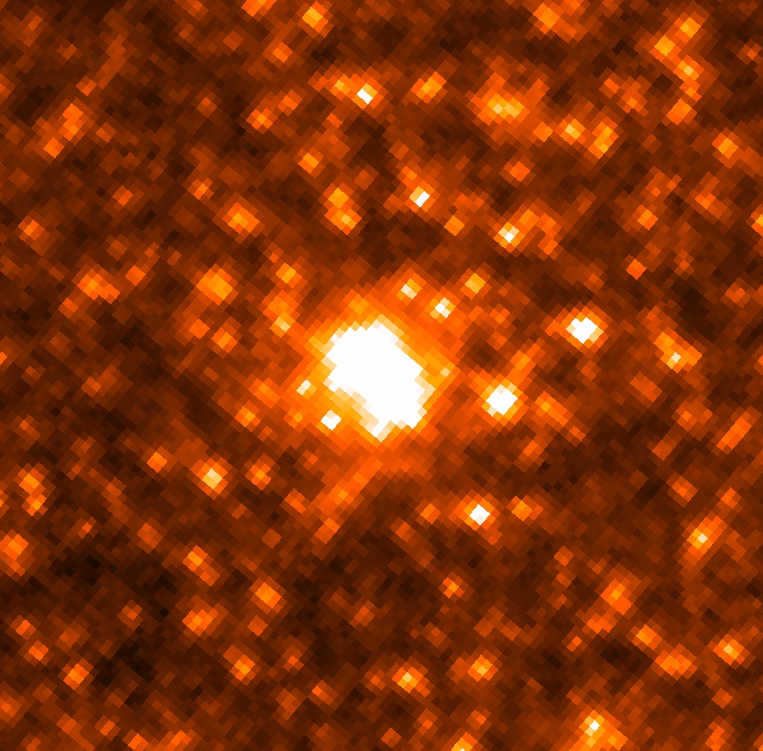}
  \includegraphics[width=0.135\hsize]{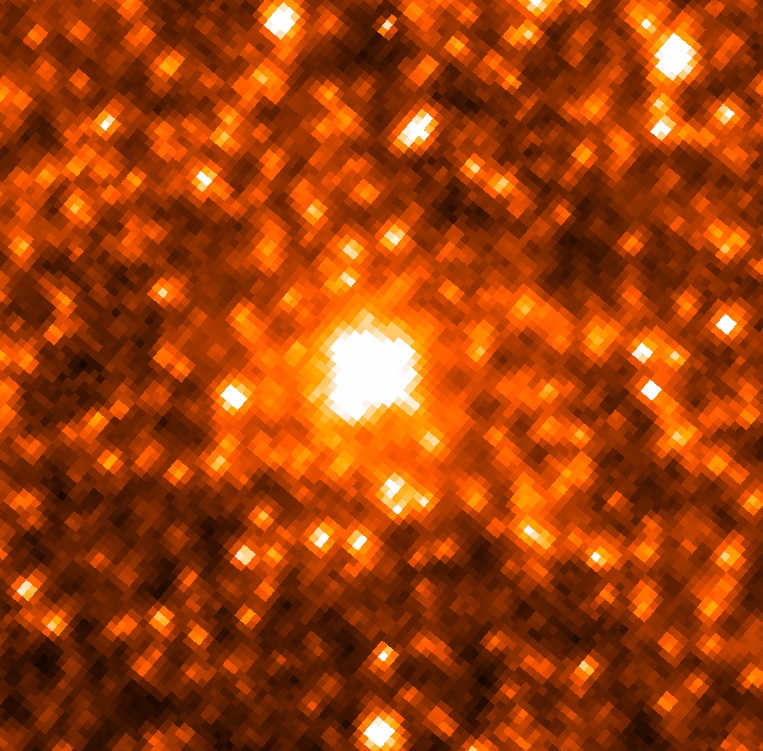}
  \includegraphics[width=0.135\hsize]{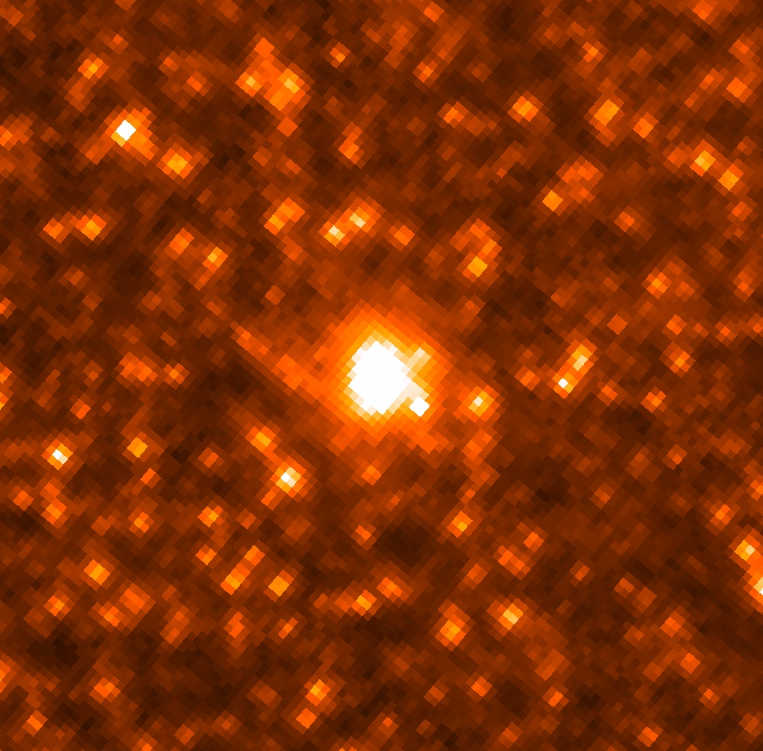}
 \caption{From left to right: Cluster candidates number \#333, \#334,
  \#335 and \#336 in Table \ref{tab_final}.} 
\end{subfigure}
  \vskip +0.2cm
 \begin{subfigure}{\textwidth}
    \centering
 \includegraphics[width=0.135\hsize]{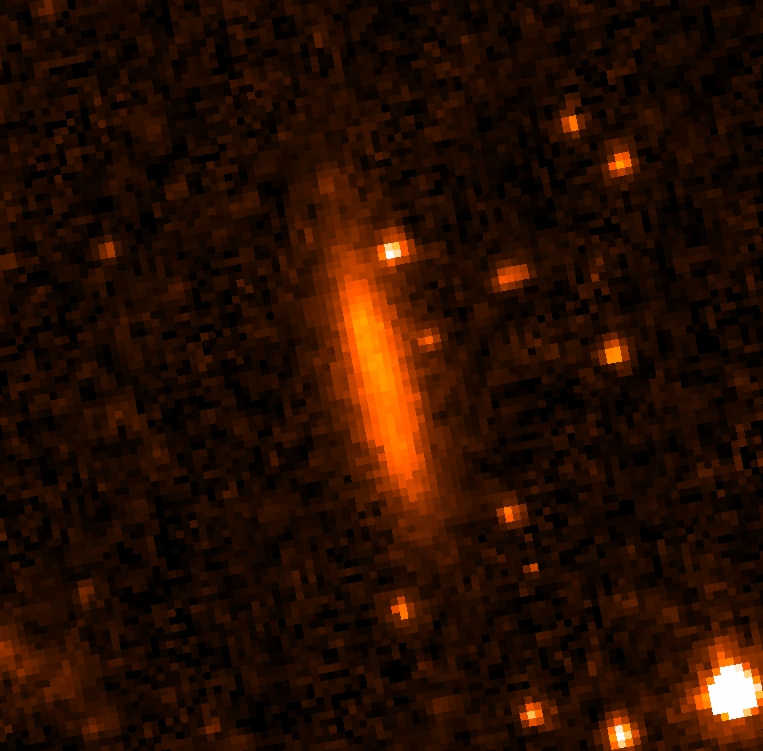}
  \includegraphics[width=0.135\hsize]{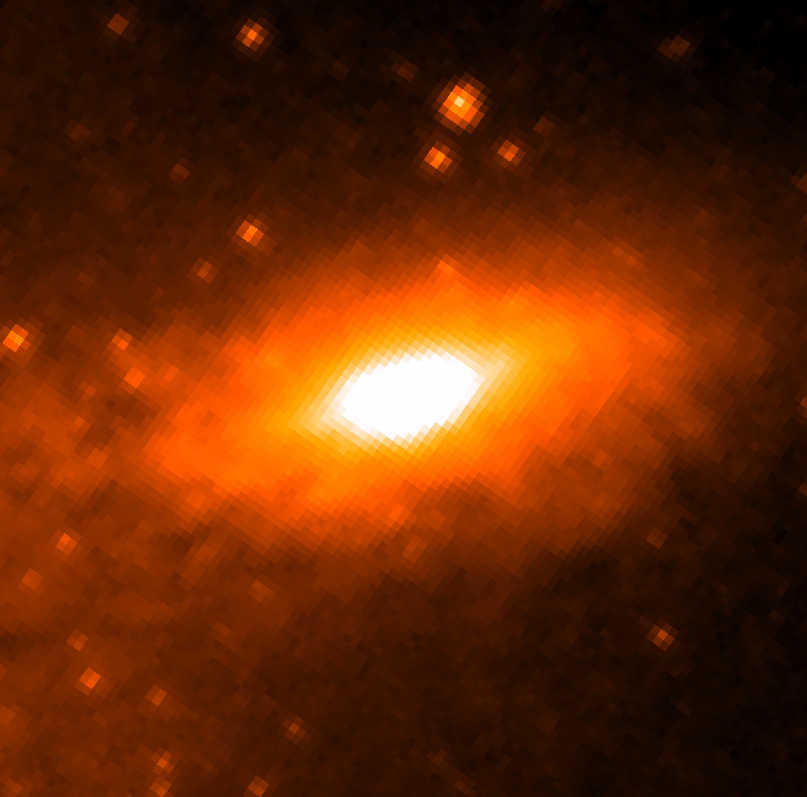}
 \includegraphics[width=0.135\hsize]{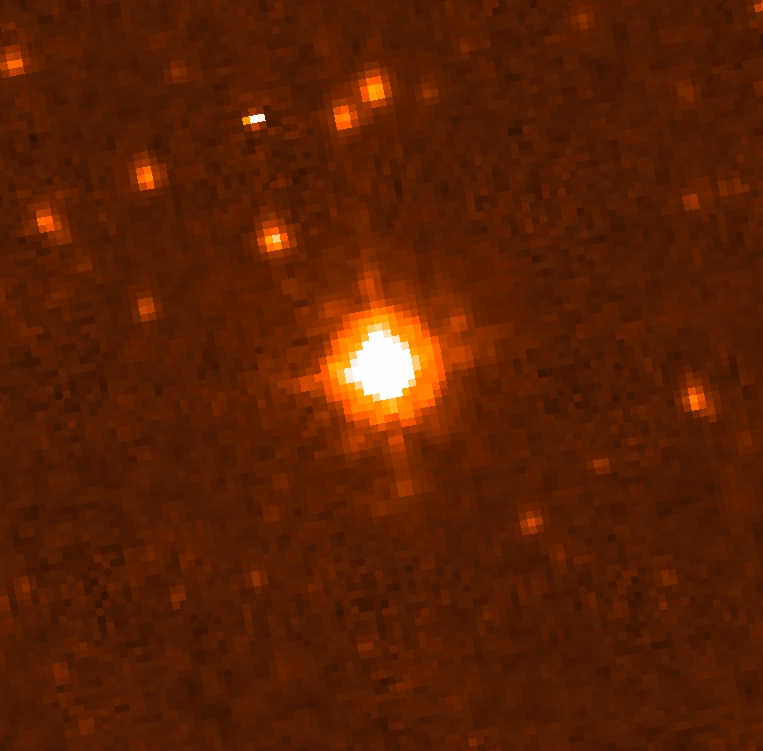}
  \includegraphics[width=0.135\hsize]{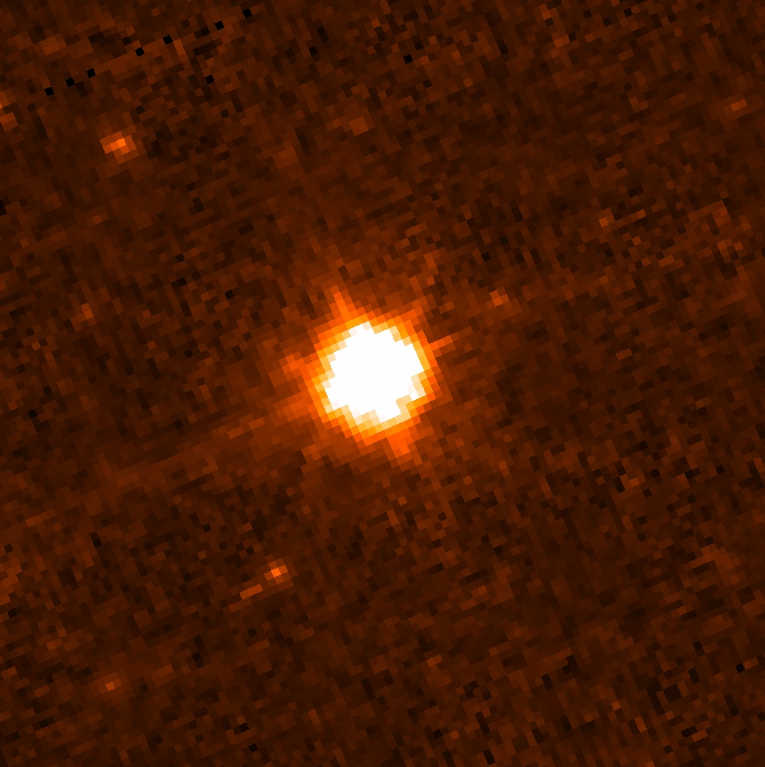}
  \caption{From left to right: Sources \#107 and \#212 in Table
    \ref{tab_final} (confirmed background galaxies), and two of
    the sources identified as stars in our selection procedure.}
 \end{subfigure}
 \begin{subfigure}{\textwidth}
    \centering
  \includegraphics[width=0.135\hsize]{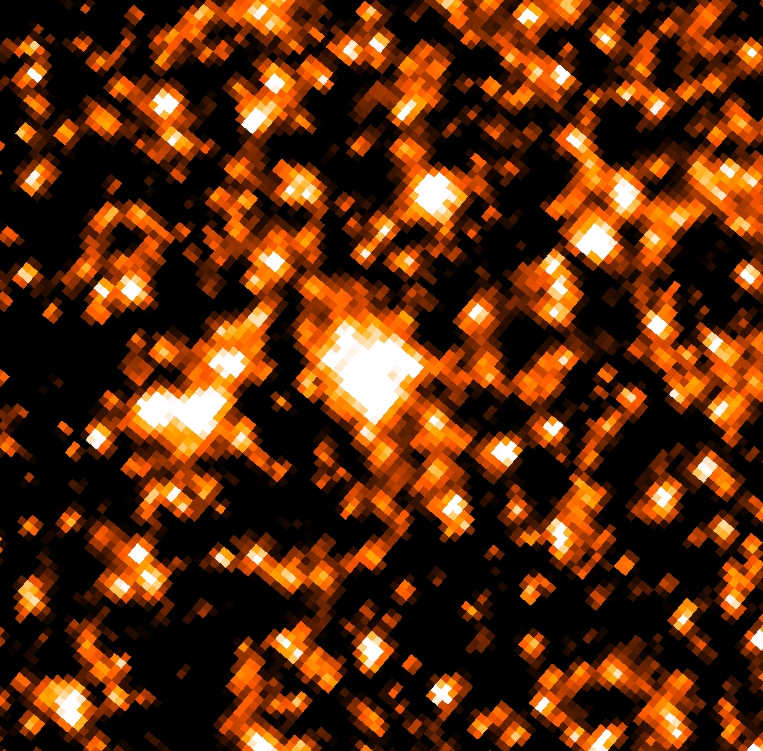}
  \includegraphics[width=0.135\hsize]{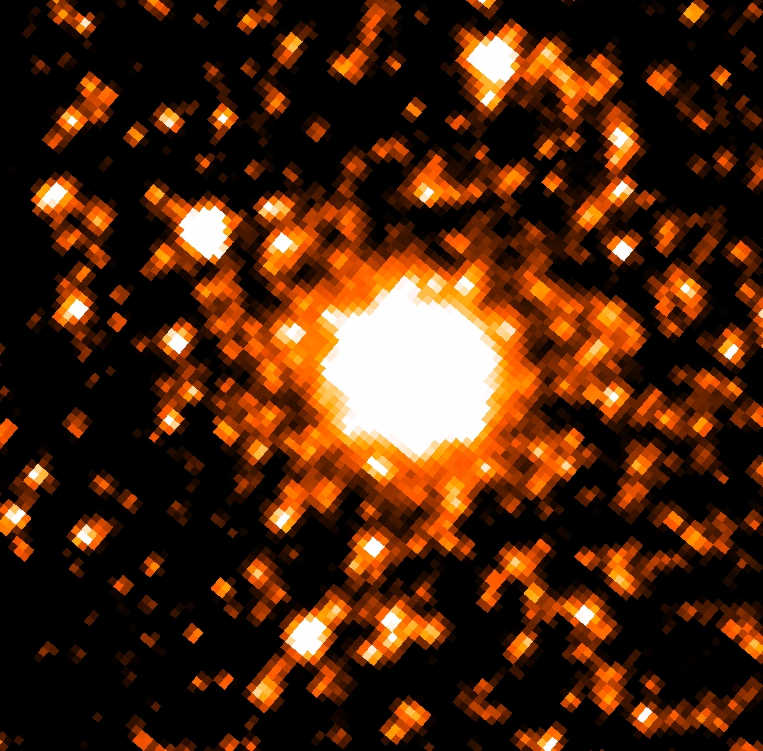}
  \includegraphics[width=0.135\hsize]{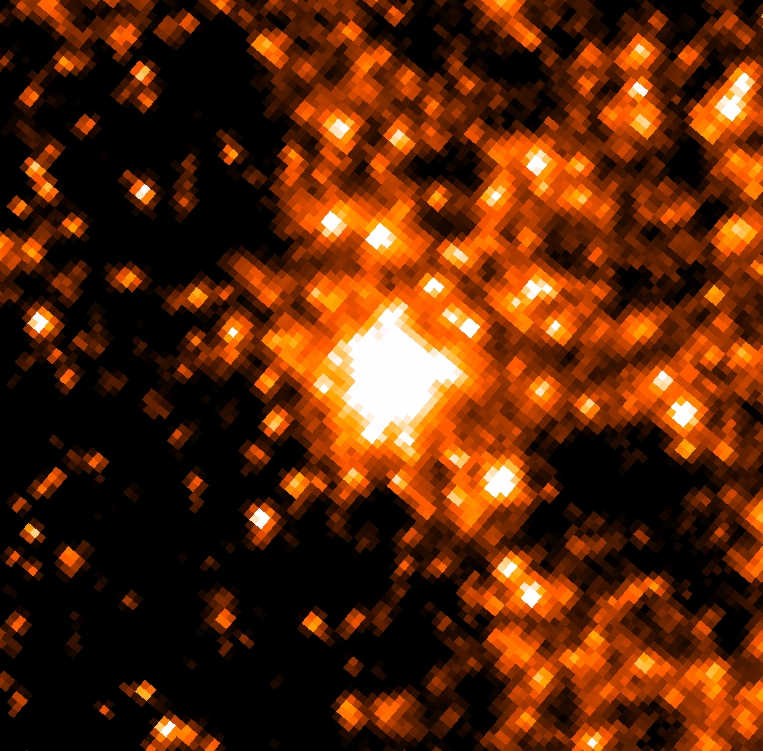}
  \includegraphics[width=0.135\hsize]{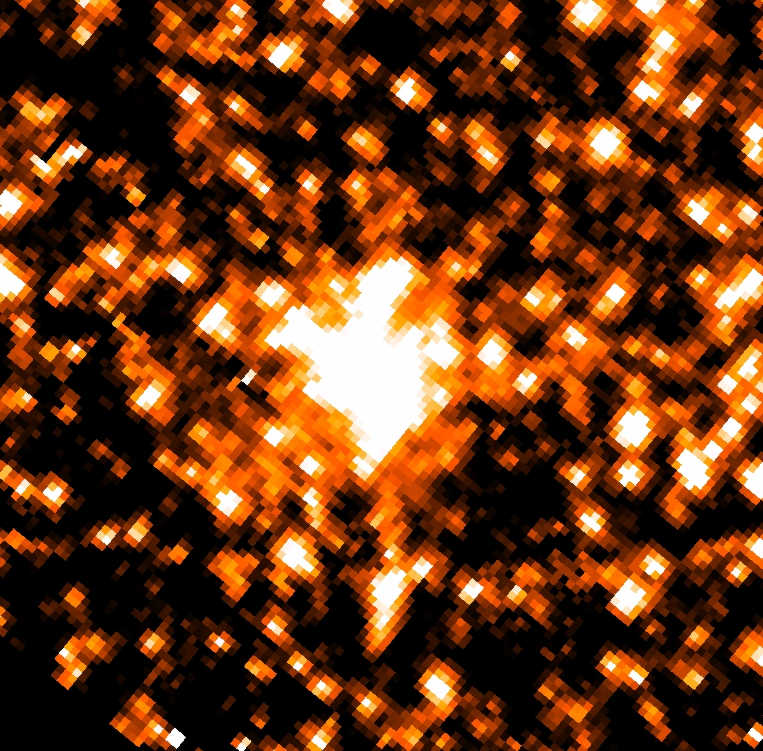}
  \includegraphics[width=0.135\hsize]{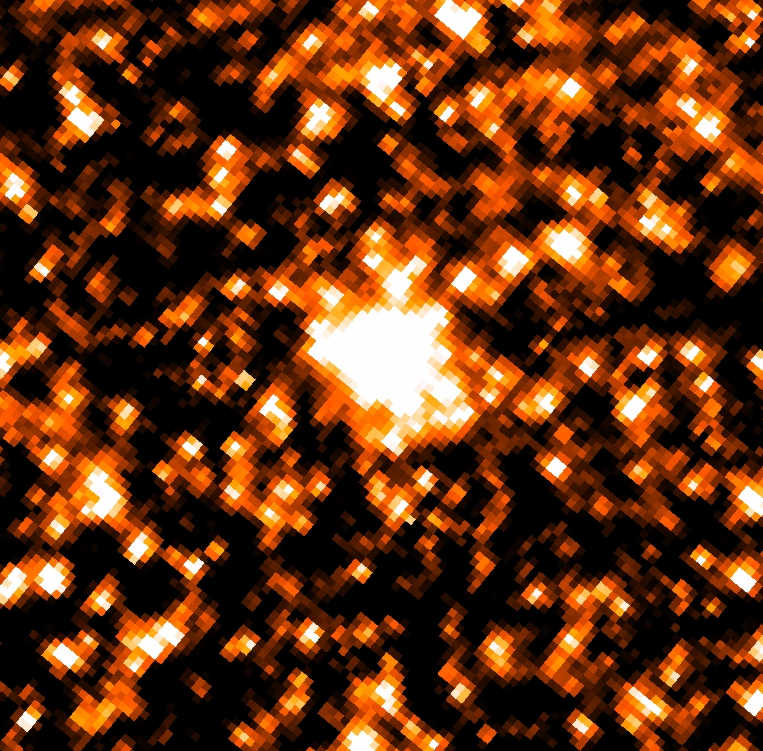}
  \includegraphics[width=0.135\hsize]{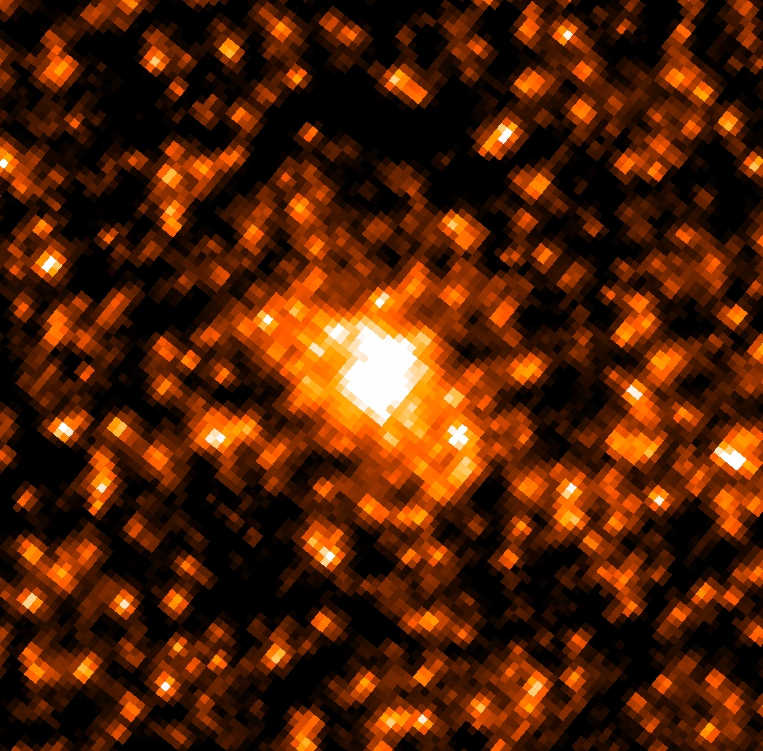}
  \includegraphics[width=0.135\hsize]{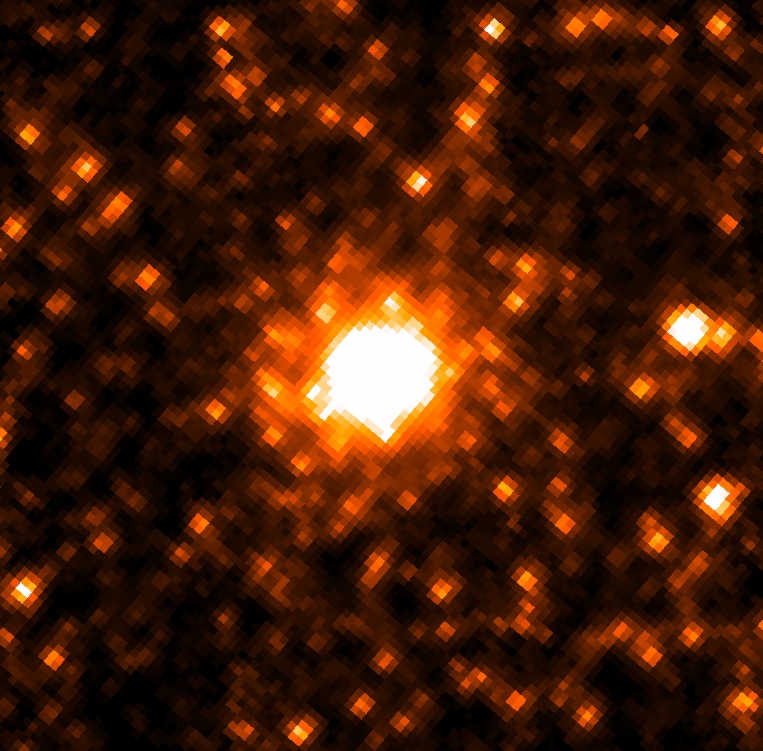}
  \caption{From left to right: Sources \#326, \#327, \#328,
    \#329,\#330, \#331 and \#332 in Table \ref{tab_final} visually
    identified as GCs, and located within the dust disk of NGC\,253.}
 \end{subfigure}
   \caption{Upper four rows (a-d): Hubble Space Telescope ACS cutouts of
     the sources selected as GCs candidates and falling in the GHOSTS
     survey footprints (F814W-band imaging data are shown). Given the
     mottled appearance, we consider all sources as star clusters in
     the galaxy. For reference, the panels in row (e) show the other
     two selected sources in the GHOSTS footprints, which are obvious
     background galaxies, and two stellar sources. The sources in the
     row ($f$) are the visually identified GC candidates (see text).}
   \label{ghosts}
  \end{figure*}

\section{Final catalog and discussion}
\label{sec_final}
\longtab{
\begin{landscape}
\miniscule
\begin{longtable}{lcccccccccccccc}
\caption{\label{tab_final} Final sample of selected GC candidates. The full table is available in electronic form at the CDS and the VEGAS survey website [NEED ROUNDOFF!].} \\
  \hline\hline
  ID & R.A.(J2000)  & Dec (J2000) &         $m_u$   &       $m_g$         &          $m_r$       &         $m_i$        &         $m_j$      &         $m_k$        &         $R_{h}$      & $b/a$    &     Other Identifications               & Spectra     &   Visual Inspection              &  Class   \\
     &    (deg)     &   (deg)     &     (mag )      &      (mag)          &         (mag)        &        (mag)         &        (mag)       &        (mag)         &         (pc)         &         &                                           & or HST      &                                  &          \\        
  (1)&    (2)       &   (3)       &     (4)         &      (5)            &         (6)          &        (7)           &        (8)         &        (9)           &         (10)         &  (11)   &        (12)                               & (13)        &       (14)                       & (15)     \\        
\hline
  1   & 11.815091 & -25.816675 & 22.094 $\pm$ 0.097 &  20.785 $\pm$ 0.008 &  20.095 $\pm$ 0.003  & 19.817 $\pm$ 0.007  & 18.686 $\pm$ 0.001  & 18.645 $\pm$ 0.008 &  8.483 $\pm$    0.1719  &  0.7433 &                                           &          &  box10 elongated			    &    No             \\
  2   & 12.125469 & -25.814489 & 23.929 $\pm$ 0.185 &  22.629 $\pm$ 0.04  &  22.06  $\pm$ 0.023  & 21.803 $\pm$ 0.047  & 21.071 $\pm$ 0.008  & 21.223 $\pm$ 0.083 &  6.422 $\pm$    0.2245  &  0.8567 &                                       	  &          &  box10 blendJ featuresJ                &   Uncertain            \\
\end{longtable}
\end{landscape}
}

  \begin{figure*}
  \centering
  \includegraphics[width=0.8\hsize]{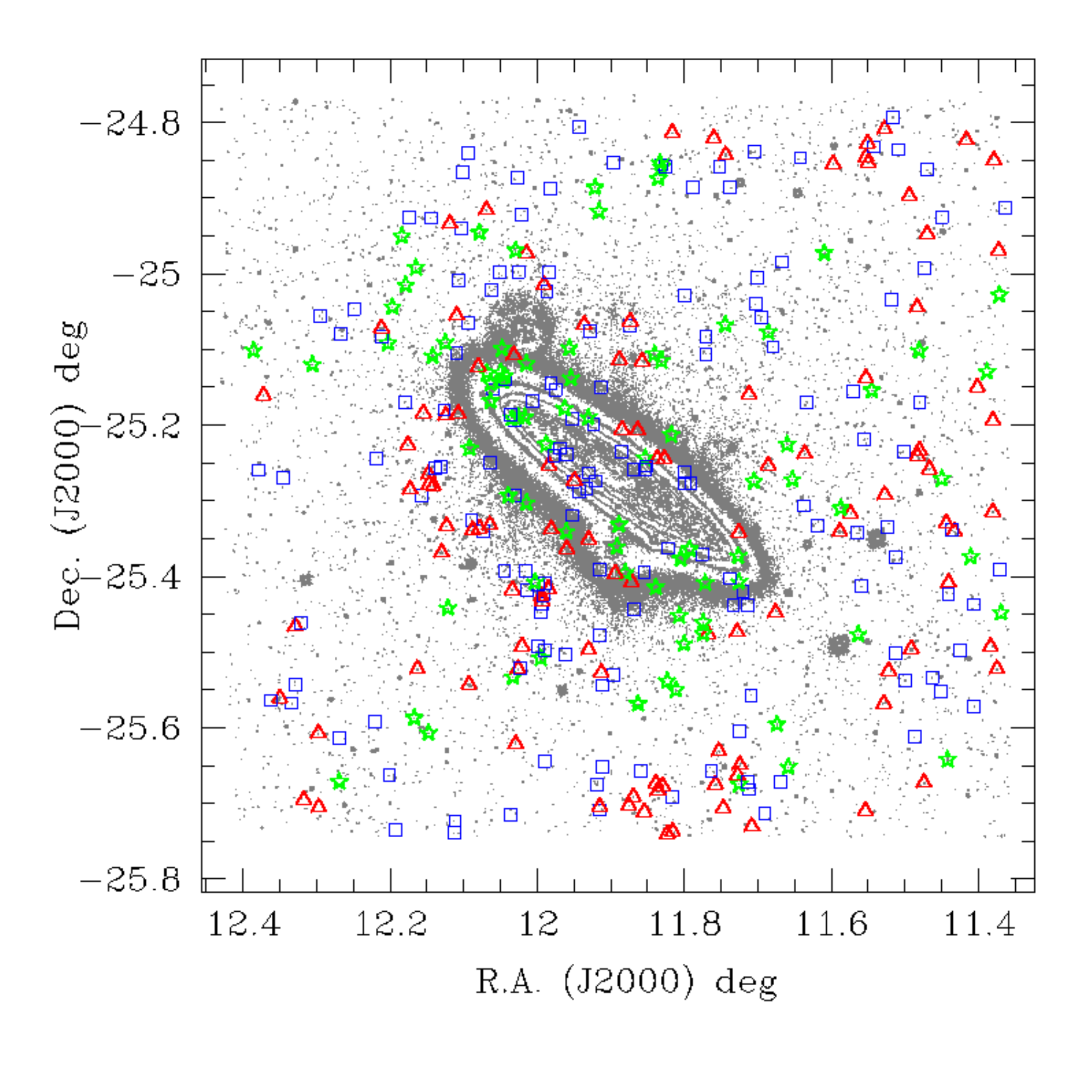}
  \vskip -0.8cm
   \caption{GC candidates overlaid to the $g$-band contour plot of
     NGC\,253. Green five-pointed stars, blue squares, and red triangles
     show the candidates flagged as Best, Uncertain, and No in Table
     \ref{tab_final}, respectively. The plotted contours show the
       $\mu_g=$23.6, 23.0, 22.4, 21.8, 21.2, 20.6 $mag/arcsec^2$
       surface brightness levels, respectively.}
   \label{ngc253_gc}
  \end{figure*}

\subsection{The catalog}

Taking advantage of the ACS Hubble Space Telescope observations of
NGC\,253 from the GHOSTS survey \citep[the GHOSTS acronym stands for:
  Galaxy, haloes, outer disks, substructures, thick disks, star
  clusters][]{rs11}, as a countercheck of our selection, we visually
inspected the GHOSTS fields containing our GC candidates. Thanks to
the exceptional resolution of ACS, star clusters at the distance of
NGC\,253 appear as obviously mottled and extended sources with respect
to the otherwise smooth background galaxies or point-like stellar
sources.  With the exception of two obvious background disk galaxies,
all of the sources selected as described in the previous section, and
falling in the ACS GHOSTS footprints, appear as star clusters. Figure
\ref{ghosts} shows the thumbnails of the 18 selected GC candidates
that also have HST data (panels from $(a)$ to $(d)$). In the figure we
also show the two background galaxies that passed our GC selections
and, for sake of comparison, two sources identified as stars in our
selection procedures (panel ($e$)).

Furthermore, some visually obvious GCs in the GHOSTS footprints, which
were not selected by our procedure, were added by hand in our final
sample, after visual inspection of GHOSTS images. Such objects, seven
in all (Figure \ref{ghosts}, panel ($f$)), although detected and
classified as extended in all cases, were rejected from the final
sample as their colors did not fit in the color-color sequences
adopted because of dust contamination. Although based on their
appearance the candidates are certain stellar clusters, in our final
Table \ref{tab_final} they are flagged as {\it Uncertain} because of
their color, and excluded from the color and magnitude distributions
analysis discussed below.

Moreover, still based on comparison with GHOSTS data, in spite of the
rich set of selection criteria adopted, including the $uiK_s$
color-color diagram that proved to be very effective for sorting GCs
out of other sources in Virgo \citep{munoz14}, the matching with HST
imaging data shows the presence of background contamination in the
final list of GC candidates. Thus, for a final characterization of the
GCs selected and to further clean the sample, we visually inspected
each one of the $\sim350$ GC candidates.

From the visual inspection we found that a substantial portion of
selected candidates are obvious galaxies for various
motivations: more or less obvious features visible in one or more
bands (tidal features, spiral arms), high elongation coupled with
closeness to a group of background galaxies, bright and elongated
structures with changing position angle at different radii, etc.

Table \ref{tab_final} lists the final sample of objects with
coordinates (Cols. 2-3); $ugriJK_s$ magnitudes and errors (Cols. 4-9);
half-light radius and axis ratio from Ishape (Cols. 10-11); existing
identifications from the literature (Col. 12); presence in GHOSTS
footprints, spectroscopic samples, or previous identifications in the
photometric samples by \citet{liller83} or \citet{blecha86} (Col. 13);
and comments from visual inspection (Col. 14). In the table we also
provide a further flag, Class (Col. 15), which defines the objects
classified as {\it bona fide} GC candidates in our list, the
candidates considered {\it uncertain} for some reason (large number of
close background galaxies, high elongation, weird residuals from
Ishape, blending features, border-line axis ratio, etc.), and sources
that are obvious galaxies ({\it no} flag), which passed
morpho-photometric selection criteria, but were rejected upon visual
inspection.

The catalog contains a total of 82 best GC candidates, 155 uncertain
candidates, and 110 sources which, although passed all our GC
selection criteria, are clearly background galaxies.

The spatial distribution of the full sample is shown in Figure
  \ref{ngc253_gc}, overlaid to the $g$-band VST contours plot.

We must note that our selection technique, based also on aperture
photometry, leaves unanswered the question about the detection
efficiency and contamination rate as a function of galactocentric
radius.  Although the majority of the globular cluster candidates are
found in the uncrowded outskirts of the galaxy, a significant number
are projected against or near the bright, crowded galaxy disk.

As is also recognizable in Figure \ref{cmdcol}, sources detected in
regions of high galaxy background suffer from a larger photometric
scatter because of the galaxy contamination and the presence of
dust.

However, of the $\sim20$ {\it bona fide} GCs candidates located in
galaxy regions with $\mu_g\leq23.6~mag/arcsec^2$, only four are new
selections, the remaining are all either spectroscopically confirmed
GCs or star clusters selected on HST/GHOSTS data, and three are
also photometric selections from \citet[][Table 6 data]{beasley00}.

\subsection{Spatial distribution and luminosity function}

\begin{figure*}
  \centering
  \includegraphics[width=0.45\hsize]{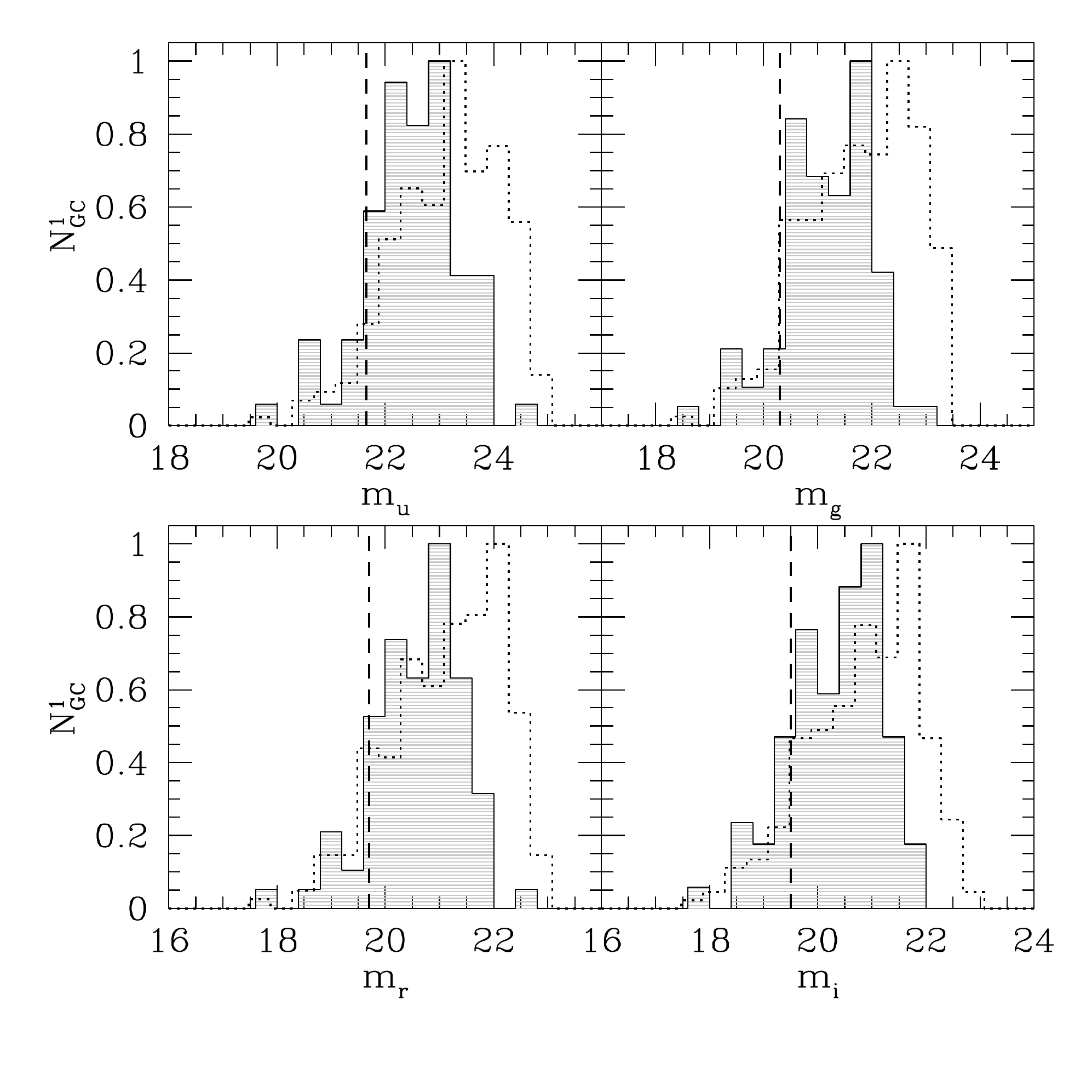}
  \includegraphics[width=0.45\hsize]{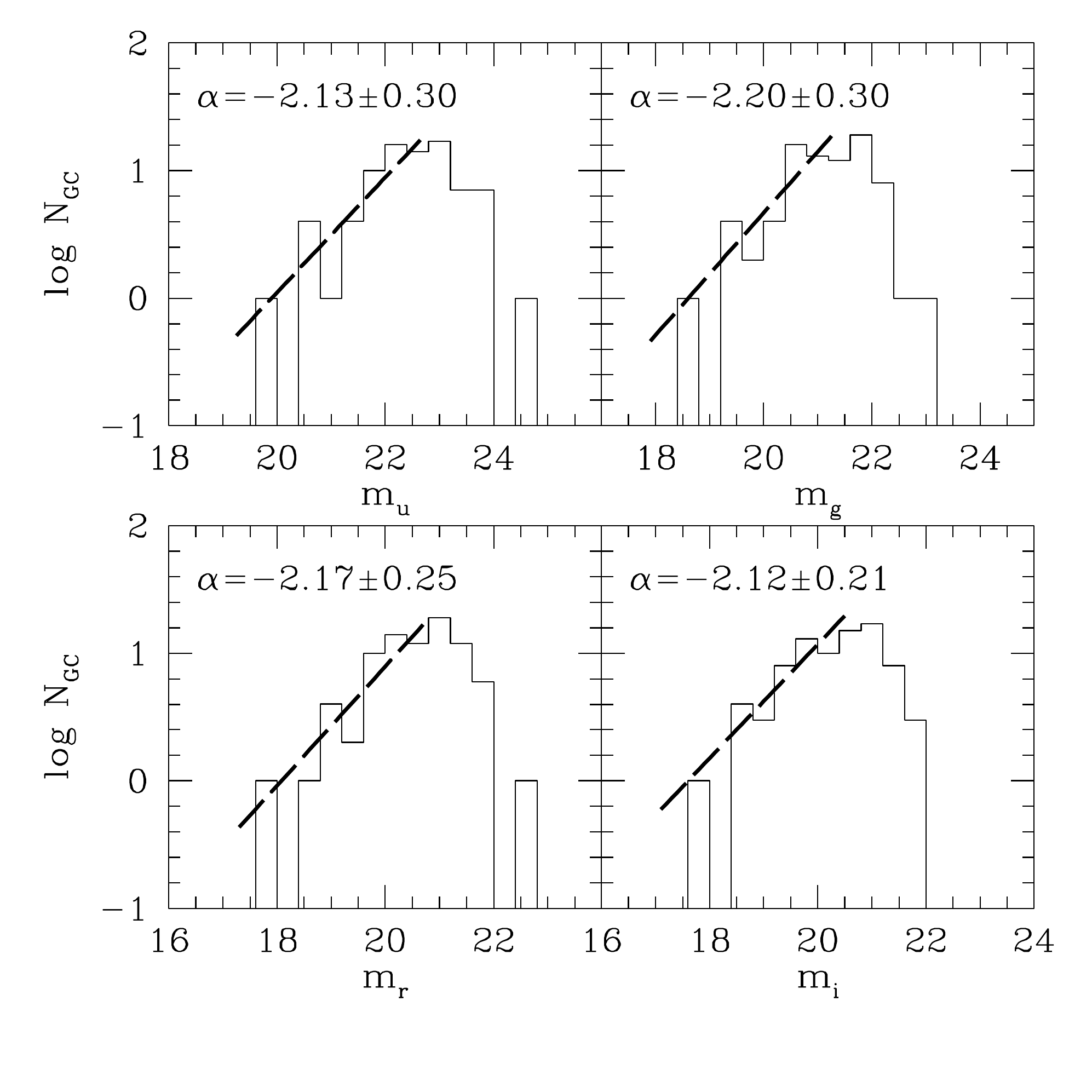}
   \caption{Left: Optical LFs of the {\it bona fide}
     sample GCs (gray shaded histogram) and of the {\it bona
       fide+uncertain} samples (dotted line). The adopted TOM in each
     band is shown with a vertical dashed line. Right: LFs of {\it bona fide} GCs, shown in logarithmic scale. The
     long-dashed line shows the power-law fit to the data. The
     exponent of the fit is reported each panel (see text).}
   \label{lumfunc}
  \end{figure*}

The optical LF of the {\it bona fide} sample and the combination
of the {\it bona fide} and {\it uncertain} samples are shown in
Figure \ref{lumfunc} (left panels). In the panels of the figure the
$M_{TOM}$ adopted, properly shifted to the galaxy distance, is also
reported. The diagrams lack the typical symmetry around the peak of
the Gaussian GCLF, which is surprising given that the bright side of
the LF appears underpopulated. By inspecting the full sample of
sources brighter than $m_g\sim20$, we found that even after adopting
reasonably broader selection criteria the list of bright candidates
does not increase. Hence, we do not have an explanation for missing
bright end of the GCLF.

Taking only the sample of spectroscopically confirmed GCs does not
improve the appearance of the GCLF because of the small size of the
sample of 21 candidates and because 7 of the candidates are
brighter than the $M_{TOM}$, and 14 are fainter than that, with the
faintest candidate at $m_g\sim21.5$ mag, i.e., at $\sim 1\sigma_{GCLF}$
the level of the faint side GCLF. If we also add the GCs identified
over the HST/GHOSTS area, the cumulative sample of HST and
spectroscopic candidates has $\sim10$ GCs that are brighter than the $M_{TOM}$
and 41 fainter than the $M_{TOM}$. Hence, whether only the spectroscopic candidates or
both spectroscopic and GHOSTS candidates are considered, again the
GCLF is highly undersampled toward bright GCs.

The incompleteness is in part due to the photometric incompleteness,
which is caused by the different depth and image quality of the imaging data
adopted. However, photometric incompleteness should only be an issue
at the faint end of the GCLF.

An alternative explanation for the asymmetric GCLF would come from
overestimated low luminosity end of GCLF that would, even in the case
of best candidates, have to be heavily contaminated. We believe this
is not the case and thus reject this (potential) explanation because
we have verified our selection criteria through a comparison with HST
GHOSTS images and with a spectroscopically confirmed sample of
GCs. Furthermore, if the low luminosity end were heavily contaminated,
the GC sample size in NGC\,253 would be too small, resulting in too
small $S_N$, as we discuss further in Section \ref{sec_sn}.

A further correction to the GCLF might come from the fact that, in
addition to photometric incompleteness, our sample is also incomplete
at large and small galactocentric radii. The largest projected
galactocentric distance of a GC candidate in the {\it bona fide}
sample is $r_{gal}\sim 35\arcmin$, or $\sim35.5$ kpc. A fraction of
$\sim 7\%$ (11 out of 158) MW GCs are located at galactocentric
distance larger than $\sim$35.5 kpc. Hence, it is reasonable to expect
that a similar fraction of GCs in NGC\,253 lies beyond the common area
of the VST and VISTA pointings.

  \begin{figure}
    \centering
    \includegraphics[bb=20 170 580 440,width=\hsize]{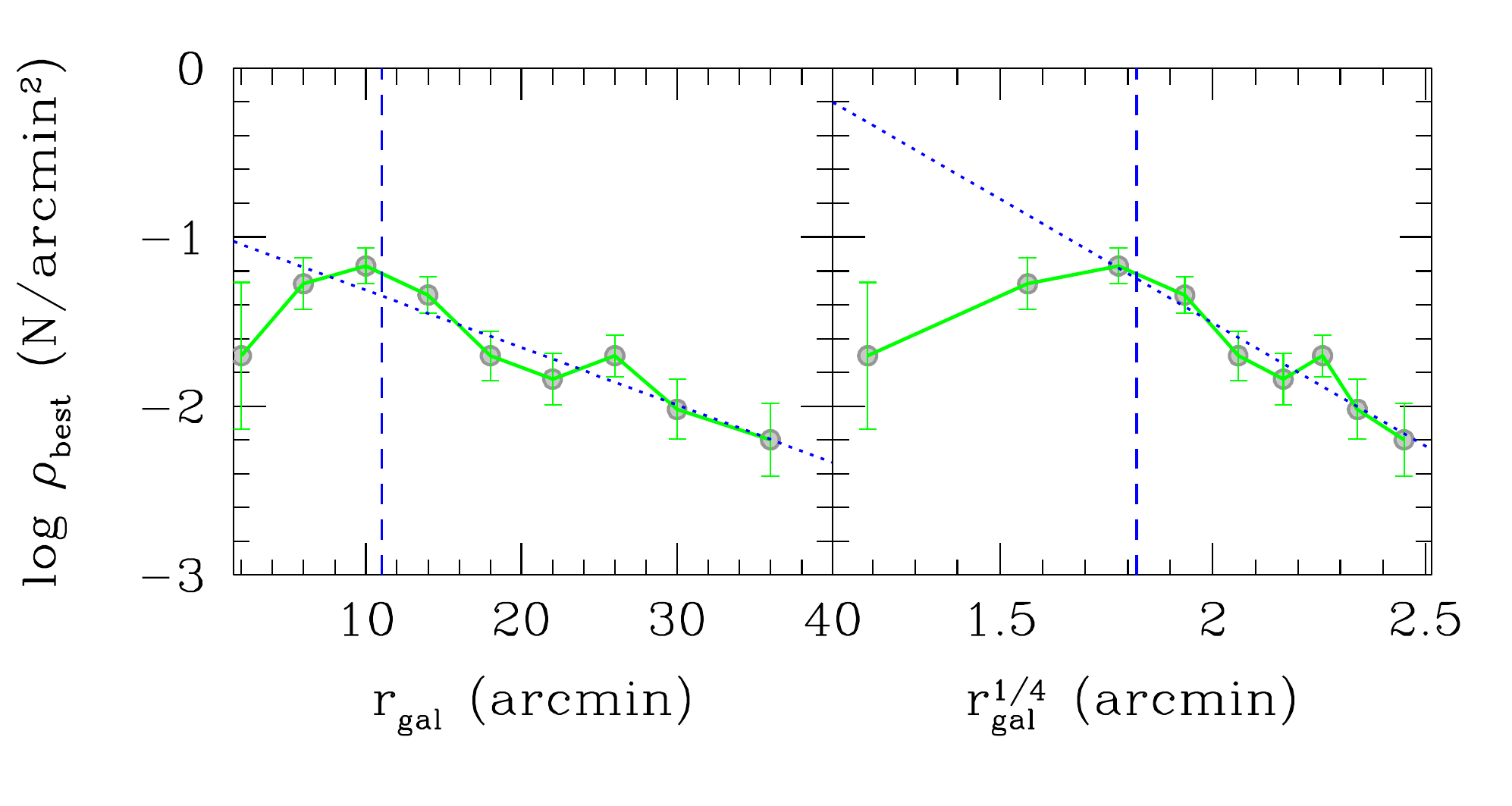}
    \vskip +3cm
   \caption{Radial density profile of the best GCs sample. Left and
     right panels show the linear and $r^{1/4}$-law profile,
     respectively. Linear fits to data are shown with blue dotted
     line. The vertical dashed line indicates the semimajor axis where
     the dusty disk begins.}
   \label{lradprof}
  \end{figure}

For the central dusty regions, as aforementioned, we partially
recovered some of the GCs by complementing our data with the ACS GCs
from GHOSTS. Nevertheless, such detections, mostly based on visual
inspection, do not necessarily allow the recovery of the entire
population of central GCs in the galaxy. As a check, we inspected the
azimuthal average of the GCs radial density profile, reported in
Figure \ref{lradprof}. The diagram shows the linear and $r^{1/4}$ fits
to the density profile, which is derived without the data for the
innermost two annuli, severely affected by dust and incompleteness. In
both panels we observe a drop of the density profile in the very
central regions, otherwise the radial (logarithmic) density profile
nicely follows the fitted density profiles.  The $r^{1/4}$-law
profile, together with the increasing or flattening of the GC density
profiles at small galactocentric radii, are well-known observational
properties of GC systems \citep{dirsch05,goudfrooij07,cantiello15}.
Consequently, it is reasonable to assume that the drop in $\log
\rho(r_{gal}),$ seen in the left panels of Figure \ref{lumfunc}, is
due to poor GCs detection in such central dusty regions. Even though
the central area dominated by dust is relatively small, $\sim80$
square arcmin, the fraction of GCs there could be significative. To
obtain an approximate estimate of the number of GCs in the central
area, we adopted the radial density profiles shown in Figure
\ref{lradprof}, assuming as lower limit to the GCs density the value
of $\rho(r_{gal})$ at $r_{gal}\sim11\arcmin$, i.e., the galactocentric
radius where the dusty disk begins. Adopting the linear or
$r^{1/4}$-law fits, the fitted GC density at $r_{gal}\sim11\arcmin$
goes from $\sim0.045$ GCs/$arcmin^2$, to $\sim0.057$
GCs/$arcmin^2$. Hence, the estimated number of GCs in the central area
is $N_{GC}^{center}\sim 5$.

In a study of RGB-tip field star population based on $V$ and $I$
Magellan/IMACS data, \citet{bailin11} found evidence for a large
shelf-like feature near the southeast side of NGC\,253 \citep[also
  confirmed by][from resolved star analyses of the VISTA imaging data
  used in this work]{greggio14}. Using GHOSTS data in two fields - one
on and one off the shelf - the authors inspected the color
distribution of RGB-tip stars and found that the feature is possibly
the remnant of a large satellite of the merging tree of NGC\,253. Yet,
the authors warned that the stellar populations in the two fields are
not dramatically different from the rest of the halo at similar
elliptical radii. Inspecting the colors of our {\it bona fide} GCs in
various regions around the galaxy, we find that the $\sim$15 GC
candidates in the projected region close to the shelf identified by
\citet{bailin11} have average colors that are bluer than the colors of GCs in
other four randomly drawn regions and than the bulk of the {\it bona fide}
sample. This might further strengthen the hypothesis of the presence
of a surface brightness feature and of a GCs subpopulation, which are both
remnants of the merging with a low-mass companion. As a matter of
fact, GCs in low-mass galaxies are typically bluer than in higher mass
galaxies \citep[e.g.,][]{peng06}. Nevertheless, because of
the small size of the GC samples in the regions inspected, the average
colors are in all regions consistent within 1$\sigma$ with the median
colors of the bulk {\it bona fide} sample.

 The presence of substructures might also help to explain the observed
 GCLF, as they imply a dynamically young
 environment. \citet{greggio14} pointed out the presence of a very
 extended (out to $\gsim30$ kpc above the disk plane), intermediate
 age AGB population in the inner halo of NGC\,253. Assuming a constant
 star formation rate, the authors estimated that the AGB population
 traces $\sim2\times 10^8~M_{\sun}$ of stars formed between 0.5 and 3
 Gyr. Hence, some intermediate age ($t\sim6~Gyr$), metal-rich
 $[Fe/H]\gsim-0.3$ star cluster, falling in a similar color interval
 of old and metal-poor GCs might be ``contaminating'' the sample of
 genuine old GCs. Indeed, the LFs in Figure \ref{lumfunc} (right
 panels), resemble the one of star clusters in the LMC as shown, for
 example, in Fig 10 of \citet{larsen02}. In the panels of the figure
 we plot the linear fit to the data obtained from the LFs down to one
 magnitude fainter than the TOM, and the slope $\alpha$ for the
 power-law fit $\frac{dN}{dL}\propto L^{\alpha}$ \citep[see eqs. 2-4
   in][]{larsen02}.  The power-law fit to the data provides exponents
 $\alpha \sim -2.1$, similar to those typically found in spirals and
 starburst galaxies
 \citep[e.g.,][]{miller97,whitmore99,larsen02,cantiello09}.

\subsection{Total GC population}
\label{sec_sn}

Including the approximate fractions of missing GCs at small, i.e., $\sim5$,
and large, i.e., $\sim7\%$ of the total population, galactocentric radii,
derived based on the properties of our best sample GCs, we
estimate a total number of GCs of $N_{GC}^{Total}\sim$100.  By using
the GC specific frequency $S_N$\footnote{A parameter relating the
  galaxy and the GC system properties, quantified as the number of GCs
  per unit galaxy luminosity $S_N\equiv N_{GC}\times 10 ^{0.4
    (M_V+15)}$, where $N_{GC}$ is the total number of clusters and
  $M_V$ is the total absolute visual magnitude of the
  galaxy \citep{harris81,harris91}.}  versus magnitude relation
\citep[e.g., from][]{peng08}, it is possible to invert the relation and
obtain an estimate of the total expected GC population for a galaxy
like NGC\,253. Adopting $S_N\sim1.8$ from the dotted curve in Figure 2
of \citet[][derived from the analysis of Virgo cluster galaxies
  covering a wide range of luminosities]{peng08}, and $M_V\sim-21$ mag
(Table \ref{tab_props}), we get $N_{GC}^{Total}\sim450$. Assuming
$\Delta S_N\sim 0.4$ and $\Delta M_B\sim 0.5$ mag, we also calculate
an uncertainty $\Delta N_{GC}^{Total}\sim260$. This number decreases to
$N_{GC}^{Total}=250\pm200$, if $S_N\sim1$ is used
\citep{harris13}. Adopting such evaluations as strictly valid leads to
the conclusion that our best sample is a factor of two to four
incomplete.

  \begin{figure*}
  \centering
 \includegraphics[width=0.43\hsize]{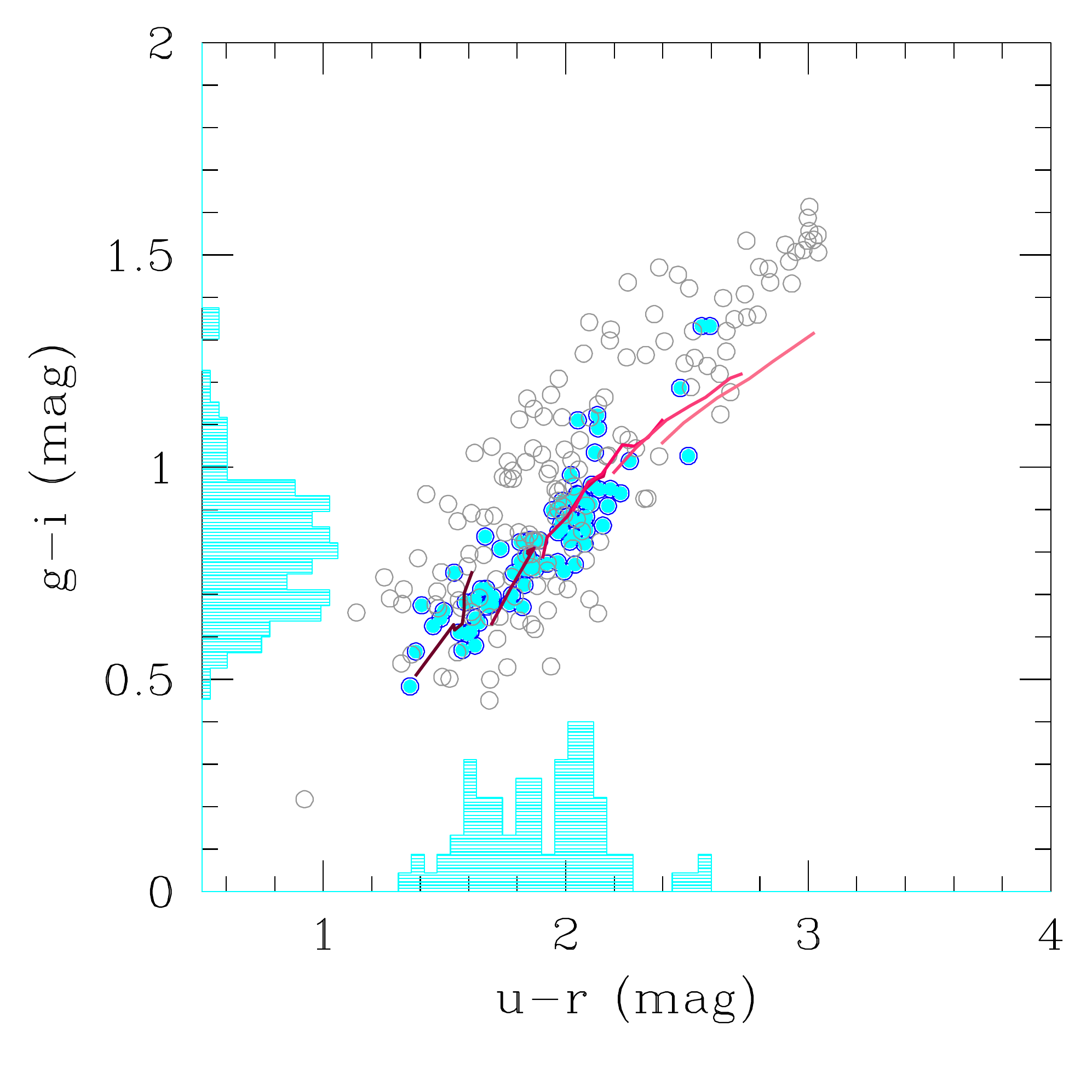}
 \includegraphics[width=0.43\hsize]{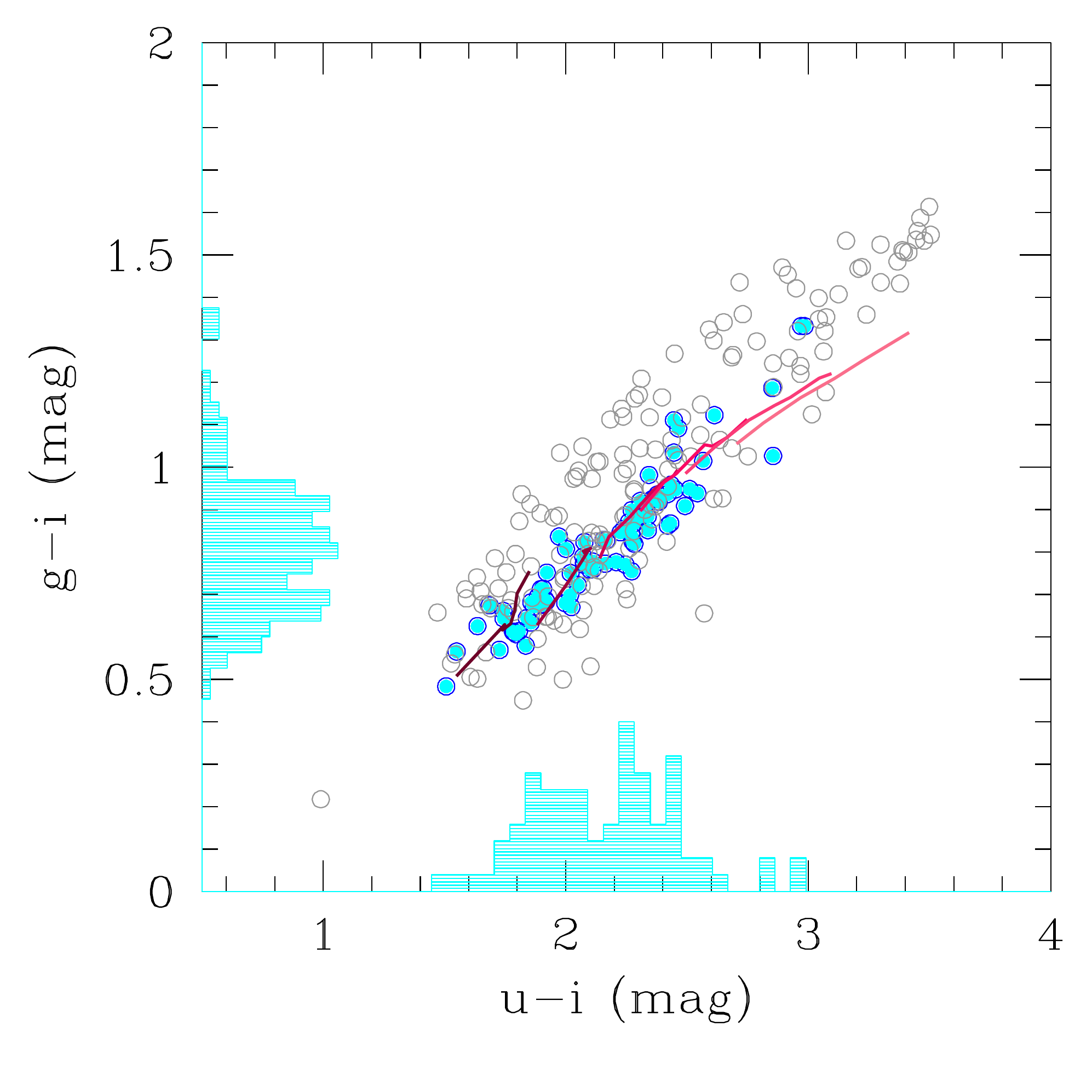}
 \includegraphics[width=0.43\hsize]{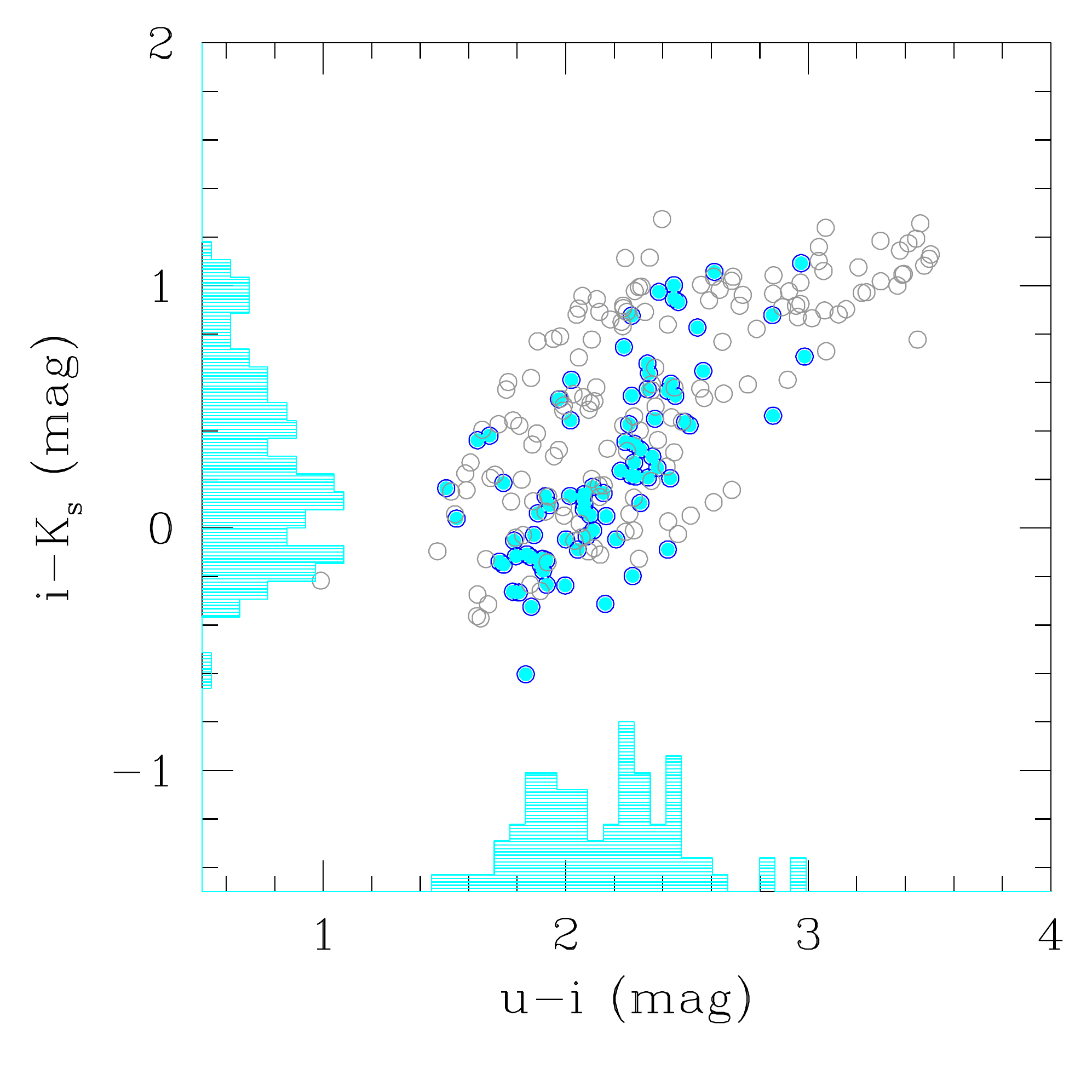}
 \includegraphics[width=0.43\hsize]{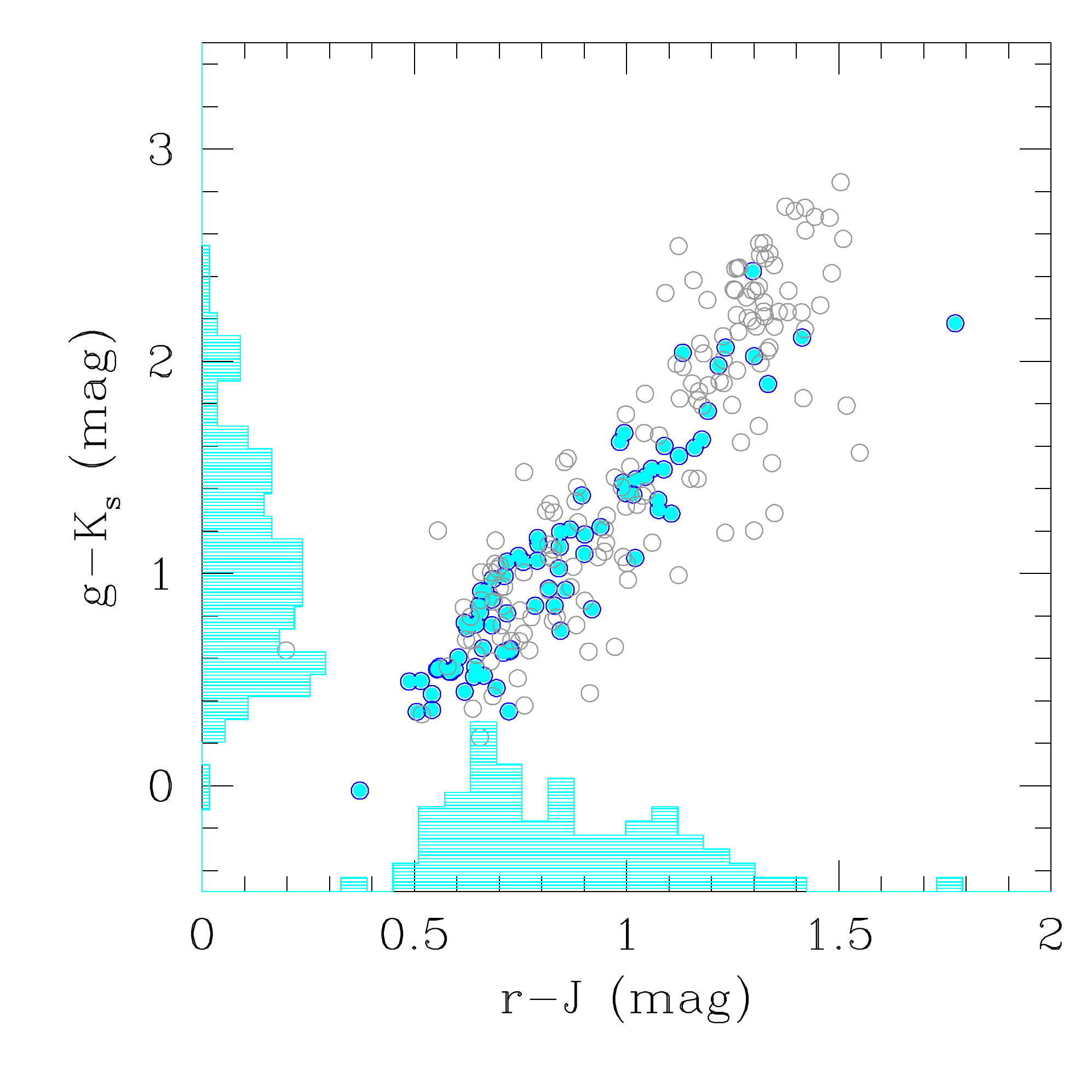}
   \caption{Color-color diagrams for the {\it bona fide} and {\it
       uncertain} samples of GCs, shown with blue filled dots and
     empty gray circles, respectively. The (arbitrarily scaled)
     histograms reported on the axes refer to the {\it bona fide}
     sample. For optical colors, the SPoT SSP models are also shown
     with lines as in Figure \ref{colors}.}
   \label{histocol}
  \end{figure*}

However, the $S_N$ versus $M_V$ relation has a substantial
scatter. For example, from the database by \citet{harris13}, it can be
seen that galaxies with $M_V=-21\pm0.2$ mag have a median population
of $\sim450$ GCs with minimum at $\sim40$ and a maximum of $\sim
7.000$. It is worth emphasizing that, based on previous literature
\citep{blecha86,olsen04}, \citet{harris13} estimates the size of the
GC system of NGC\,253 to be $N_{GC}=90\pm40$. If only galaxies with
morphological classes and magnitudes similar to NGC\,253 are drawn from the
\citet{harris13} sample, the median is $N_{GC}\sim80$. The two
galaxies with closest morphological class and total magnitude to
NGC\,253 (M\,101 and NGC\,6956) have a total population of detected
GCs with $N_{GC}\sim90\pm40$ and $150\pm40$, which is consistent with
the total (coverage corrected) population we estimated.

Inspecting the radial profiles shown in Figure \ref{lradprof}, we note
a slight enhancement of GCs density at $r_{gal}\sim25$ kpc. Although
the estimated errors are relatively high, such enhancement roughly
corresponds to the radial distance of the feature first found and
discussed by \citet{greggio14}, identified as a possible stellar
stream remnant. However, we must highlight that our radial profile is
derived from the azimuthal average of GC counts, while the RGB star
counts excess found in \citet{greggio14} is specifically located on
the northwestern side of the galaxy.

\subsection{Color bimodality}

A further typical element of discussion in extragalactic GC systems is
the color distribution. In particular, the presence of a nearly
universal color bimodality in GC systems, and the astrophysical
processes underlying such ubiquitous feature, have generated a
prolific and still open debate in the last decade
\citep{spitler06,yoon06,cantiello07d,chiessantos12,usher12,brodie14,cantiello14}. We
adopted the {\it bona fide} sample, to verify the presence of color
bimodality (i.e., to identify two well-separated blue and red
peaks) and to estimate the differences between the two GC
subpopulations in terms of the color distributions.  To do so, we
used the Gaussian mixture modeling code \citep[GMM;][]{muratov10}. The GMM
code uses the likelihood-ratio test to compare the goodness of fit for
double-Gaussian versus a single-Gaussian.  For the best-fit double
model, it estimates the means and widths of the two components, their
separation DD in terms of combined widths, and the kurtosis of the
overall distribution. Values of DD larger than $\sim2$, and negative
kurtosis are necessary but not sufficient conditions for
bimodality. Also, GMM provides uncertainties based on bootstrap
resampling.  In addition, the GMM analysis provides the positions,
relative widths, and fraction of objects associated with each peak.

Inspecting the results reported in Table \ref{tab_gmm2} (upper section
of the table), GMM analysis shows that our {\it bona fide} GCs sample is
poorly fitted by two-Gaussian distributions in most of the inspected
colors. This is testified by either the high $p$-values (Cols. 10-12
in the table), the changing fractions of red GCs ($f_2$) when
different colors are used, the positive kurtosis or the low values for
the peak separation estimator (DD).

\begin{table*}
\caption{\label{tab_gmm2} Results of GMM runs}
\centering
\begin{tabular}{lcccccccccccc}
\hline
\hline
Color &     Peak1         &       Peak2      &   $\sigma$1      &    $\sigma$2     & $N_{GC}$ & $f_{2}$ & DD & Kurtosis & $p(\chi^2)$ & $p(DD)$ & $p(kurt)$ & bi?\\ 
 (1)  &     (2)           &       (3)        &   (4)            &    (5)           &  (6)     & (7)    & (8)& (9)      & (10)       &(11)     & (12)    &   (13)  \\  
\hline
$u{-}i$ &  1.84 $\pm$ 0.14 &  2.20 $\pm$ 0.29 &  0.08 $\pm$ 0.07 &  0.30 $\pm$ 0.10 &   81 & 0.88 &   1.62 &  -0.04 &   0.525 &    0.73  &   0.675  &   N   \\ 
$g{-}i$ &  0.79 $\pm$ 0.06 &  1.22 $\pm$ 0.19 &  0.13 $\pm$ 0.04 &  0.11 $\pm$ 0.05 &   81 & 0.05 &   3.44 &   0.93 &   0.050 &    0.14  &   0.974  &  Y?    \\ 
$u{-}r$ &  1.59 $\pm$ 0.09 &  1.97 $\pm$ 0.13 &  0.12 $\pm$ 0.04 &  0.23 $\pm$ 0.06 &   81 & 0.81 &   2.06 &  -0.09 &   0.634 &    0.61  &   0.630  &   N   \\ 
$i{-}k$ &  0.03 $\pm$ 0.10 &  0.63 $\pm$ 0.21 &  0.23 $\pm$ 0.07 &  0.26 $\pm$ 0.10 &   82 & 0.35 &   2.49 &  -0.59 &   0.063 &    0.42  &   0.157  &  Y?    \\ 
$r{-}j$ &  0.65 $\pm$ 0.04 &  0.95 $\pm$ 0.12 &  0.08 $\pm$ 0.03 &  0.25 $\pm$ 0.06 &   81 & 0.64 &   1.62 &   0.77 &   0.004 &    0.74  &   0.961  &   N   \\ 
$g{-}k$ &  0.55 $\pm$ 0.14 &  1.19 $\pm$ 0.30 &  0.12 $\pm$ 0.09 &  0.49 $\pm$ 0.12 &   81 & 0.79 &   1.79 &  -0.35 &   0.026 &    0.70  &   0.378  &  Y?    \\ 
$r{-}i$ &  0.22 $\pm$ 0.02 &  0.29 $\pm$ 0.03 &  0.03 $\pm$ 0.01 &  0.06 $\pm$ 0.01 &   82 & 0.68 &   1.41 &   0.42 &   0.075 &    0.83  &   0.895  &  N   \\ 
$j{-}k$ & -0.37 $\pm$ 0.06 & -0.03 $\pm$ 0.08 &  0.18 $\pm$ 0.03 &  0.05 $\pm$ 0.04 &   82 & 0.10 &   2.53 &  -0.45 &   0.226 &    0.42  &   0.271  &  N    \\ 
\hline
\multicolumn{12}{c}{Without the red GCs}\\
\hline
$g{-}i$ & 0.67 $\pm$ 0.05 &  0.87 $\pm$ 0.05 &  0.07 $\pm$ 0.02 &  0.09 $\pm$ 0.02 &   76 & 0.61 &   2.39 &  -0.70 &   0.444 &    0.48  &   0.086  &  N    \\ 
$u{-}i$ & 1.94 $\pm$ 0.06 &  2.36 $\pm$ 0.04 &  0.18 $\pm$ 0.03 &  0.11 $\pm$ 0.03 &   78 & 0.43 &   2.77 &  -0.91 &   0.024 &    0.31  &   0.013  &  Y    \\ 
$u{-}r$ & 1.71 $\pm$ 0.06 &  2.07 $\pm$ 0.03 &  0.17 $\pm$ 0.03 &  0.08 $\pm$ 0.03 &   77 & 0.41 &   2.67 &  -0.97 &   0.005 &    0.34  &   0.006  &  Y    \\ 
$i{-}k$ & 0.04 $\pm$ 0.09 &  0.56 $\pm$ 0.13 &  0.23 $\pm$ 0.06 &  0.13 $\pm$ 0.06 &   74 & 0.23 &   2.80 &  -0.70 &   0.208 &    0.32  &   0.081  &  Y?   \\ 
$r{-}j$ & 0.70 $\pm$ 0.04 &  1.07 $\pm$ 0.08 &  0.13 $\pm$ 0.03 &  0.07 $\pm$ 0.04 &   74 & 0.25 &   3.47 &  -0.95 &   0.004 &    0.14  &   0.012  &  Y    \\ 
$g{-}k$ & 0.82 $\pm$ 0.12 &  1.51 $\pm$ 0.17 &  0.32 $\pm$ 0.10 &  0.14 $\pm$ 0.09 &   73 & 0.19 &   2.83 &  -0.85 &   0.136 &    0.29  &   0.032  &  Y?    \\ 
$r{-}i$ & 0.23 $\pm$ 0.01 &  0.32 $\pm$ 0.01 &  0.04 $\pm$ 0.01 &  0.01 $\pm$ 0.01 &   74 & 0.25 &   3.10 &  -0.97 &   0.004 &    0.22  &   0.011  &  Y    \\ 
$j{-}k$ &-0.44 $\pm$ 0.10 & -0.22 $\pm$ 0.06 &  0.15 $\pm$ 0.04 &  0.09 $\pm$ 0.03 &   72 & 0.26 &   1.74 &  -0.35 &   0.705 &    0.71  &   0.395  &  N    \\ 
\hline 
\end{tabular}
\tablefoot{Columns list: (1) color; (2-3) mean and uncertainty of the
  first and second peaks in the double-Gaussian model; (4-5) width and
  uncertainty of the first and second peaks; (6) number of GC
  candidates selected; (7) fraction of GC candidates associated with
  the second, red, peak; (8) separation of the peaks in units of the
  two Gaussian widths; (9) kurtosis of the distribution (DD$ \geq$2
  and negative kurtosis are required for significative split between
  the two Gaussian distributions); (10-12) GMM $p$-values based on the
  likelihood-ratio test $p(\chi^2)$, peak separation $p(DD)$, and
  kurtosis $p(kurt)$, indicating the significance of the preference
  for a double-Gaussian over a single-Gaussian model (lower $p$-values
  are more significant); (13) assessment of the evidence for
  bimodality.}
\end{table*}

\begin{table*}
\caption{\label{tab_gmm3} Results of GMM runs for the three Gaussian model} 
\begin{tabular}{lcccccccc}
\hline
\hline
 Color  & Peak1 ($N_{GC}$, $\sigma$) & Peak2 ($N_{GC}$, $\sigma$)  & Peak3 ($N_{GC}$, $\sigma$) &   DD & Kurtosis & $p(\chi^2)$ & $p(DD)$ & $p(kurt)$ \\
\hline
     $u{-}i$ &   1.94 ( 43.70 ,  0.19 ) &  2.36 ( 33.30 ,  0.12 ) &  2.92 (  4.00 ,  0.06 ) &   2.73 &  -0.04 &   0.024 &   0.351  &   0.675 \\  
     $g{-}i$ &   0.69 ( 39.10 ,  0.08 ) &  0.88 ( 30.20 ,  0.06 ) &  1.07 ( 11.60 ,  0.17 ) &   2.54 &   0.93 &   0.111 &   0.511  &   0.974 \\  
     $u{-}r$ &   1.71 ( 45.40 ,  0.17 ) &  2.07 ( 31.60 ,  0.08 ) &  2.53 (  4.00 ,  0.05 ) &   2.67 &  -0.09 &   0.002 &   0.459  &   0.630 \\  
     $i{-}k$ &   0.04 ( 56.80 ,  0.23 ) &  0.56 ( 16.60 ,  0.12 ) &  0.96 (  8.60 ,  0.08 ) &   2.80 &  -0.59 &   0.071 &   0.345  &   0.157 \\  
     $r{-}j$ &   0.67 ( 44.60 ,  0.11 ) &  1.04 ( 35.40 ,  0.17 ) &  1.77 (  1.00 ,  0.04 ) &   2.61 &   0.77 &   0.002 &   0.470  &   0.961 \\  
     $g{-}k$ &   0.69 ( 43.00 ,  0.25 ) &  1.33 ( 30.40 ,  0.26 ) &  2.09 (  7.60 ,  0.16 ) &   2.49 &  -0.35 &   0.122 &   0.581  &   0.597 \\  
     $r{-}i$ &   0.23 ( 53.40 ,  0.04 ) &  0.32 ( 16.10 ,  0.01 ) &  0.36 ( 12.50 ,  0.06 ) &   3.15 &   0.42 &   0.047 &   0.221  &   0.895 \\  
     $j{-}k$ &   0.44 ( 53.20 ,  0.15 ) & -0.24 ( 14.20 ,  0.04 ) & -0.03 ( 14.50 ,  0.07 ) &   1.87 &  -0.45 &   0.474 &   0.618  &   0.271 \\ 
\hline 
\end{tabular}
\tablefoot{Columns list: (1) color; (2-4) first, second and third
  peaks in the three Gaussian model, numbers within parentheses are
  the number of GCs associated with each peak, and the width of the
  distribution; (5) separation of the peaks in units of the three
  Gaussian widths; (6) kurtosis of the distribution; (7-9) GMM
  $p$-values, as in Table \ref{tab_gmm2}.}
\end{table*}

We also run GMM in three Gaussian peaks mode. The results, reported in
Table \ref{tab_gmm3}, are more satisfactory than the previous, as
testified by the generally lower $p$-values.  The presence of a third
color peak is also visible as a red tail of GCs in the panels of
Figure \ref{histocol}. The very red candidates are scattered around
the frame, i.e., they are not sources close to the galaxy disk and
reddened by the dust. In the Figure we report the color-color
diagrams, together with color histograms for various colors and for
the {\it bona fide} GCs sample. Taking advantage of the results on the
presence of such third peak, we rerun GMM in double Gaussian mode
after rejecting the candidate GCs in the third reddest peak; GMM also
provides as output a table with the probability membership of each GCs
to one of the fitted Gaussian. The GMM results of such a {\it culled
  best} sample, again with a two Gaussian model, are presented in the
lower section of Table \ref{tab_gmm2}.

Although the new GMM results for the selected GCs sample appear, with
respect to the full bona fide sample, more consistent with a
bimodal-color scenario there is no homogeneity of the various
inspected colors, as color bimodality is not coherently observed in
all colors. In this respect, then, NGC\,253 should certainly not be
considered a galaxy with a bimodal GC population, showing coherent
and consistent color bimodality.

\section{Conclusions}

We summarize the conclusion of the present work, dedicated to the
definition of an updated catalog of GC candidates in NGC\,253, as
follows:

\begin{itemize}
\item The adoption of an even richer collection of color,
  photometric, and morphologic parameters make the selection of GC
  candidates more robust, but does not guarantee the complete removal
  of all contaminants.
  
\item Depending on the spatial resolution of the imaging data used,
  for sources close enough, such as the galaxy studied here, the
  foreground MW stars can be separated from GCs, as the latter appear
  extended, with FWHM definitely larger than a point source. In cases
  such as that presented here, background galaxies are the
    primary source of contamination.
\item About 10 of our candidates were already photometrically selected
  as GC candidates by other authors. Conversely, the largest portion
  of GC candidates with photometric selection in the literature are
  mostly identified as stars in our sample.
\item Even the adoption of the $uiK_s$ diagram, which proved to be
  extremely efficient in \citet{munoz14} to identify GCs in the Virgo
  galaxy cluster, does not ensure a clean sample of GCs.
\item Spectroscopy by itself is not sufficient to define a
  contamination-free GC sample. In both the spectroscopic catalogs we
  adopted as reference, we found GCs that in our analysis are rather
  classified as contaminants: two background galaxies for
  \citet{beasley00} catalog and two foreground stars for
  \citet{olsen04} catalog. This is equivalent to a level of $\sim
  20\%$ contamination in the full sample of spectroscopically
  confirmed GCs.
\item We estimate a total population of $\sim100$ GC, such a value
  should be regarded as a lower limit. The literature data for a
  couple of galaxies similar in morphology and luminosity to NGC\,253
  show that they have similar populations of detected GCs.
\item The LF of {\it bona fide} sample does not show the typical
  symmetry around the GCLF $M^{TOM}$ peak. A fraction of bright GCs
  might be undetected because of large galactocentric radii or because
  these GCs are along the line of sight of the dusty disk. We do not
  have a clear explanation for the missing bright candidates. However,
  our results support previous studies that found evidence for recent
  interactions for NGC\,253. Hence, one possible explanation of the
  observed LF might be the presence of a fraction of intermediate age,
  $t\sim6$ Gyr, GCs contaminating the population of old GCs.
  \item The radial profile and color distributions of our {\it bona fide} GC
  sample show properties similar to other well-studies galaxies,
  i.e., a radial profile fit by a $r^{1/4}-law$.
  \item As for color bimodality, the statistical preference of
    bimodal over unimodal color distribution is not strong because this bimodality is
    evident with some colors and not in others.
  \item Finally, we provide an updated list of candidate GCs.
\end{itemize}

Moreover, as a byproduct of our analysis, thanks to the very wide
wavelength interval of imaging data used, in addition to GCs we have
been able to identify and distinguish the sequences of stars and the
(various classes of) background galaxies (see Appendix).

The GCs in NGC\,253, will be an interesting target for next generation
large aperture, $\geq$30m-class telescopes, supported by adaptive
optics modules that will facilitate reaching the diffraction limit.

We take as an example the MICADO/MAORY configuration at the E-ELT
\citep{diolaiti16,davies16}, which has an expected optimal resolution limit of
$\sim10$mas at near-IR wavelengths; i.e., $\sim5-10$ times better than ACS
and WFC3 on board  HST and the forthcoming NIRCAM on JWST; with very
high Strehl ratios on a field of view of $\sim20\arcsec$ in the case
of SCAO, the instrument will allow for the first time the study of the
resolved color-magnitude diagrams of stars in the GCs in the Sculptor
group.  Our work is an effort to provide a large census of GCs in
NGC\,253, which is the brightest galaxy in the group.

\begin{acknowledgements}

 We gratefully acknowledge INAF for financial support to the VSTceN.

This research was made possible through the use of the AAVSO
Photometric All-Sky Survey (APASS), funded by the Robert Martin Ayers
Sciences Fund.
  
We acknowledge the usage of the HyperLeda database \citep{makarov14},
\url{http://leda.univ-lyon1.fr}.  This research has made use of the
NASA Astrophysics Data System Bibliographic Services, the NASA
Extragalactic Database, and the SIMBAD database, operated at CDS,
Strasbourg.

It is a pleasure to thank W. Harris and M. Hilker for enlightening
discussions.
\end{acknowledgements}


\bibliographystyle{aa}
\bibliography{cantiello_mar17}


\begin{appendix}
\section{Color-color diagrams}
\label{appendix}
\begin{figure*}
  \centering
  \includegraphics[width=0.33\hsize]{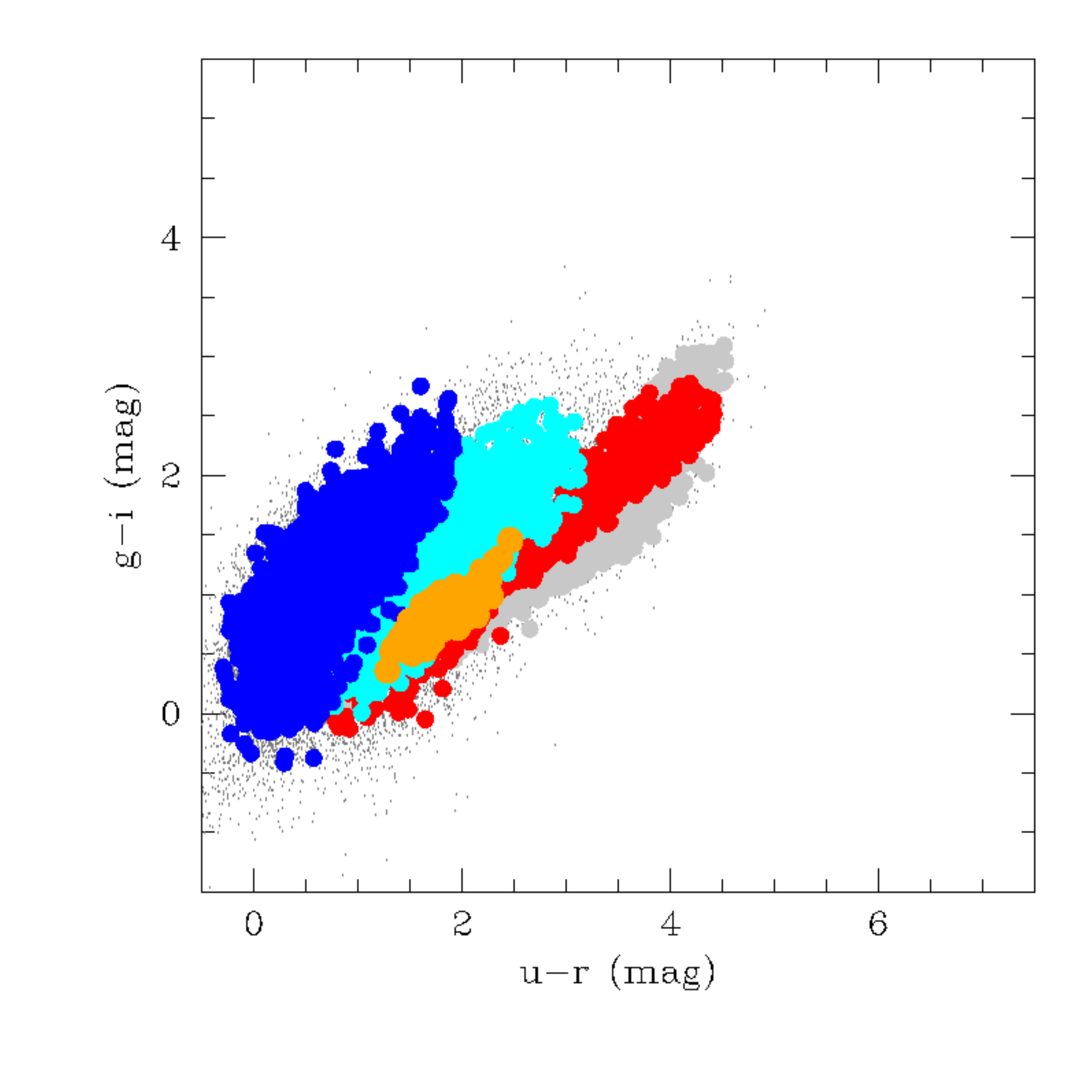}
  \includegraphics[width=0.33\hsize]{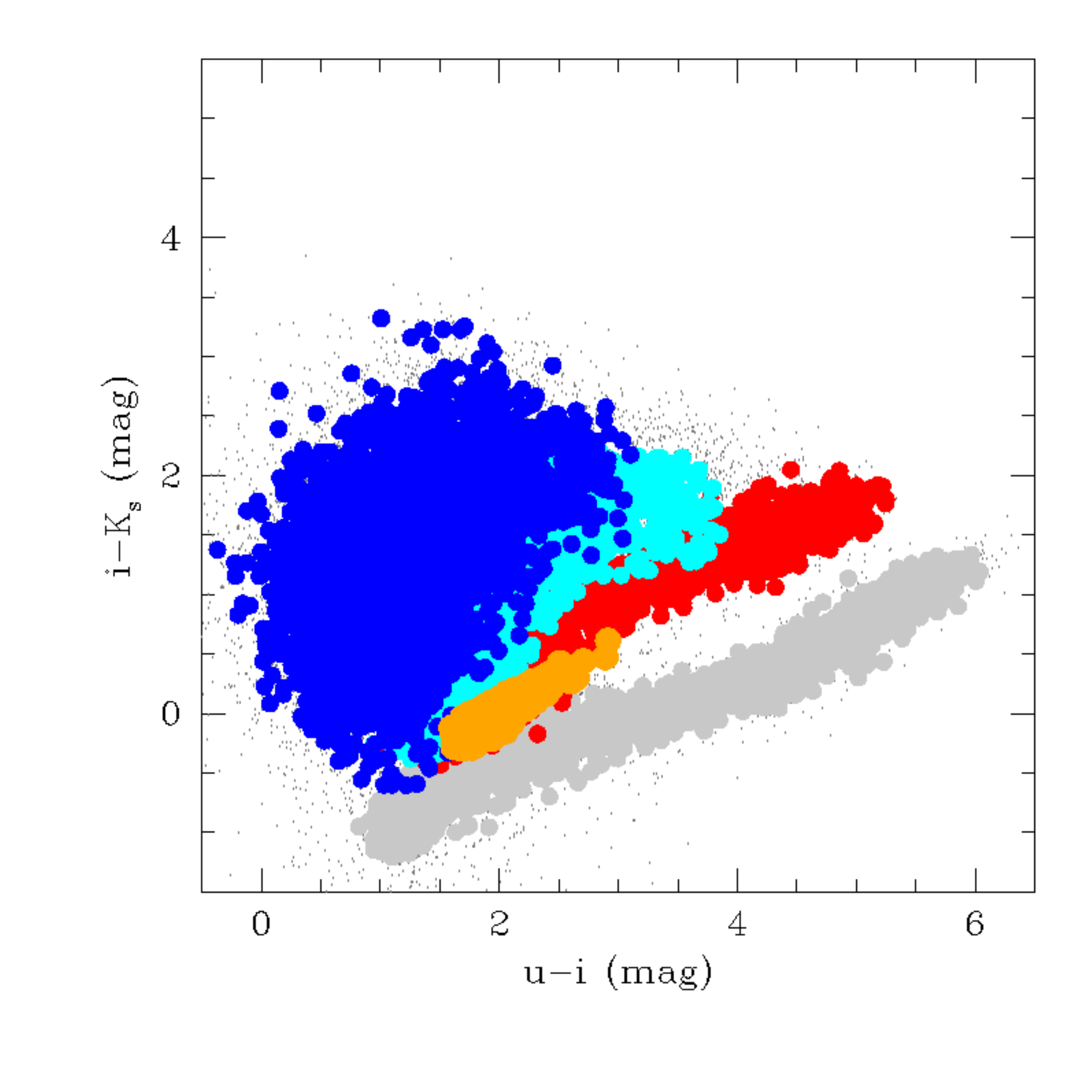}
  \includegraphics[width=0.33\hsize]{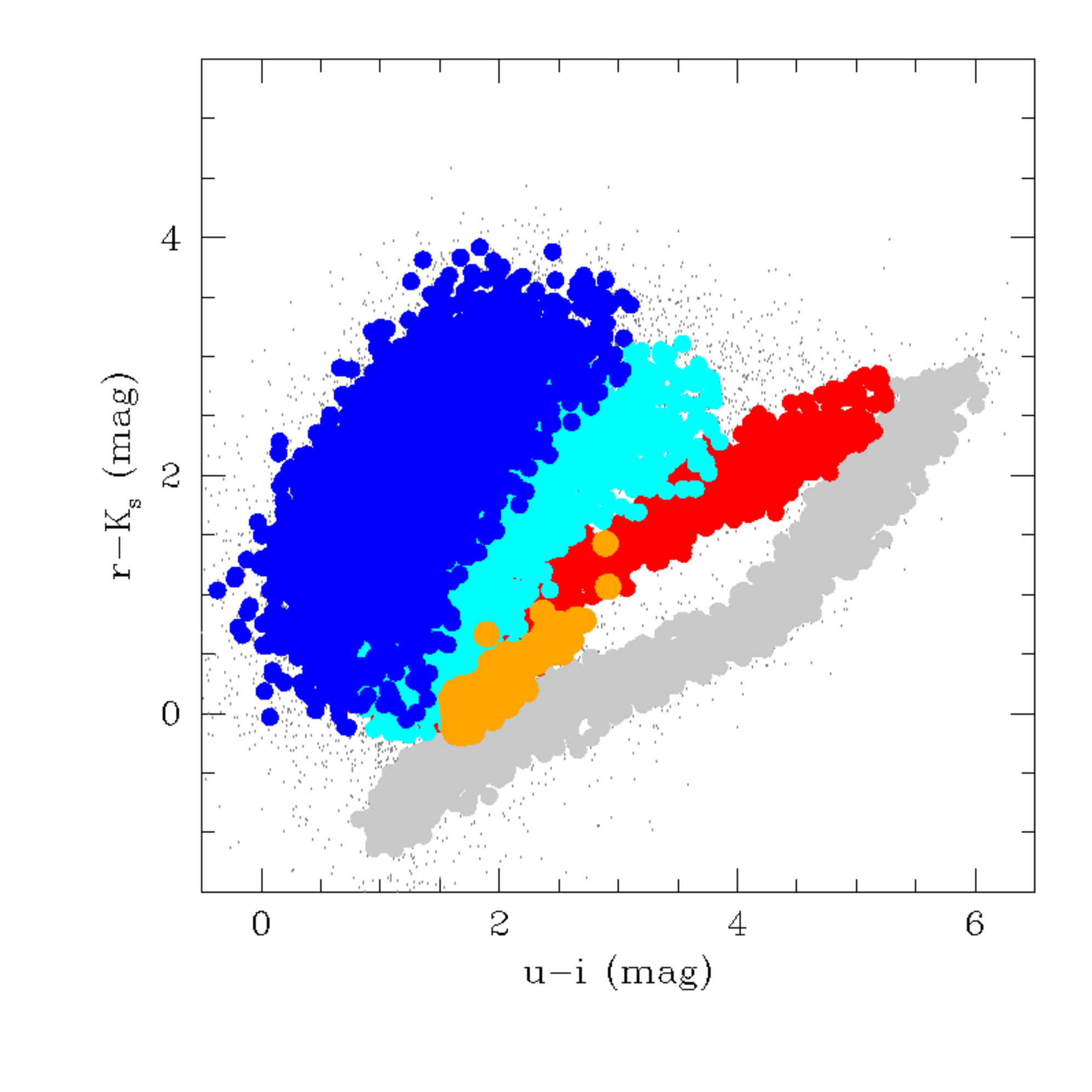}
  \includegraphics[width=0.33\hsize]{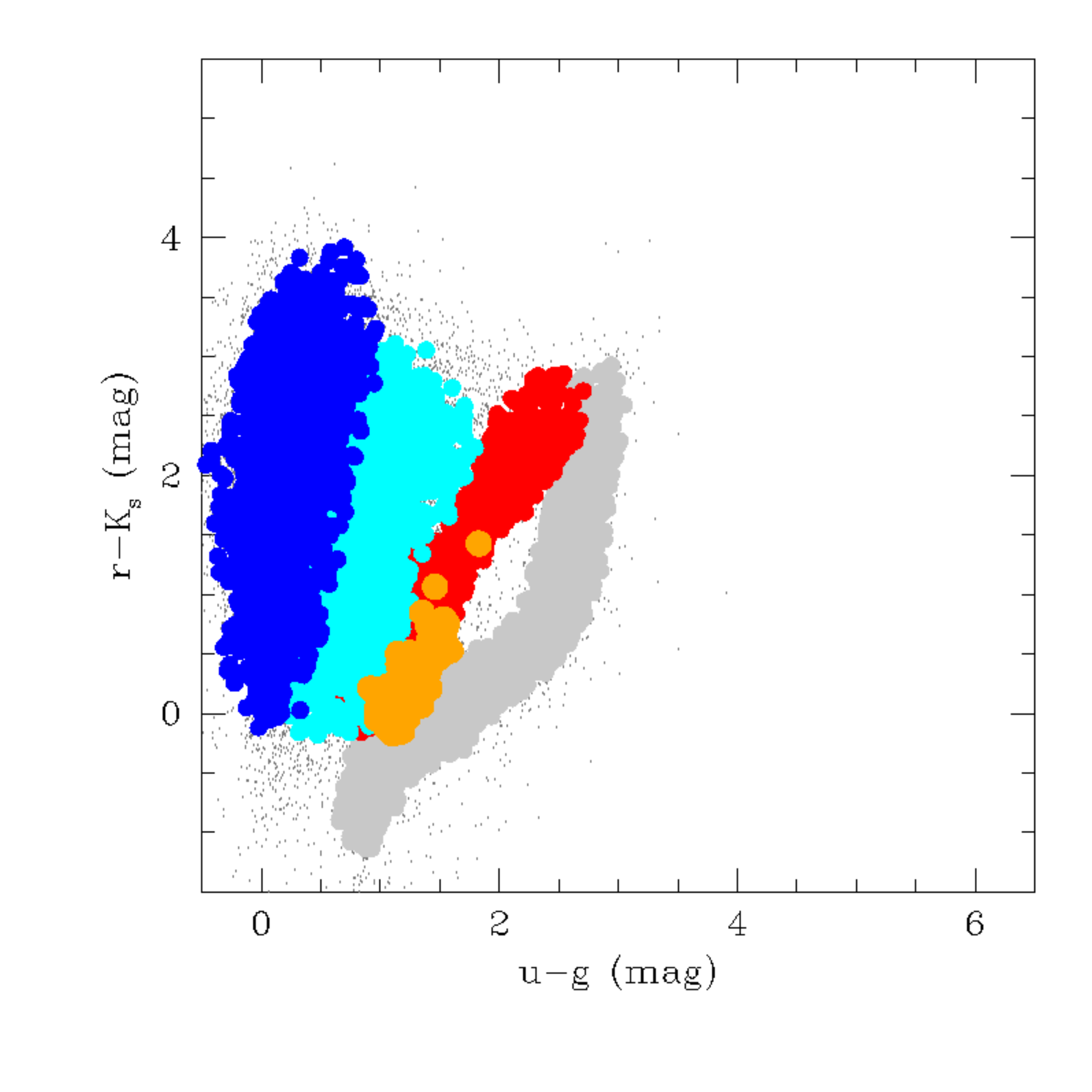}
  \includegraphics[width=0.33\hsize]{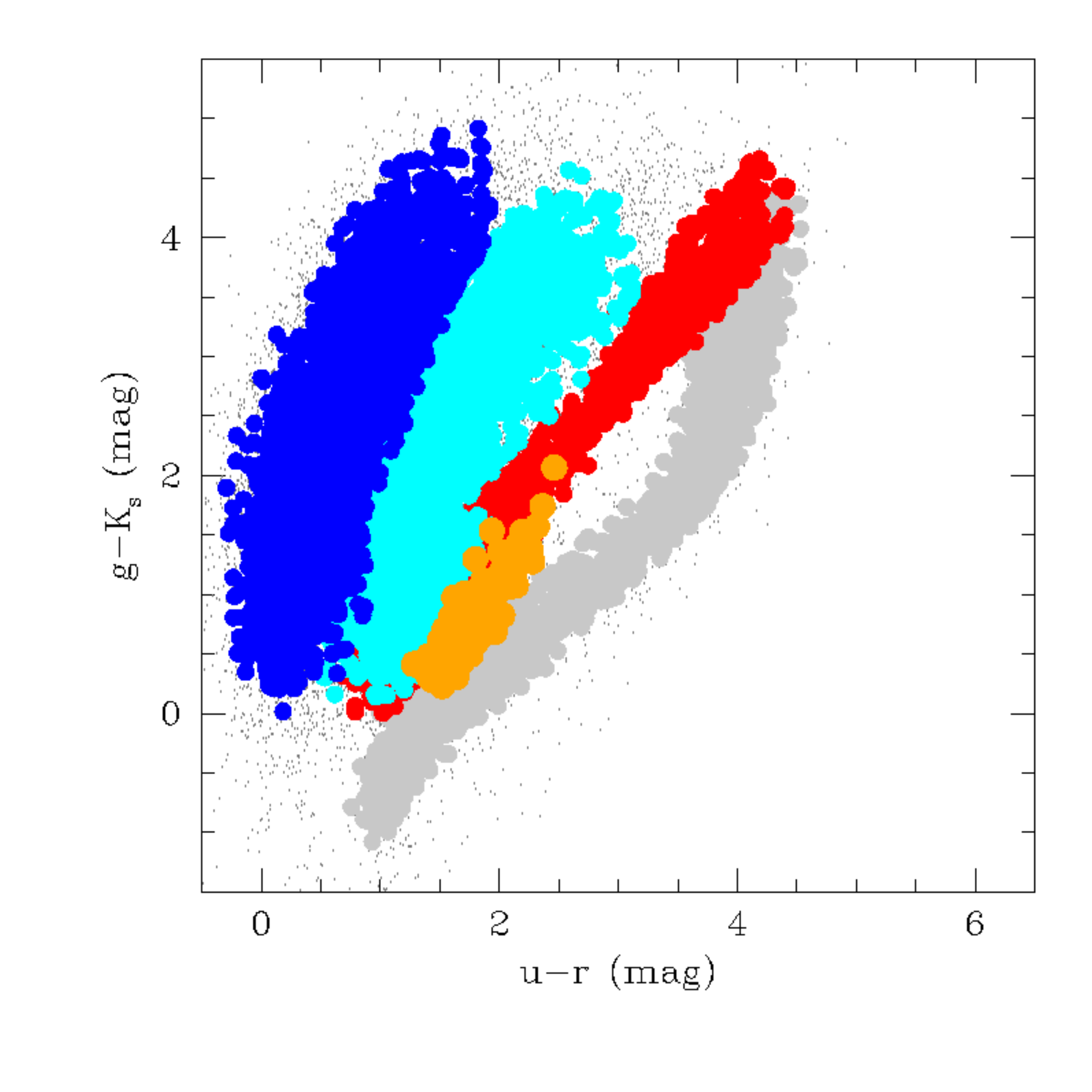}
  \includegraphics[width=0.33\hsize]{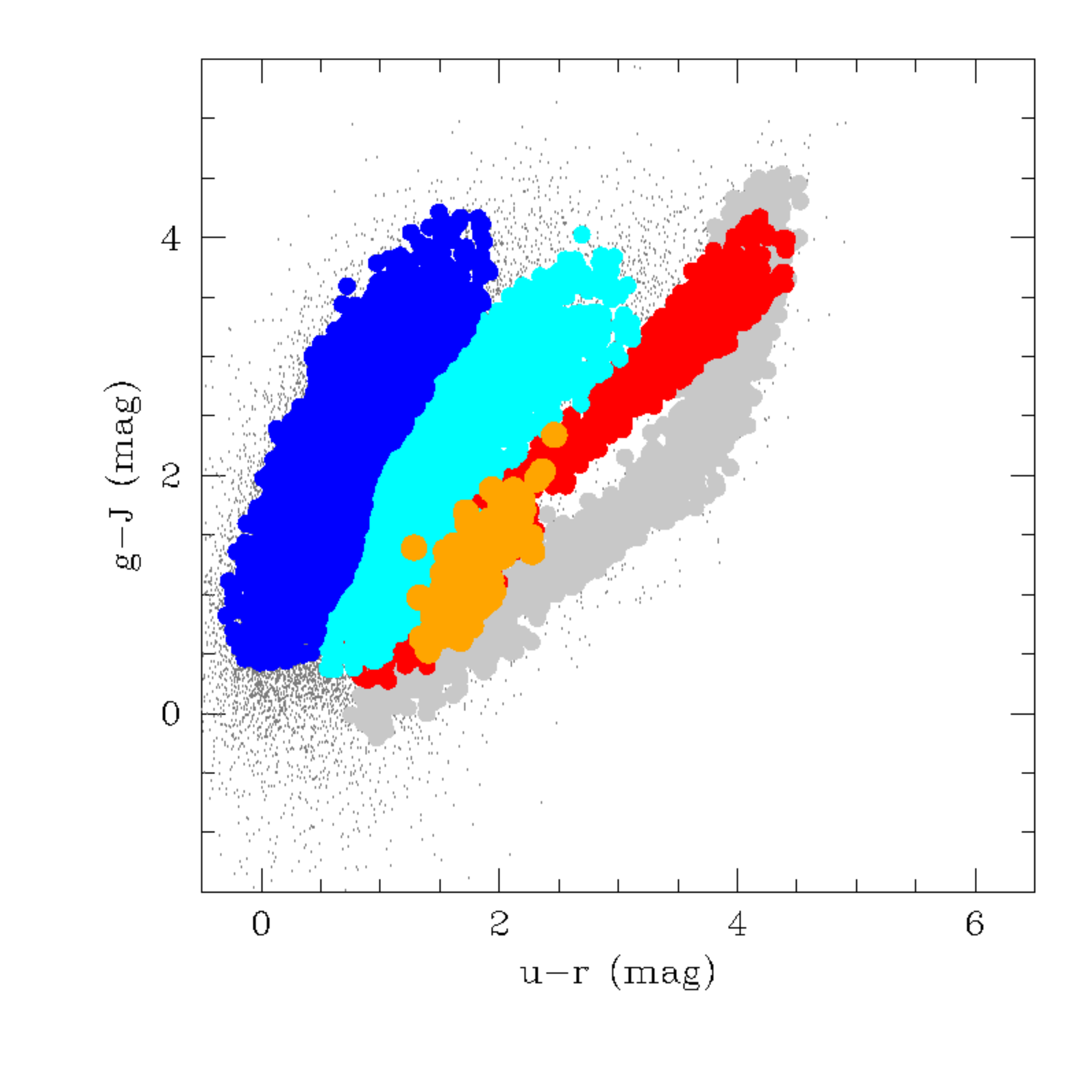}
  \includegraphics[width=0.33\hsize]{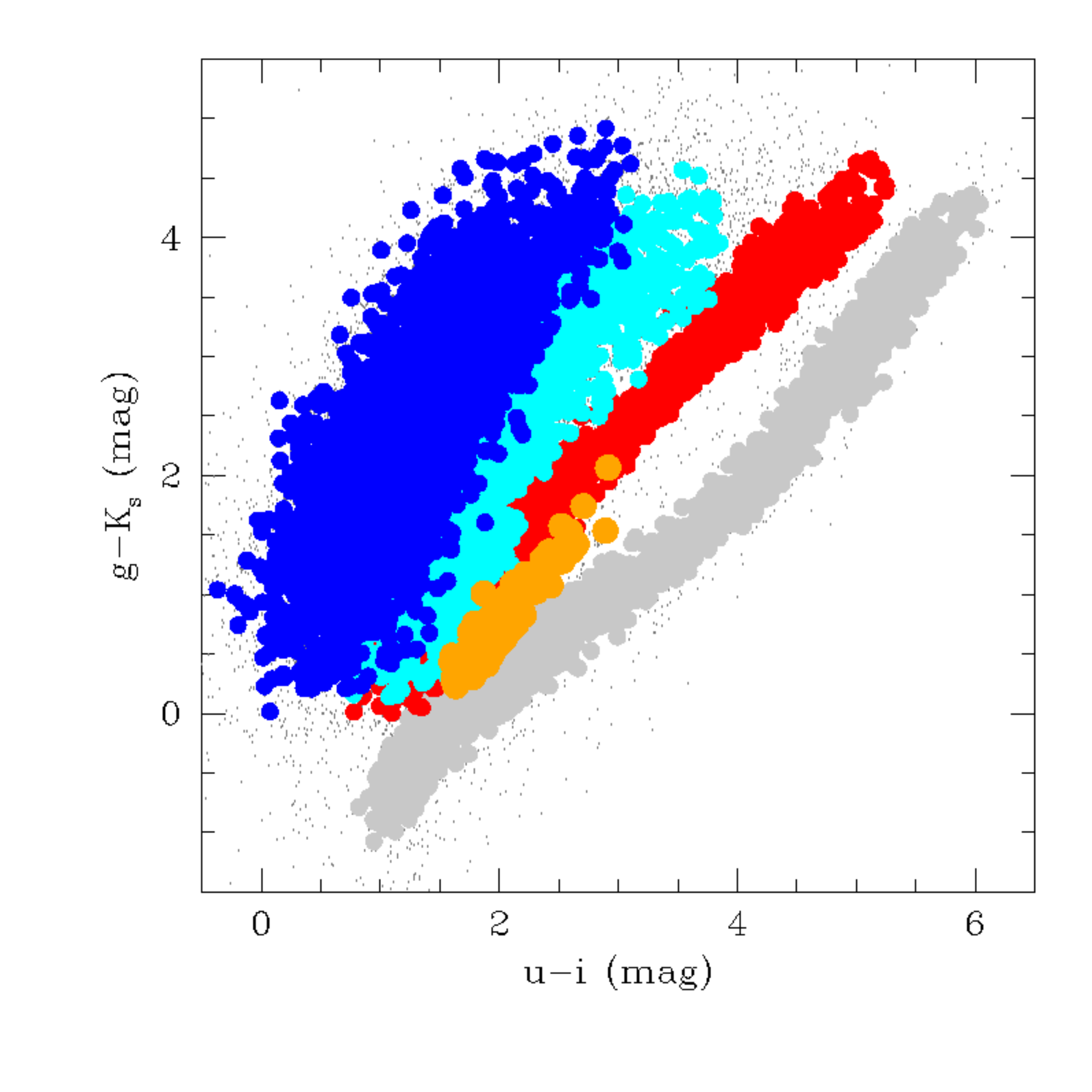}
  \includegraphics[width=0.33\hsize]{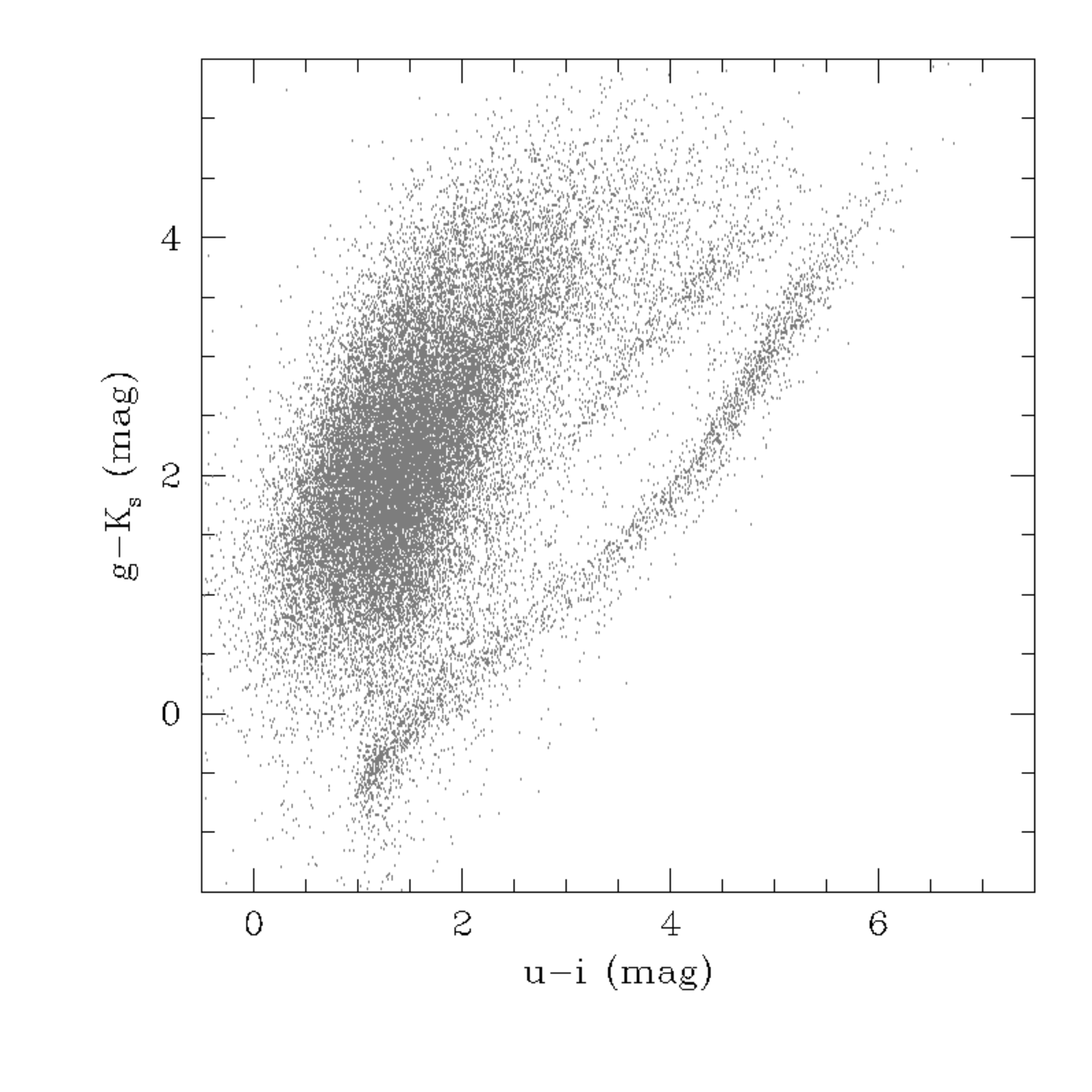}
  \includegraphics[width=0.33\hsize]{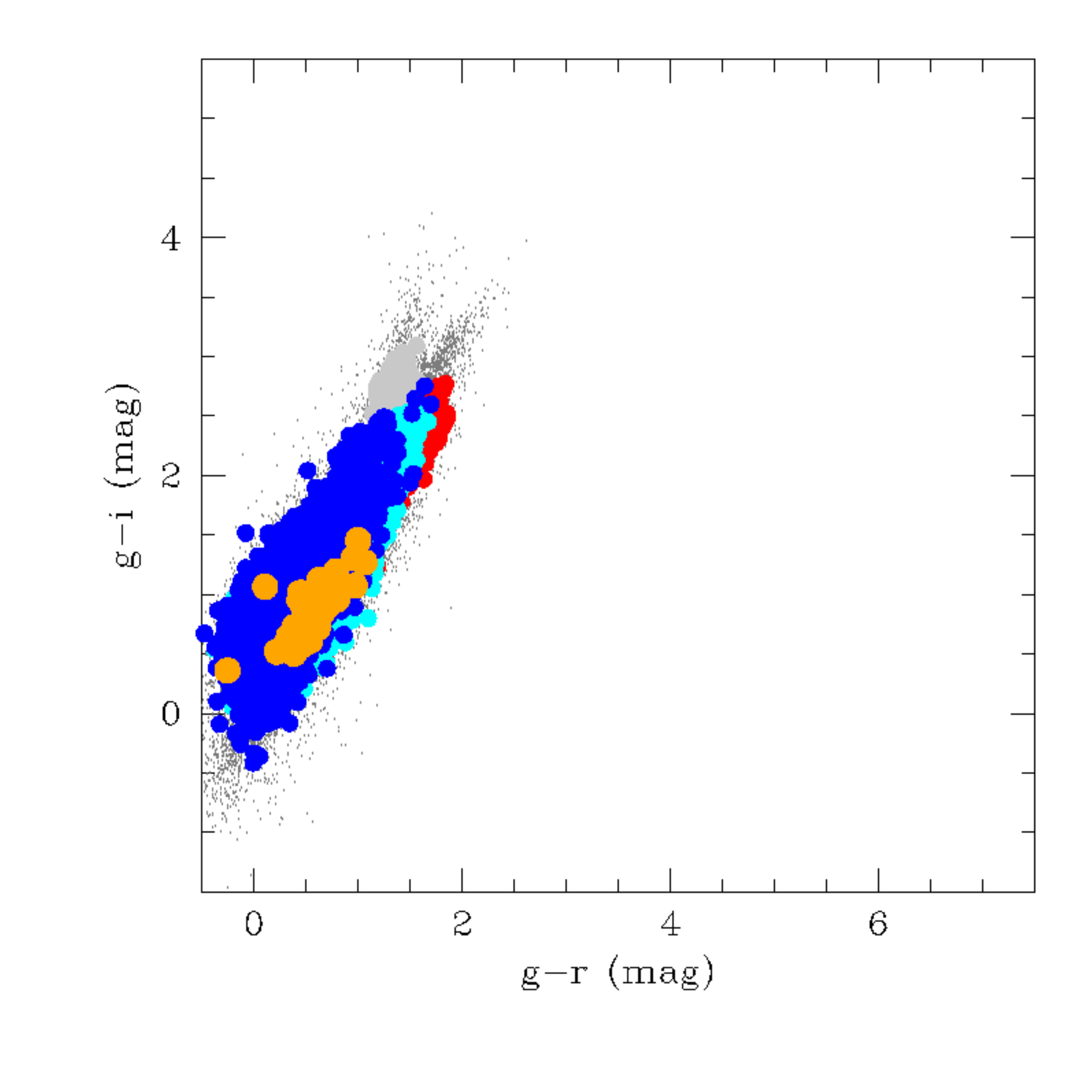}
   \caption{Color-color diagrams with the various sequences
     highlighted in different colors. The full sample of matched
     sources is indicated with black
     dots. Orange, gray, red, light blue, and blue dots indicate the approximate
     loci of GCs, MW stars, passive galaxies, blue galaxies, and star-forming
     galaxies.  The left and middle panels in the lowermost row show
     the same color-color diagram, although in the second row we do not use
     colors for the sequences identified. The total overlap of
     the identified sequences in the lower right panel is notable.}
    \label{marina}
 \end{figure*}

  The large wavelength coverage of our dataset provides enough
  leverage for evidencing the presence of several sequences in the
  color-color diagrams inspected. Because of the different properties
  of the observed targets, some sequences merge or are relatively
  well separated with respect to others, depending on the color-color
  plane inspected. Using the same color-color selection procedure
  described in Section \ref{sec_sel} for selecting GC candidates, we
  derived the approximate loci of the various sequences observed in
  the color-color diagrams. As an example, candidate stars were
  identified from the $uiK_s$ diagram. Then, the identified stars were
  analyzed in other color-color panels, and all sources scattered
  around the sequence were removed from the list. In a similar way, we
  identified the loci for passive galaxies (red early-type
  candidates), blue galaxies (e.g., spirals), star-forming
  (e.g., irregular galaxies) beyond the stellar and GCs loci.  A
  selection of the color-color diagrams used is reported in Figure
  \ref{marina}. In the figure we plot full sample of matched sources
  (black dots), and highlight with different colors five different
  sequences: orange, gray, red, light blue, blue filled circles indicate,
  respectively, the approximate loci of GCs, MW stars, passive
  galaxies, blue galaxies, and star-forming galaxies. To show how sequences
  appear before separating these sequences, we also plotted one of the
  color-color panels, $(g{-}K_S)$ versus \ui, with and without the
  sequences defined.  Also, to highlight the limits of color selection
  when a narrow range of wavelengths is available, the third panel in
  the lower row shows the sequences for the $g{-}i$ versus $g{-}r$
  diagram.

\end{appendix}  

\end{document}